\documentclass[12pt]{article}

\usepackage{mathrsfs}
\usepackage[T1]{fontenc}
\usepackage{mathpazo}
\usepackage{setspace}
\usepackage{amsfonts}
\usepackage{amssymb}
\usepackage{amsmath}
\usepackage{epsfig}
\usepackage{latexsym}
\usepackage{color}
\usepackage{graphicx}
\usepackage{nicefrac}
\usepackage[latin1]{inputenc}
\usepackage{pstricks}
\usepackage{slashed}
\newcommand{\bea}{\begin{eqnarray}}
\newcommand{\eea}{\end{eqnarray}}
\newcommand{\be}{\begin{equation}}
\newcommand{\ee}{\end{equation}}
\newcommand{\beq}{\begin{equation}}
\newcommand{\eeq}[1]{\label{#1}\end{equation}}
\newcommand{\ber}{\begin{eqnarray}}
\newcommand{\eer}[1]{\label{#1}\end{eqnarray}}
\newcommand{\eqn}[1]{(\ref{#1})}

\def\hybrid{\topmargin -20pt    \oddsidemargin 0pt
        \headheight 0pt \headsep 0pt
        \textwidth 6.25in       % A4 paper
        \textheight 9.5in       % A4 paper
        \marginparwidth .875in
        \parskip 5pt plus 1pt   \jot = 1.5ex}

\hybrid

\catcode`\@=11

\def\marginnote#1{}
\newcount\hour
\newcount\minute
\newtoks\amorpm
\hour=\time\divide\hour by60 \minute=\time{\multiply\hour by60
\global\advance\minute by-\hour}
\edef\standardtime{{\ifnum\hour<12 \global\amorpm={am}%
        \else\global\amorpm={pm}\advance\hour by-12 \fi
        \ifnum\hour=0 \hour=12 \fi
        \number\hour:\ifnum\minute<10 0\fi\number\minute\the\amorpm}}
\edef\militarytime{\number\hour:\ifnum\minute<10
0\fi\number\minute}
\def\draftlabel#1{{\@bsphack\if@filesw {\let\thepage\relax
   \xdef\@gtempa{\write\@auxout{\string
      \newlabel{#1}{{\@currentlabel}{\thepage}}}}}\@gtempa
   \if@nobreak \ifvmode\nobreak\fi\fi\fi\@esphack}
        \gdef\@eqnlabel{#1}}
\def\@eqnlabel{}
\def\@vacuum{}
\def\draftmarginnote#1{\marginpar{\raggedright\scriptsize\tt#1}}

\def\draft{\oddsidemargin -.5truein
        \def\@oddfoot{\sl preliminary draft \hfil
        \rm\thepage\hfil\sl\today\quad\militarytime}
        \let\@evenfoot\@oddfoot \overfullrule 3pt
        \let\label=\draftlabel
        \let\marginnote=\draftmarginnote
   \def\@eqnnum{(\theequation)\rlap{\kern\marginparsep\tt\@eqnlabel}%
\global\let\@eqnlabel\@vacuum}  }

\def\preprint{\twocolumn\sloppy\flushbottom\parindent 2em
        \leftmargini 2em\leftmarginv .5em\leftmarginvi .5em
        \oddsidemargin -.5in    \evensidemargin -.5in
        \columnsep .4in \footheight 0pt
        \textwidth 10.in        \topmargin  -.4in
        \headheight 12pt \topskip .4in
        \textheight 6.9in \footskip 0pt
        \def\@oddhead{\thepage\hfil\addtocounter{page}{1}\thepage}
        \let\@evenhead\@oddhead \def\@oddfoot{} \def\@evenfoot{} }

\def\numberbysection{\@addtoreset{equation}{section}
        \def\theequation{\thesection.\arabic{equation}}}

\def\underline#1{\relax\ifmmode\@@underline#1\else
        $\@@underline{\hbox{#1}}$\relax\fi}

\def\titlepage{\@restonecolfalse\if@twocolumn\@restonecoltrue\onecolumn
     \else \newpage \fi \thispagestyle{empty}\c@page\z@
        \def\thefootnote{\fnsymbol{footnote}} }

\def\endtitlepage{\if@restonecol\twocolumn \else \newpage \fi
        \def\thefootnote{\arabic{footnote}}
        \setcounter{footnote}{0}}  %\c@footnote\z@ }

\catcode`@=12 \relax

\def\figcap{\section*{Figure Captions\markboth
        {FIGURECAPTIONS}{FIGURECAPTIONS}}\list
        {Figure \arabic{enumi}:\hfill}{\settowidth\labelwidth{Figure
999:}
        \leftmargin\labelwidth
        \advance\leftmargin\labelsep\usecounter{enumi}}}
 \relax
\def\tablecap{\section*{Table Captions\markboth
        {TABLECAPTIONS}{TABLECAPTIONS}}\list
        {Table \arabic{enumi}:\hfill}{\settowidth\labelwidth{Table
999:}
        \leftmargin\labelwidth
        \advance\leftmargin\labelsep\usecounter{enumi}}}
 \relax
\def\reflist{\section*{References\markboth
        {REFLIST}{REFLIST}}\list
        {[\arabic{enumi}]\hfill}{\settowidth\labelwidth{[999]}
        \leftmargin\labelwidth
        \advance\leftmargin\labelsep\usecounter{enumi}}}
 \relax
\makeatletter
\newcounter{pubctr}
\def\publist{\@ifnextchar[{\@publist}{\@@publist}}
\def\@publist[#1]{\list
        {[\arabic{pubctr}]\hfill}{\settowidth\labelwidth{[999]}
        \leftmargin\labelwidth
        \advance\leftmargin\labelsep
        \@nmbrlisttrue\def\@listctr{pubctr}
        \setcounter{pubctr}{#1}\addtocounter{pubctr}{-1}}}
\def\@@publist{\list
        {[\arabic{pubctr}]\hfill}{\settowidth\labelwidth{[999]}
        \leftmargin\labelwidth
        \advance\leftmargin\labelsep
        \@nmbrlisttrue\def\@listctr{pubctr}}}
 \relax
\makeatother
\newskip\humongous \humongous=0pt plus 1000pt minus 1000pt

\newif\ifdtup

\relax

\def\be{\begin{equation}}
\def\ee{\end{equation}}
\def\ba{\begin{eqnarray}}
\def\ea{\end{eqnarray}}

\renewcommand{\theequation}{\thesection.\arabic{equation}}

\author{
  \begin{minipage}{.97\linewidth}
    \vspace{0cm}
    \begin{center}
      \begin{small}
        \textbf{Ioannis Bakas}\footnote{bakas@ajax.physics.upatras.gr} ${\ }^1$,
        \textbf{Fran\c{c}ois Bourliot}\footnote{bourliot@cpht.polytechnique.fr} ${\ }^2$,
        \textbf{Dieter L\"ust}\footnote{dieter.luest@lmu.de} ${\ }^{3,4}$ and
         \textbf{Marios Petropoulos}\footnote{marios@cpht.polytechnique.fr} ${\ }^2$
      \end{small}
    \end{center}
    \vspace{0.5cm}
    \hspace{2cm}\begin{minipage}{.7\linewidth}
     {\it \begin{footnotesize}
    \begin{itemize}
      \item[${}^1$] Department of Physics, University of Patras\\
       26500 Patras, Greece
      \item[${}^2$] Centre de Physique Th\'eorique\\
        Ecole Polytechnique,  CNRS UMR 7644\\
        91128 Palaiseau Cedex, France
        \item[${}^3$] Max-Planck-Institut f\"ur Physik\\
       F\"ohringer Ring 6, 80805 M\"unchen, Germany
                    \item[${}^4$] Arnold-Sommerfeld-Center f\"ur Theoretische Physik\\
        Department f\"ur Physik, Ludwig-Maximilians-Universit\"at M\"unchen\\
        Theresienstra\ss e 37, 80333 M\"unchen, Germany
        \end{itemize}
     \end{footnotesize}}
    \end{minipage}
    \vspace{0.5cm}
  \end{minipage}
}

\date{\today}

\title{\vspace{0.1cm}
 \boldmath \begin{large}
    \textbf{
    GEOMETRIC FLOWS IN HO\v{R}AVA--LIFSHITZ GRAVITY
    }
  \end{large} \unboldmath
}

\begin{document}

\renewcommand{\thepage}{\arabic{page}}
\setcounter{page}{1}
%THIS IS PAGE 1 (INSERT TEXT OF REPORT HERE)

%%%%%%%%%%%%%%%%%%%%%%%%%%%%%%%%%%%%%%%%%%%%%%%%%%%%%

\begin{titlepage}
  \maketitle
  \thispagestyle{empty}

  \vspace{-14.5cm}
  \begin{flushright}
    CPHT-RR078.0709\\
  MPP-2009-179\\
    LMU-ASC 47/09
  \end{flushright}

  \vspace{11cm}

  \begin{center}
    \textsc{Abstract}\\
  \end{center}
We consider instanton solutions of Euclidean Ho\v{r}ava--Lifshitz gravity in
four dimensions satisfying the detailed balance condition. They are described
by geometric flows in three dimensions driven by certain combinations of the
Cotton and Ricci tensors as well as the cosmological-constant term. The deformation
curvature terms can have competing behavior leading to a variety of fixed points.
The instantons interpolate between any two fixed points, which are vacua of
topologically massive gravity with $\Lambda > 0$, and their
action is finite. Special emphasis is placed on configurations with $SU(2)$
isometry associated with homogeneous but generally non-isotropic Bianchi IX model
geometries. In this case, the combined Ricci--Cotton flow reduces
to an autonomous system of ordinary differential equations whose properties are
studied in detail for different couplings. The occurrence and stability of
isotropic and anisotropic fixed points are investigated analytically
and some exact solutions are obtained. The corresponding instantons are classified
and they are all globally $\mathbb{R} \times S^3$ and complete spaces. Generalizations
to higher-dimensional gravities are also briefly discussed.

\end{titlepage}

\vskip1cm

\tableofcontents

%\newpage

\section{Introduction}
\setcounter{equation}{0}

Based on ideas that were originally developed  in condensed matter physics \cite{lifshitz:1941}
and later applied to the description of aspects of particle interactions
\cite{Chadha:1982qq, Iliopoulos:1980zd, Antoniadis:1983ek, Petrini:1997kk}, a modification
of general relativity was recently proposed in \cite{Horava:2008ih, Horava:2009uw}
and further studied under the name of Ho\v{r}ava--Lifshitz gravity. In this theory, which
includes higher-order curvature terms on spatial slices,
the diffeomorphism invariance of general relativity is broken explicitly setting a
privileged time direction. This affects the ultra-violet behavior of the quantum theory,
which, hence, looks power-counting renormalizable. Some efforts have been made to prove
consistency of the quantum theory \cite{Orlando:2009en, wu},
but  the number of propagating degrees of freedom seems to invalidate the matching
with Einstein's gravity in the infrared regime, and hence seems to disprove this theory
as a viable alternative to general relativity \cite{Charmousis:2009tc, pang, blas, koyama}.
The investigation of these issues is still going on. Despite the difficulties and
reservations, the Ho\v{r}ava--Lifshitz gravity still provides an interesting classical and
quantum field theory framework, where one can address some interesting
questions and explore several connections to ordinary gravity or string theory. These
also include the appearance and relevance of geometric flows, which is the main subject
of the present work.

Geometric flows, and, in particular, Ricci flows are interesting in their own right. In
mathematics they turned out to play a crucial role in implementing Hamilton's program for
proving Poincar\'e's and Thurston's conjectures \cite{hamilt, perel} (but see also
\cite{yau} and \cite{tian} and references therein). In physics they originally appeared in
off-critical string theory via the renormalization-group equations of two-dimensional
non-linear sigma models, where the evolution of the metric under the Ricci flow equations
provides the running of the bulk coupling to lowest order in perturbation theory
(see \cite{Friedan:1980jm} for the original result). In this context,
the renormalization-group time is provided by the logarithmic length scale of the world-sheet,
but in some cases it can also assume the role of genuine time, describing real-time evolution in
string theory in regimes where the friction due to the motion of the dilaton effectively
reduces the second-order evolution equations to the first-order renormalization-group flow
equations \cite{Schmidhuber:1994bv, Bakas:2006bz}.

Ricci flow models also appear in the framework of four-dimensional gravitational instantons
of general relativity.  Solving Einstein's equations is, in general, an impossible task. It is
substantially simplified under the assumption of self-duality as a sufficient condition to find
vacuum solutions in the Euclidean sector of the theory. Homogeneity of spatial sections is
often a further simplification to find explicit solutions.
Although what we call space is somewhat arbitrary in Euclidean gravity,
the latter statement can be made precise by assuming a foliation in three-dimensional leaves
that are invariant under an isometry group of motions. For these particular vacuum solutions,
it turns out that the Euclidean time evolution of the homogeneous leaves inside the gravitational
instanton can be recast as Ricci flow equations for the corresponding geometry on the homogeneous
model spaces \cite{Sfetsos:2006,Bakas:2006bz,Bourliot:2009fr}.

The modification of gravity proposed by Ho\v{r}ava in \cite{Horava:2008ih, Horava:2009uw}
shares some features with the previous setting that allow to define the analogue of
gravitational-instanton configurations. In particular, a foliation
of the four-dimensional space is assumed from the very beginning with a privileged time direction
at the level of the action. Furthermore, a condition called \emph{detailed balance}, which is
borrowed from non-equilibrium thermodynamics, requires that the dynamics follows from an
appropriately chosen three-dimensional superpotential action. In the Euclidean version of the theory,
this resembles the self-duality condition with similar consequences: for a class of configurations
that minimize the action, the time evolution becomes  first-order and describes a geometric flow
on the leaves of the foliation. Unlike general relativity, where Ricci flow is equivalent to self-duality
only for
configurations with homogeneous leaves, the description of instanton-like solutions by geometric
flows is generic in Ho\v{r}ava--Lifshitz gravity. The nature of the corresponding flow depends
on the choice of the three-dimensional action used for detailed balance and it is
driven, in general, by a certain combination of Ricci and Cotton tensors as well as the
cosmological constant term. It is not our concern, and it will not be pursued here at all,
to find whether such combinations of curvature tensors can also arise from the renormalization-
group equations of some quantum field theory.

Our aim in this paper is to investigate aspects of the Euclidean dynamics of Ho\v{r}ava--Lifshitz
gravity with detailed balance. Our motivations are diverse. First, classical instanton-like
solutions are important for the determination of transition amplitudes in quantum gravity. They
are also useful in the Hartle--Hawking formulation of quantum cosmology \cite{harhaw},
even though classical
cosmology per se requires the analysis of real-time equations. The geometric flows that
emerge in this framework involve tensors with higher-order spatial derivatives terms, and, as
such, they are new in the literature; they reduce to previously studied examples only for some
special values of their parameters. Thus, it is instructive to formulate the flow equations
in all generality, determine the nature of the fixed points and their stability properties,
obtain explicit solutions, as well as study general questions such as the monotonicity of the
evolution, the possible formation of singularities, the occurrence of bounces and so on.
These questions arise naturally in the
general theory of geometric flows and they are bound to be relevant for the space--time
interpretation of the analogue of gravitational instantons in Ho\v{r}ava--Lifshitz theory.

The answer to these questions will be accomplished partially using some ansatz
for the underlying
three-dimensional spaces, leading to mini-superspace truncation of the flow equations.
Otherwise, it does not be seem possible to draw general conclusions for the general
system of equations, at least at the current level of our understanding of this problem.
Still, the results that will be described are indicative of what should be expected
in general. More systematic investigation of the
infinite-dimensional dynamical system at hand requires substantial mathematical work that
is not contained in this paper. Following the paradigm of gravitational instantons
with isometry groups in Euclidean Einstein gravity, we will consider homogeneous geometries
on the three-dimensional spatial slices of Ho\v{r}ava--Lifshitz gravity and focus,
in particular, to the case of Bianchi IX geometry as a class of homogeneous
but generally non-isotropic deformation of $S^3$; this model is often referred to as
mixmaster universe in the Lorentzian (real-time) approach to cosmology
\cite{Misner:1969, bkl, Barrow:1981sx}, and was recently discussed in the framework of
Ho\v{r}ava--Lifshitz gravity \cite{Bakas:2009ku, Myung:2009if}.
Other Bianchi classes as well as more general inhomogeneous deformations of the three-sphere
(under appropriate ansatz) can also be studied along similar lines, but they will not be
discussed.

We set up the general problem using the Bianchi IX model geometry and study in detail
some specific examples of the flow for different couplings in Ho\v{r}ava--Lifshitz gravity.
Even in this
case the resulting equations in mini-superspace are not easily tractable for generic values
of the couplings. First, we will consider the Ricci flow and some of its variants that
describe solutions of the modified theory of gravity with anisotropy scaling
parameter $z=2$ (see next section for this and other definitions) and compare them to
instantons with $SU(2)$ isometry in ordinary gravity. We will also consider the Cotton flow,
separately, and use it to construct solutions of Ho\v{r}ava--Lifshitz theory with anisotropy
scaling parameter $z=3$ by ignoring all Ricci curvature terms that become subdominant
when the volume of $S^3$ is very small. We will also consider the combined Ricci--Cotton flow
and explore the equations in detail first in the limit that the speed of light vanishes
or equivalently Newton's constant becomes infinite (it is often called Carroll limit after \cite{Carroll}) .
The normalized Ricci--Cotton flow and the unnormalized
variant of it with vanishing cosmological constant provide the relevant equations in this
limit. Finally, we will consider the general Ricci--Cotton flow with arbitrary couplings and
obtain several qualitative results for its solutions.

In all cases it is assumed that the parameter of the superspace metric of the theory is
restricted to values $\lambda < 1/3$ (in which case the cosmological constant will also be
taken non-negative) so that the flow equations extremize the classical action, up to
important boundary terms. Proper account of the boundary terms leads to the definition of
instanton solutions as finite-action trajectories that interpolate between fixed points. Our
analysis shows that the Ricci and the Cotton tensor terms can compete with each other, and,
depending on the relative sign between the two, the flow equations can exhibit symmetric as well
as anisotropic fixed points. The nature of these fixed points and their stability properties
also have implications for the space--time interpretation of the corresponding gravitational
instanton solutions. Axisymmetric solutions are associated with spaces with
$SU(2) \times U(1)$ isometries, and, hence they are easier to describe in closed form.

Instanton solutions of Ho\v{r}ava--Lifshitz gravity, as they are defined, are rather
special configurations that rely on the existence of multiple-degenerate vacua and correspond
to special flow lines, which guarantee finiteness of their action. As it will turn out, they
are also free of singularities and their space--time metrics are regular and complete.
Note, however, that other flow lines, possibly with infinite action, also describe solutions
of the second-order equations of motion, but they may have singularities. Although we are
primarily interested in the class of instanton solutions, one should be open-minded for other
more general possibilities too. For this
reason, as well as for mathematical completeness, we will investigate the phase portraits
of the flow equations in all generality. The selection of special trajectories that
correspond to instantons will be made much later together with their space--time
interpretation.

In section 2, we first briefly review the formulation of Ho\v{r}ava--Lifshitz gravity with
detailed balance condition putting emphasis on the structure of the potential term and its
associated superpotential. This analysis is then carried to the Euclidean regime where
``zero-energy'' (i.e., self-dual-like) configurations exist satisfying the flow equations.
Restrictions on the parameter $\lambda$ are also obtained together with an entropy
functional that changes monotonically along the flow lines. The results are then used to
define instanton solutions as in ordinary point particle systems. The Bianchi IX model
geometry is introduced in section 3 where the truncation of the Ricci and Cotton flows are
studied separately in detail. Section 4 is entirely devoted to the analysis of the
normalized Ricci--Cotton flow and the explicit construction of its axisymmetric solutions.
Section 5 discusses the case of unnormalized Ricci--Cotton flow obtained for general
couplings. It contains the case study of positive and zero cosmological constant for
axisymmetric configurations with $\lambda < 1/3$.
Section 6 is devoted to the space--time interpretation of the flow line, as gravitational
instantons, making comparisons with the analogous instanton solutions arising in
general relativity. Complete classification of all gravitational-instanton metrics
with $SU(2)$ isometry is also obtained. In section 7 we outline generalizations of the
framework to higher-dimensional Ho\v{r}ava--Lifshitz gravities, which, for instance, in $4+1$
dimensions give rise to a new system of flow equations on four-manifolds driven by the Bach
tensor. Finally, section 8 contains our conclusions and poses several questions for future work.

\section{Non-relativistic gravity, detailed balance and flows}
\setcounter{equation}{0}

\subsection{Non-relativistic gravity: a reminder}

The theory of non-relativistic gravity developed in \cite{Horava:2008ih, Horava:2009uw} is
valid for general space--time dimension $D+1$. It has three main features:
\begin{itemize}
\item Space-time is assumed to be topologically $\mathcal{M}_{D+1}=\mathbb{R} \times\mathcal{M}_D$,
leading to a natural codimension-one foliation. Diffeormorphism invariance is broken down to the
subgroup of foliation-preserving transformations. This breaking is controlled by a parameter $\lambda$.
\item Scaling properties of space and time are different and captured by an integer $z$. Power
counting renormalizability of the theory requires $z=D$.
\item The interactions are determined by a detailed balance condition following from a Euclidean
$D$-dimensional diffeomorphism-invariant action, which gives rise to marginal and relevant terms
in $D+1$ dimensions.
\end{itemize}

\noindent
The last item above is not generic in Ho\v{r}ava--Lifshitz gravity and it can be relaxed by allowing
more arbitrary coefficients for the various marginal and relevant terms. However, it is a necessary
ingredient in our study to connect it naturally with the theory of geometric flows. Thus, detailed
balance will be assumed in the following.

Let us adopt the ADM (Arnowitt--Deser--Misner) decomposition of the metric (see, for instance,
\cite{misner6}), which
is suitable for the $D+1$ foliation of space--time,
\begin{equation}
\label{genmet}
\mathrm{d}s^2 = -N^2 \mathrm{d}t^2 +g_{ij}\left(\mathrm{d}x^i +N^i \mathrm{d}t\right)
\left(\mathrm{d}x^j +N^j \mathrm{d}t\right) ,
\end{equation}
where $N^i$ and $N$ are the shift and lapse functions respectively. Here, $i,j,\ldots$ run in $D$
dimensions and all tensors that will appear in the following are $D$-dimensional.

Using this decomposition, the Einstein--Hilbert action in $D+1$ dimensions (up to total derivative
terms that may contribute in topologically non-trivial spaces) reads as follows,
\begin{equation}
\label{SEH}
S_{\mathrm{EH}}=\frac{1}{16 \pi G_{\mathrm{N}}}\int\mathrm{d}^{D+1} x\,\sqrt{g}N\left(K_{ij}K^{ij}
-K^2+R-2\Lambda\right) ,
\end{equation}
where $\Lambda$ is the genuine cosmological constant in $D+1$ dimensions. In this expression,
$K_{ij}$ is the second fundamental form that measures the extrinsic curvature of the leaves at
constant $t$,
\begin{equation}
\label{kext}
K_{ij}=\frac{1}{2N}\left(\partial _t g_{ij}-\nabla_iN_j-\nabla_jN_i\right) ,
\end{equation}
and its trace $K = g^{ij}K_{ij}$ is the mean curvature. The first two terms in equation (\ref{SEH})
provide the kinetic energy, since they include time derivatives of the field $g_{ij}$. Their specific
combination can be recast in the form
\begin{equation}
\label{EHkin}
K_{ij}G^{ijk\ell}_{\mathrm{DW}}
K_{k\ell} =K_{ij}K^{ij}
-K^2
\end{equation}
using the DeWitt metric in superspace
\begin{equation}
\label{dwm-orig}
G^{ijk\ell}_{\mathrm{DW}}=\frac{1}{2}\left(g^{ik}g^{j\ell}+g^{i\ell}g^{jk}\right)-
g^{ij}g^{k\ell} .
\end{equation}
The potential term in Einstein gravity is provided by the three-dimensional Ricci scalar
curvature $R$ and the four-dimensional cosmological constant term $\Lambda$ (when it is
present), as shown in (\ref{SEH}).

In non-relativistic gravity, space and time scale as  $[t]=-z$, $[x]=-1$ and it is further
assumed\footnote{Proper restoration of the speed of light, which scales as
$[c]=z-1$, explains the various dimensions, as described in detail in the original works
\cite{Horava:2008ih, Horava:2009uw}.} that $[N_i]=z-1$, $[N]=0$ and $[g_{ij}] = 0$
so that $[K^2]=2z$; it should be contrasted to general
relativity where space and time scale the same with $z=1$. This asymmetry is further
implemented in the action, both in the kinetic and the potential terms by requiring
foliation preserving covariance. The kinetic term is generalized as
\begin{equation}
\label{GENkin}
S_{\mathrm{K}}=\frac{2}{\kappa^2}\int \mathrm{d}t\,\mathrm{d}^D x\,\sqrt{g}N K_{ij}G^{ijk\ell}
K_{k\ell} =  \frac{2}{\kappa^2}\int \mathrm{d}t\,\mathrm{d}^D x\,\sqrt{g}N\left(K_{ij}K^{ij}
-\lambda K^2\right),
\end{equation}
where
$\left[\kappa^2\right]=z-D$ and  $\lambda$ is a dimensionless coupling measuring the breaking of the
full diffeomorphism group. Here, $G^{ijk\ell}$ is the generalized metric in superspace
\begin{equation}
\label{dwm}
G^{ijk\ell}=\frac{1}{2}\left(g^{ik}g^{j\ell}+g^{i\ell}g^{jk}\right)-\lambda
g^{ij}g^{k\ell}.
\end{equation}
that coincides with the DeWitt metric when $\lambda=1$.
It is worth stressing that this metric can be positive-definite or indefinite depending on
$\lambda$.
Indeed, $g_{k\ell}$ is an ``eigenvector'',
\begin{equation}
%\label{}
G^{ijk\ell}g_{k\ell} = (1-\lambda D)g^{ij} ~,
\end{equation}
with eigenvalue  $1-\lambda D$. The sign of the latter changes at $\lambda =1/D$ where the
inverse no longer exists. Thus, for $\lambda < 1/D$ the metric is positive-definite and it
becomes indefinite for all $\lambda > 1/D$ that include, in particular, $\lambda = 1$.
This behavior and the fact that $\lambda$ is ultimately an unprotected parameter of the theory,
which, in principle can take any real value, should be kept in mind when considering quantum
corrections.

The potential term of the theory has the general form
\begin{equation}
\label{pot}
S_V=-\int \mathrm{d}t\,\mathrm{d}^Dx\,\sqrt{g}N\,V[g]
\end{equation}
and can also contribute in various ways to the breaking of diffeomorphism invariance.
Note that $[V]=z+D$ and there is a large freedom to choose $V$ so that it includes operators of
dimension less than or equal to $z+D$ (called relevant and marginal operators, respectively).
In order to reduce this freedom and take advantage of the renormalization properties of a
$D$-dimensional system, it was proposed in \cite{Horava:2008ih, Horava:2009uw} to introduce a
detailed balance condition that allows to express the potential in terms of a ``superpotential''
as follows:
\begin{equation}
\label{Vdbc}
V=\frac{\kappa^2}{2} E^{ij}\mathcal{G}_{ijk\ell}
E^{k\ell},
\end{equation}
where
\begin{equation}
\label{eom}
E^{ij}=-\frac{1}{2 \sqrt{g}}\frac{\delta W[g]}{\delta g_{ij}}
\end{equation}
and $W$ a $D$-dimensional action so that $[E^{ij}]=D$. The tensor $\mathcal{G}_{ijk\ell}$ is
defined as
\begin{equation}
\mathcal{G}_{ijk\ell}=\frac{1}{2}\left(g_{ik}g_{j\ell}+g_{i\ell}g_{jk}\right)-
\frac{\lambda}{D\lambda-1} g_{ij}g_{k\ell}
\end{equation}
and coincides with the inverse of the metric in superspace with generic $\lambda$, i.e.,
\begin{equation}
G^{ijk\ell}\mathcal{G}_{k\ell mn}= \frac{1}{2}(\delta^i_m \delta^j_n +\delta^i_n
\delta^j_m ) ~.
\end{equation}

The resulting theory is not invariant under general coordinate transformations of space--time.
Indeed, since $\mathcal{M}_{D+1}$ is topologically $\mathbb{R} \times\mathcal{M}_D$, it is
only appropriate to consider invariance of the action under the restricted class of
foliation-preserving diffeomorphisms,
\begin{equation}
\tilde{t} = \tilde{t} (t) ~, \quad {\tilde{x}}^i = {\tilde{x}}^i (t, x) ~.
\end{equation}
Then, the lapse function $N$ associated with the freedom of time reparametrization is
restricted to be a function of $t$ alone, whereas the shift functions $N_i$
associated with diffeomorphisms of $\mathcal{M}_D$ can depend on all space--time
coordinates. This is often called the {\em projectable case} of Ho\v{r}ava--Lifshitz
gravity and it will be assumed in the following. The non-projectable version of the
theory leads to dynamical inconsistencies \cite{henneaux}.

The choice of $W$ depends on the dimension $D$. Here, we recall the choice for $D=3$ with
$z=3$ that ensures power-counting renormalizability of the four-dimensional theory;
generalization to higher dimensions will be discussed later in section 7. Then, the
marginal operators in question are obtained from the three-dimensional gravitational
Chern--Simons action, which is familiar from topologically massive gravity \cite{cs},
\begin{equation}
\label{gcs}
W_{\mathrm{CS}}=\frac{1}{w_{\mathrm{CS}}}
\int \omega_3(\omega) ~,
\end{equation}
with density given in terms of the connection one-form $\omega$ by
\begin{equation}
\omega_3(\omega)= {1 \over 2} \mathrm{Tr}\left(\omega\wedge \mathrm{d}\omega+
\frac{2}{3}\omega\wedge\omega\wedge\omega\right) .
\end{equation}
The corresponding variation gives
\begin{equation}
\label{cotton}
E^{k\ell}_{\mathrm{CS}}=-\frac{1}{w_{\mathrm{CS}}} \frac{\varepsilon^{ijk}}{\sqrt{g}}
\nabla_i\left(R^{\ell}_j-\frac{1}{4}R\delta^{\ell}_j \right)
\equiv-\frac{1}{w_{\mathrm{CS}}}   C^{k\ell},
\end{equation}
where $C^{k\ell}$ is the Cotton tensor and $\epsilon^{123} =1$. The Cotton tensor is traceless,
conserved and it vanishes identically for conformally flat metrics.

Relevant operators in four dimensions are generated by the Einstein--Hilbert three-dimensional
action
\begin{equation}
W_{\mathrm{EH}}=\frac{2}{\kappa_W^2}\int \mathrm{d}^Dx \sqrt{g} (R-2\Lambda_W).
\end{equation}
Note that neither $\kappa_W^2$ is the four-dimensional Newton's constant nor $\Lambda_W$ is the
four-dimensional cosmological constant, but they will be identified shortly. The variation of this
action leads to
\begin{equation}
\label{riccion}
E^{k\ell}_{\mathrm{EH}}=
\frac{1}{\kappa_W^2}\left(
R^{k\ell} -\frac{R}{2}g^{k\ell}+\Lambda_Wg^{k\ell}\right) .
\end{equation}

Combining the Chern--Simons and Einstein--Hilbert contributions to $E^{k\ell}$, with their
respective couplings, the full potential of Ho\v{r}ava--Lifshitz gravity reads
\begin{eqnarray}
\label{potCSEH}
V&=&
\frac{\kappa^2}{2w_{\mathrm{CS}}^2}
C^{ij}C_{ij} -\frac{\kappa^2}{w_{\mathrm{CS}}\kappa_W^2}
C^{ij}R_{ij}+ \nonumber\\
& & \frac{\kappa^2}{2\kappa_W^4}\left(
R^{ij}R_{ij}-\frac{4\lambda-1}{4(3\lambda-1)}
R^2 \right) +
\nonumber\\
&& \frac{\kappa^2\Lambda_W}{2(3\lambda -1)\kappa_W^4}\left(
R-3\Lambda_W \right).
\end{eqnarray}
The ultra-violet behavior of the resulting theory is dictated by the quadratic Cotton curvature
term, which is marginal with dimension $2z = 6$, and corresponds to $z=3$. It improves a lot
the ultra-violet behavior of ordinary Einstein gravity at the expense of breaking general
covariance of the theory at short distances. When the Cotton term is absent, the resulting
theory has a potential with quadratic Ricci curvature terms that become dominant in the
ultra-violet regime and so $z=2$.
In either case, in the infrared limit one expects to flow by the most
relevant operators (of dimension 2 and zero), which correspond to the last terms in equation
(\ref{potCSEH}), and recover general relativity provided that $\lambda$ also flows to 1.
However, no rigorous proof of any of these statements is yet available in the literature.
Also, the counting of physical degrees of freedom of the theory, which is crucial for viewing
it as viable modification of general
relativity, is obscured by the outcome of local invariances and their potential restoration.

We also recall for completeness, using the relativistic coordinate $x^0 = ct$, that the
effective speed of light for general $\lambda$ is given by
\begin{equation}\label{c}
c={\kappa^2\over 2\kappa_W^2}\sqrt{{\Lambda_W\over 1-3\lambda}}
\end{equation}
with $[c] = z-1$.
This shows that $\Lambda_W$ must be negative when $\lambda>1/3$ to ensure reality of $c$;
likewise, $\Lambda_W$ must be positive when $\lambda<1/3$. Also, the four-dimensional
effective cosmological constant is given by
\begin{equation}
\Lambda ={3 \over 2}\Lambda_W
\end{equation}
and, therefore, the range  $\lambda>1/3$ does not allow for de Sitter-like backgrounds in
Ho\v{r}ava--Lifshitz gravity. These identifications are necessary in order to compare the
infrared limit of the deformed theory to ordinary gravity so that the effective Newton
constant reads as
\begin{equation}\label{gn}
G_N={\kappa^2\over 32\pi c}\, .
\end{equation}
Furthermore, $\lambda$ should approach (flow to) $1$ in the infrared limit in order to
recover the full reparametrization invariance of general relativity, but this particular
problem will not concern us at all here.

The search for classical solutions requires the use of the potential (\ref{potCSEH}) and it
is impossible to solve in full generality. Symmetry ansatz such as spatial homogeneity makes the
problem more tractable, but still not exactly solvable. This includes, for example, the case of
Bianchi IX geometry leading to the mixmaster universe model in four space--time dimensions with
Lorentzian signature.  We will not pursue this line of investigation here (see \cite{Bakas:2009ku}
for a detailed analysis and comparison with the mixmaster universe in general relativity
\cite{Misner:1969, bkl}), but elaborate on the Euclidean version of
Ho\v{r}ava--Lifshitz gravity and then analyze its instanton solutions for Bianchi IX spatial
geometries.

\subsection{Euclidean action and flow equations }

Besides the various physical motivations pertaining to the analysis of the Euclidean version
of Ho\v{r}ava--Lifshitz theory, there is also a technical advantage for constructing solutions
that satisfy first-order equations in time. This possibility is also encountered in general
relativity when self-duality on the Riemann (and more generally on the Weyl) tensor is imposed
leading to gravitational-instanton solutions in the Euclidean regime\footnote{Self-duality is
best described in terms of the curvature two-form, as $\mathcal{R}_{ab}=
\pm\tilde{\mathcal{R}}_{ab}$ in an orthonormal frame \cite{Eguchi:1980jx}.
These equations are second-order in time,
but they can be integrated once to yield first-order equations \cite{Gibbons:1979xn}
that will be paralleled to the instanton solutions of
Ho\v{r}ava--Lifshitz gravity; for further details see also \cite{Bourliot:2009fr}, where this
analogy is made even sharper for gravitational instantons with homogeneous spatial sections.}.
Although there is no
direct analogue of self-duality in gravitational theories with anisotropic scaling, the detailed
balance condition offers the appropriate replacement for defining instanton-like configurations.
This is in fact possible in all dimensions unlike gravitational instantons of ordinary gravity
that are only defined in four space--time dimensions.

The Euclidean action is obtained by setting $t\to -it$, $N^j \to iN^j$, whereas $iS$ is traded
for $-S$. Using equations (\ref{GENkin}) and (\ref{pot}) one obtains
\begin{equation}
\label{EucS}
S=\int \mathrm{d}t\,\mathrm{d}^Dx\,\sqrt{g}N\left(\frac{2}{\kappa^2}\left( K_{ij}K^{ij}
-\lambda K^2\right)+V\right).
\end{equation}
The expression for the potential (\ref{Vdbc}) allows to rewrite the Euclidean action (\ref{EucS})
in the form $S=S'+S''$, where
\begin{equation}
\label{EucStd}
S^{\prime} =\frac{2}{\kappa^2}\int \mathrm{d}t\,\mathrm{d}^Dx\,\sqrt{g}N
\left(K_{ij}\pm \frac{\kappa^2}{2}\mathcal{G}_{ijmn} E^{mn}\right){G}^{ijk\ell}
\left(K_{k\ell}\pm \frac{\kappa^2}{2}\mathcal{G}_{k\ell rs} E^{rs}\right)
\end{equation}
and $S''$ is a total-derivative contribution to the action \cite{Horava:2008ih}.
This boundary term  will be considered later in detail (see equation (\ref{sprssec})).
The different signs correspond to the choice of time direction.

The action \eqn{EucStd} is bounded below by zero provided that the superspace metric
${G}^{ijk\ell}$ is positive-definite. Then, configurations that obey the first-order
differential equations
\begin{equation}
\label{foeq}
K_{ij}=\mp \frac{\kappa^2}{2}\mathcal{G}_{ijk\ell} E^{k\ell},
\end{equation}
are extrema of the action and as such they provide solutions of the Euclidean theory
$S^{\prime}$; the leaves of the corresponding space--time foliations have prescribed
extrinsic curvature.
This possibility arises only when $\lambda < 1/D$, in which case we must also demand that
$\Lambda_W$ is non-negative so that the speed of light in the Lorentzian version of the
theory is real. Otherwise, for $\lambda > 1/D$, the action is non-bounded below by zero and
the first-order equations are not guaranteed to provide classical solutions. Thus, from
now on, we restrict ourselves to $\lambda < 1/D$ and $\Lambda_W \geq 0$, where the first-order
equations (\ref{foeq}) provide extrema of $S'$. They are also extrema of the action $S$
provided that the boundary term $S''$ is properly accounted. This problem will be treated
carefully in section \ref{sec23} and lead to the precise definition of instantons.

The solutions that we will investigate can be expressed in the form of geometric gradient
flow equations for
the metric $g_{ij}$ modulo reparametrizations generated by the shift functions,
\begin{equation}
\label{first}
\partial _t g_{ij}=
\mp \kappa^2 N\mathcal{G}_{ijk\ell} E^{k\ell}
+ \nabla_iN_j+\nabla_jN_i ~.
\end{equation}
Since we are only considering the projectable case of Ho\v{r}ava--Lifshitz gravity, $N$ is
only a function of $t$ and can be absorbed by redefining time, as
$N(t) \mathrm{d}t \rightarrow \mathrm{d}t$.
It is also natural to define vector fields with components $\xi_i = N_i / N$ that generally
depend on space and time coordinates. Then, the geometric-flow equations assume the more
standard form that will be used in the following,
\begin{equation}
{1 \over N(t)} \partial _t g_{ij}=
\pm \frac{\kappa^2}{2\sqrt{g}}\mathcal{G}_{ijk\ell}
\frac{\delta W[g]}{\delta g_{k\ell}}
+ \nabla_i\xi_j+\nabla_j\xi_i ~.
\end{equation}

Specializing to $D=3$, we write down explicitly the flow equations obtained by combining the
variation of the Chern--Simons and Einstein--Hilbert actions,
\begin{equation}
\label{fullflow}
{1 \over N} \partial_t g_{ij}= -
\frac{\kappa^2}{\kappa_W^2}\left(
R_{ij} - {2\lambda - 1 \over 2(3\lambda -1)} R g_{ij} +\frac{\Lambda_W}{1-3\lambda}  g_{ij}
\right)+
\frac{\kappa^2 }{w_{\mathrm{CS}}} C_{ij} +
\nabla_i\xi_j+\nabla_j\xi_i
\end{equation}
choosing for definiteness one of the two sign options; the other follows by time reversal.
These equations describe the parametric evolution of a family of three-dimensional geometries
\begin{equation}
\label{m3dgen}
\mathrm{d}s^2_{\rm t} = g_{ij}(x; t) \mathrm{d}x^i \mathrm{d}x^j
\end{equation}
driven by the Ricci and Cotton tensors and the cosmological constant term, and, as such,
they will be called {\em Ricci--Cotton flow} equations\footnote{Perhaps a more appropriate
name is Ricci--Yamabe--Cotton flow, since $Rg_{ij}$ is the driving term of the so called
Yamabe flow, $\partial_t g_{ij} = - R g_{ij}$. The latter was introduced in the literature
\cite{Hamilton} to solve Yamabe's conjecture stating that any metric is conformally equivalent
to a metric of constant curvature. Its effect is complementary to the Cotton term of the flow,
which changes the conformal class of the metric. Thus, the combined flow equations we have
obtained contain several competing deformations of the metric.}.

The fixed points are determined
(modulo reparametrization terms) by the solutions of three-dimensional topologically
massive gravity:
\begin{equation}
{\kappa_W^2 \over w_{\mathrm CS}} C_{ij} = R_{ij} - 2 \Lambda_W g_{ij} ~, \quad
{\rm with} ~~ R = 6 \Lambda_W ~.
\end{equation}
They include Einstein metrics with vanishing Cotton tensor, like the round sphere metric
on $S^3$ (for $\Lambda_W > 0$), which is homogeneous and isotropic. There are other
fixed points, however, with
constant scalar curvature but with non-vanishing Cotton tensor. We will see later, as
example, that they correspond to particular homogeneous but non-isotropic metrics on $S^3$.
The coexistence of fixed points from different conformal classes of the metric make this
flow particularly complex.

The driving terms of the Ricci--Cotton flow involve, in general, third-order derivatives in
space coordinates (originating from the Cotton tensor), and, therefore, it is not possible
to apply standard results from the mathematics literature to prove even the short-time
existence of solutions. Nevertheless, the mini-superspace models that will be studied later
show that these flow equations are well-behaved and the trajectories converge to fixed points
after sufficiently long time.

Some special cases are worth noting, since they have already appeared in the literature
for different reasons:
\begin{itemize}
\item
$w_{\mathrm CS}  \rightarrow  \infty$: the Cotton tensor contribution drops out
and one obtains a variant of the Ricci flow on three-manifolds, which is second-order and
well studied in the literature. Its trajectories describe solutions of $z=2$
Ho\v{r}ava--Lifshitz gravity in $3+1$ dimensions, whereas the fixed points are Einstein
metrics $R_{ij} = 2 \Lambda_W g_{ij}$.
\item
$\kappa_W^2  \rightarrow  \infty$: the Ricci and cosmological constant terms
drop out and one obtains the pure Cotton flow that was recently introduced in the
literature \cite{Kisisel:2008jx}. Its trajectories describe solutions of $z=3$
Ho\v{r}ava--Lifshitz gravity in $3+1$ dimensions,
in the limit under consideration. The fixed points are conformally flat metrics,
$C_{ij} = 0$.
\end{itemize}

\noindent
Even these simpler cases are impossible to solve in all generality. Mini-superspace
models have been used to study the long time behavior of the Ricci and
Cotton flows for homogeneous geometries \cite{Isenberg:1992, Kisisel:2008jx}.

Apart from the Ricci and Cotton flows that will studied separately in the next section,
there are also some other special cases of the combined Ricci--Cotton flow that are
relatively easier to study. First, by considering that limiting case $\lambda \rightarrow
- \infty$, which lies in the allowed range $\lambda < 1/3$, one obtains the normalized
Ricci--Cotton flow, which is driven by a traceless tensor, and, thus, preserves the
volume of space; the cosmological constant decouples in this case. It becomes relevant
in the Carroll limit of Ho\v{r}ava--Lifshitz gravity, where the effective speed of light
vanishes\footnote{When
$\lambda \rightarrow -\infty$, the metric in superspace becomes singular as its inverse
has zero eigenvalues. Yet the flow equations are well-defined and so is the potential
term $S_{V}$ of the gravity action. It is opposite to the case $\lambda = 1/3$ for
which the metric in superspace has zero eigenvalues and its inverse becomes
singular; the latter case corresponds to the limit of infinite speed of light, where
the theory develops anisotropic Weyl invariance.}. The effective speed of light
vanishes also when $\Lambda_W = 0$, irrespective of $\lambda$, and the corresponding
flow equations will be studied separately. The general case, with arbitrary coefficients,
is much more complex. The pattern of fixed points and specific trajectories will
only be discussed for axially symmetric deformations of $S^3$, which correspond to
solutions with $SU(2) \times U(1)$ isometry.

\subsection{Entropy functional and action bound} \label{sec23}

From now on, and in all examples that will be studied later, we consider flows
without the effect of space reparametrizations, setting $N_i = 0$. Also, we will take
advantage of time reparametrizations to set $N(t) =1$ for convenience. We will
also assume that the spatial slices are compact spaces without boundary. Here, we
provide an entropy functional for the geometric flows arising in Ho\v{r}ava--Lifshitz
gravity in arbitrary dimensions. This functional is also be related to the lower
bound of the Euclidean action $S$ (rather than $S^{\prime}$) when boundary terms $S''$
are properly taken into account.

When the metric in superspace is positive-definite (choosing $\lambda < 1/D$ in $D$
spatial dimensions), the superpotential functional $W$ changes monotonically along
the flow. This follows easily by considering
\begin{equation}
\label{bpsb}
{\mathrm{d}W \over \mathrm{d}t} = -2 \int \mathrm{d}^Dx \sqrt{g} E^{ij}
\partial_t g_{ij} = \pm 2 \kappa^2
\int \mathrm{d}^Dx \sqrt{g} E^{ij} \mathcal{G}_{ijk\ell} E^{k\ell} ~,
\end{equation}
which is the integral of a quadratic quantity, and, therefore, increases or
decreases monotonically depending on the overall sign. Thus, $W$ is an entropy
functional for the flows under consideration.

With this in mind, let us revisit the original Euclidean action $S$ of the theory
and its lower bound taking into proper account the boundary terms. Equations
\eqn{EucS} and (\ref{Vdbc}) yield
\begin{eqnarray}
S & = & {2 \over \kappa^2} \int \mathrm{d}t \mathrm{d}^Dx \sqrt{g} K_{ij} G^{ijk\ell} K_{k\ell} +
{\kappa^2 \over 2} \int \mathrm{d}t \mathrm{d}^Dx \sqrt{g} E^{ij} \mathcal{G}_{ijk\ell} E^{k\ell}
\nonumber\\
& = & \frac{2}{\kappa^2}\int \mathrm{d}t\,\mathrm{d}^Dx\,\sqrt{g}
\left(K_{ij}\pm \frac{\kappa^2}{2}\mathcal{G}_{ijmn} E^{mn}\right){G}^{ijk\ell}
\left(K_{k\ell}\pm \frac{\kappa^2}{2}\mathcal{G}_{k\ell rs} E^{rs}\right) \nonumber\\
& & \mp 2 \int \mathrm{d}t \mathrm{d}^Dx \sqrt{g} K_{ij} E^{ij}. \label{sprssec}
\end{eqnarray}
The first term in (\ref{sprssec}) is $S'$, given in equation (\ref{EucStd}), and the last term
is the advertised boundary contribution $S''$.
For positive-definite superspace metric, the Euclidean action $S$ is bounded from
below by the boundary term $S''$, because $S' \geq 0$. Thus,
\begin{equation}
S \ge \mp 2 \int \mathrm{d}t \mathrm{d}^Dx \sqrt{g} K_{ij} E^{ij} =
\mp \int \mathrm{d}t \mathrm{d}^Dx \sqrt{g} E^{ij}
\partial_t g_{ij} = \pm {1 \over 2} \int \mathrm{d}t {\mathrm{d}W \over \mathrm{d}t},
\end{equation}
having set $N(t)=1$. The time integral of equation (\ref{bpsb}) shows that the lower bound of $S$ is
always positive, as expected.

The flow equations (\ref{foeq}) provide time-dependent extrema of the action $S'$. They are
actually its ground states, since they make $S'$  vanish. Since $S''$ is a boundary term, these
ground states are also extrema of $S$ under appropriate boundary conditions that make the variational
problem well-posed. This can be easily verified for the class of flows with finite action (i.e.
finite $S''$). Note for this purpose that the fixed points of the flow are
static solutions of both $S$ and $S^{\prime}$, since they are, by construction, critical points
of the $D$-dimensional action functional $W$ sitting at the minima of the
Ho\v{r}ava--Lifshitz potential. If different minima exist, they will be all degenerate with zero
potential energy. Hence, time-dependent solutions that
interpolate between any two fixed points are guaranteed to satisfy the equations
of motion following from the Ho\v{r}ava--Lifshitz action $S$.
These solutions have finite action, given by the value of the boundary
term, and is natural to call them instantons as they interpolate between two different
static minima, which are connected by
trajectories of the geometric flow. Their action is simply given by
\begin{equation}
\label{finitac}
S_{\rm instanton} = {1 \over 2} |\Delta W| ~,
\end{equation}
where $\Delta W$ is the difference of the corresponding values of $W$ at the
two critical points. Note that $\Delta W \neq 0$, in general, since $W$ changes
monotonically along the flow lines and the instanton action is finite. Then,
this yields the standard description of instanton solutions of a point particle
moving in Euclidean time, but the number of degrees of freedom is infinite now,
as the evolution takes place in superspace.

Finally, let us consider the evolution of the volume of spatial slices under the
flow. In general, it takes the form
\begin{equation}
{\mathrm{d} \over \mathrm{d}t} {\rm vol}(\mathcal{M}_D) = {1 \over 2} \int
\mathrm{d}^Dx \sqrt{g} g^{ij}
\partial_t g_{ij} = \mp {\kappa^2 \over 2 (1-3\lambda)} \int \mathrm{d}^Dx \sqrt{g}
g_{ij} E^{ij}
\end{equation}
and, therefore, the trace-free part of the driving curvature terms do not
contribute to the evolution. Otherwise, the volume changes without definite
sign. Thus, in principle, the volume can bounce along the flow. At the fixed
points, where $g_{ij} E^{ij} = 0$, the volume reaches a local maximum or
minimum depending on circumstances.

As an example, let us consider the Ricci--Cotton flow described by equation
(\ref{fullflow}). Then, since the Cotton tensor is traceless, the volume of
space changes as
\begin{equation}
{\mathrm{d} \over \mathrm{d}t} {\rm vol}(\mathcal{M}_3) = {\kappa^2 \over 4(1-3\lambda)
\kappa_W^2} \int \mathrm{d}^Dx \sqrt{g} (R-6 \Lambda_W)
\end{equation}
and it can have either sign. Of course, it is possible to normalize this
(or any other flow that arises in this context) by rescaling the metric with a
function of time followed by a suitably chosen time reparametrization
$t \rightarrow \tilde{t}(t)$ so that the volume is preserved in time $\tilde{t}$.
This does not resolve the problem, however, since $\tilde{t}(t)$ is not
a monotonic function of $t$ in general. Thus, the volume does not provide
an entropy functional.

Other entropy functionals might also exist for these flows, generalizing
Perelman's functional for the Ricci flow \cite{perel}, but we have not
been able to find them.

\section{Bianchi IX model geometry}\label{mixmas}
\setcounter{equation}{0}

All homogeneous space geometries in three dimensions provide consistent ansatz for the
mini-superspace truncation of the Ricci and Cotton flows and their combination thereof.
Such spaces follow the Bianchi classification, but for practical reasons we will only
consider the case of Bianchi IX model geometries that describe homogeneous but generally
non-isotropic metrics on $S^3$. The corresponding gravitational instantons of the
four-dimensional Euclidean theory are special in that they admit an $SU(2)$ isometry group
and they provide the simplest examples in our study. In this section, we set up the notation
and present some useful formulas that will enable us to formulate the problem as an
autonomous system of ordinary differential equations. The Ricci and Cotton flows are
studied separately here for Bianchi IX geometries. Comparison with the gravitational
instantons of ordinary gravity will also be made at the appropriate places.

\subsection{Some basic facts}

We consider four-dimensional Riemannian manifolds that are foliated by homogeneous
three-dimensional spaces of the form
\begin{equation}
\label{m4dHL}
\mathrm{d}s^2 = \mathrm{d}t^2   +\sum_i \gamma_i(t) \left(\sigma^i\right)^2 ~,
\end{equation}
setting $N=1$ and $N_i = 0$. The coefficients $\gamma_i$ are taken to depend only on $t$
and $\sigma^i$ are the left-invariant Maurer--Cartan one-forms of $SU(2)$
\begin{eqnarray}
\sigma^1 & = &\sin\vartheta \sin\psi \, \mathrm{d}\varphi+\cos \psi \,
\mathrm{d}\vartheta \nonumber\\
\sigma^2 &=& \sin\vartheta\cos\psi\, \mathrm{d}\varphi-\sin\psi\, \mathrm{d}\vartheta\\
\sigma^3 &=& \cos\vartheta\, \mathrm{d}\varphi+\mathrm{d}\psi \nonumber
\end{eqnarray}
with Euler angles ranging as $0\leq\vartheta\leq \pi$, $0\leq\varphi\leq 2\pi$, $
0\leq\psi\leq {4\pi}$, which satisfy
\begin{equation}
\mathrm{d}\sigma^i + \frac{1}{2} \epsilon^i_{\hphantom{i}jk}\sigma^j \wedge \sigma^k=0.
\end{equation}

The three-dimensional leaves are, in general, homogeneous but non-isotropic three-spheres.
The isometry group is enhanced to $SU(2) \times U(1)$ when any two $\gamma_i$'s coincide by
imposing axial symmetry. Full isotropy requires all $\gamma_i$'s to be equal, in which case
the symmetry of the model is promoted to $SU(2)\times SU(2)$. The volume of the three-sphere
is
\begin{equation}
\label{V3d}
V=16\pi^2 \sqrt{\gamma_1 \gamma_2 \gamma_3}
\end{equation}
and when all coefficients $\gamma_i$ are equal to $L^2/4$ the volume is expressed in terms
of the radius $L$ as $V=2\pi^2L^3$.

The Ricci and Cotton tensors are diagonal for all homogeneous geometries and this ensures
consistency of the reduced models. They take the following form for the Bianchi IX class,
\begin{eqnarray}
R_{11}&=& \frac{1}{2\gamma_2 \gamma_3} \left[
\gamma_1^2 -(\gamma_2 - \gamma_3)^2 \right]\label{ric} ~, \label{R11} \\
C_{11}&=& -\frac{\gamma_1}{2( \gamma_1 \gamma_2 \gamma_3)^{\nicefrac{3}{2}}}
\left[\gamma^2_1\left(2\gamma_1-\gamma_2-\gamma_3\right)
-\left(\gamma_2+\gamma_3\right)\left(\gamma_2-\gamma_3\right)^2 \right] ~,\label{C11}
\label{cot}
\end{eqnarray}
and similarly for the other two components that follow by cyclic permutation of the indices.
Also, the Ricci scalar curvature is given by
\begin{equation}
R = \frac{1}{2\gamma_1\gamma_2 \gamma_3} \left[
2\gamma_1\gamma_2 +2\gamma_2 \gamma_3+2\gamma_3 \gamma_1-\gamma_1^2 -\gamma_2^2-\gamma_3^2
\right]\label{Rsc} ~,
\end{equation}
whereas the trace of the Cotton tensor vanishes, as it can be readily checked.

With these explanations in mind, we arrive at the following system of ordinary
differential equations for the metric coefficients $\gamma_i (t)$
\begin{equation}
\label{m4dHLfullflow}
\frac{\mathrm{d}\gamma_i}{\mathrm{d}t}=-\frac{\kappa^2}{\kappa_W^2}\left(R_{ii}
-\frac{2\lambda-1}{2(3\lambda-1)}R \gamma_{i} +\frac{\Lambda_W}{1-3\lambda}  \gamma_{i}
\right) + \frac{\kappa^2}{w_{\rm CS}}C_{ii}
\end{equation}
as the Bianchi IX mini-superspace model of the combined Ricci--Cotton flow with
general couplings.

\subsection{Ricci flow}

When $w_{\rm CS} \rightarrow \infty$ the Cotton term decouples and one arrives at a
variant of the Ricci flow as the relevant equation for Ho\v{r}ava--Lifshitz gravity
with anisotropy scaling parameter $z=2$. In this case, the system becomes
\begin{equation}
\label{halfi}
\frac{\mathrm{d}\gamma_i}{\mathrm{d}t}=-\frac{\kappa^2}{\kappa_W^2}\left(R_{ii}
-\frac{2\lambda-1}{2(3\lambda-1)}R \gamma_{i} +\frac{\Lambda_W}{1-3\lambda}  \gamma_{i}
\right)
\end{equation}
and its properties resemble the ordinary Ricci flow on $S^3$
\begin{equation}
\label{ordiricci}
\frac{\mathrm{d} \gamma_i}{\mathrm{d} t}=- R_{ii} (\gamma) ~.
\end{equation}
Formally, one follows from the other by rescaling the metric with a function of time
and changing time variable by suitable reparametrization; in such case, the
components of the Ricci tensor remain invariant and they assume the
same form for the rescaled components of the metric.

The Ricci flow equations \eqn{ordiricci} for homogeneous geometries are well
studied in the literature following the original work \cite{Isenberg:1992}.
For Bianchi IX geometries they take the form
\begin{eqnarray}
\label{halphen}
{2 \over \gamma_1} {\mathrm{d} \gamma_1 \over \mathrm{d}t} & = &
{1 \over \gamma_1 \gamma_2 \gamma_3}
\left[\left(\gamma_2 - \gamma_3 \right)^2 - \gamma_1^2 \right] ~,\nonumber\\
{2 \over \gamma_2} {\mathrm{d} \gamma_2 \over \mathrm{d}t} & = &
{1 \over \gamma_1 \gamma_2 \gamma_3}
\left[\left(\gamma_3 - \gamma_1 \right)^2 - \gamma_2^2 \right] ~,\\
{2 \over \gamma_3} {\mathrm{d} \gamma_3 \over \mathrm{d}t} & = &
{1 \over \gamma_1 \gamma_2 \gamma_3}
\left[\left(\gamma_1 - \gamma_2 \right)^2 - \gamma_3^2 \right] \nonumber
\end{eqnarray}
and coincide with the celebrated \emph{Darboux--Halphen} system that was introduced by
Darboux in the nineteenth century \cite{Darboux} and subsequently solved by Halphen
\cite{halph}\footnote{For the comparison one must consider the variables
$\omega_1=\gamma_2\gamma_3$, $\omega_2 = \gamma_1 \gamma_3$, $\omega_3 = \gamma_1
\gamma_2$ and change the time coordinate to $\mathrm{d}T = \mathrm{d}t/\gamma_1 \gamma_2
\gamma_3$. Then, the equations take the equivalent form
\begin{equation}
{\mathrm{d} \omega_1 \over \mathrm{d}T} = \omega_2 \omega_3 - \omega_1
(\omega_2 + \omega_3) \nonumber
\end{equation}
with cyclic permutations for $\omega_2$ and $\omega_3$. The identification of the
Ricci flow equations for Bianchi IX geometry with the Darboux--Halphen system has
escaped attention in the mathematics literature.}. In principle, solutions
of these equations can be translated into solutions of the original system
\eqn{halfi}.

It can be shown quite generally that for given initial data $\gamma_i^{(0)}$ the
metric will evolve towards the configuration $\gamma_1 = \gamma_2 = \gamma_3 = 0$
by making $S^3$ rounder and rounder until the whole space collapses to a
point. A particularly simple solution that exhibits this behavior is provided by
\begin{equation}
\label{triviasola}
\gamma_1 (t) = \gamma_2 (t) = \gamma_3 (t) = {1 \over 2} (t_0 - t)
\end{equation}
and describes an isotropic metric on $S^3$ whose radius evolves from infinitely
large size to zero as $t$ varies from $-\infty$ to $t_0$.

Actually, following the literature \cite{Isenberg:1992}, the convergence of the flow
lines is best described in terms of the normalized Ricci flow equation on $S^3$
\begin{equation}
{\mathrm{d} \gamma_i \over \mathrm{d}t} = - R_{ii} + {1 \over 3} R \gamma_i ~,
\end{equation}
which follows directly from equation (\ref{halfi}) in the limit $\lambda\rightarrow
-\infty$, and
which can be obtained from the ordinary Ricci flow by (yet another) suitable rescaling
of the metric and time reparametrization. Then, the volume is preserved along the flow
and the round metric (fully isotropic model with finite radius) arises as fixed point
that is exponentially reached after infinitely long time, regardless of initial
conditions. The normalized Ricci flow will also be in focus later in section 4 for
different reasons.

The Darboux--Halphen system is not algebraically integrable when $\gamma_1 \neq
\gamma_2 \neq \gamma_3$. All its solutions, however, can be expressed in terms of
modular forms (see, for instance, \cite{Takhtajan:1992qb}).
The system becomes algebraically integrable when two $\gamma_i$'s are equal. Then,
the corresponding three-dimensional space is a three-sphere with axially symmetric
metric. Setting $\gamma_1 = \gamma_2 \geq \gamma_3$, the system \eqn{halphen}
simplifies and exhibits a first integral
\begin{equation}
{1 \over \gamma_3^2} - {1 \over \gamma_1 \gamma_3} = {1 \over 4m^2}
\end{equation}
with arbitrary parameter $m$. The solution is subsequently described as
\begin{equation}
\label{spezisol}
{\gamma_1 \over m} + {\rm arcsinh} {\gamma_1 \over m} = {-t + t_0 \over m} =
{\rm log} {2m + \gamma_3\over 2m - \gamma_3} - 2m \left({1 \over 2m + \gamma_3}
- {1 \over 2m - \gamma_3} \right) ~,
\end{equation}
using another integration constant $t_0$. The fully isotropic solution with
$SU(2) \times SU(2)$ isometry is obtained by taking the limit
$m \rightarrow \infty$.

Remarkably, the same system of equations arises in the description of a class of
self-dual instantons with $SU(2)$ isometry in ordinary gravity \cite{Gibbons:1979xn},
as well as in the description of the moduli space of $SU(2)$ BPS monopoles with
magnetic charge 2 \cite{Atiyah1, Atiyah2}. In these two cases,
which share many features with each other, the ansatz for the Bianchi IX
geometries takes the form
\begin{equation}
\label{GRgravinst}
\mathrm{d}s^2 = \mathrm{d}t^2 + a^2 (t) \left(\sigma^1\right)^2 + b^2 (t)
\left(\sigma^2\right)^2 + c^2(t)  \left(\sigma^3\right)^2
\end{equation}
with $a(t) = \gamma_1 (t)$, $b(t) = \gamma_2 (t)$ and $c(t) = \gamma_3 (t)$
satisfying the same system of equations \eqn{halphen} above. Then, the
corresponding axially symmetric gravitational instanton with $\gamma_1 = \gamma_2$
is the self-dual Taub--NUT metric with NUT parameter $m$. It can be brought into
standard form using a radial coordinate $r \geq m$ with  $m ~ {\rm arcosh}(r/m) +
\sqrt{(r-m)(r+m)} = -t + t_0$ so that the solution \eqn{spezisol} becomes
\begin{equation}
\gamma_1 = \gamma_2 = \sqrt{(r-m)(r+m)} ~, \quad \gamma_3 = 2m
\sqrt{{r-m \over r+m}} ~.
\end{equation}
The fully anisotropic instantons of Einstein gravity with $SU(2)$ isometry
correspond to the so called Atiyah--Hitchin metric \cite{Atiyah1, Atiyah2}, which
is also the metric on the moduli space of charge 2 BPS $SU(2)$ monopoles in general
position. These metrics will be discussed further in section 6 while comparing
instanton solutions of general relativity with those of Ho\v{r}ava--Lifshitz theory.

\subsection{Cotton flow}

Next, we consider the pure Cotton flow equations\footnote{The Cotton flow was
originally introduced in the literature \cite{Kisisel:2008jx} as an alternative
to the Ricci flow for studying the existence of constant curvature metrics on
three-manifolds. So far it has only been applied to homogeneous geometries and its
general utility for proving the Poincar\'e conjecture remains questionable.}
that arise in the limit $\kappa_W \rightarrow \infty$. They describe solutions of
$z=3$ Ho\v{r}ava--Lifshitz theory when the volume of space is very small, i.e., in
the deep ultra-violet regime where the Cotton term dominates and all subleading
relevant operators can be safely dropped from the potential. Then, the equations
for Bianchi IX geometries take the form
\begin{eqnarray}
\label{thecottfle}
{2 w_{\rm CS} \over \kappa^2 \gamma_1} {\mathrm{d} \gamma_1 \over \mathrm{d}t} & = &
-\frac{1}{( \gamma_1 \gamma_2 \gamma_3)^{\nicefrac{3}{2}}}
\left[\gamma^2_1\left(2\gamma_1-\gamma_2-\gamma_3\right)
-\left(\gamma_2+\gamma_3\right)\left(\gamma_2-\gamma_3\right)^2 \right] , \nonumber\\
{2 w_{\rm CS} \over \kappa^2 \gamma_2} {\mathrm{d} \gamma_2 \over \mathrm{d}t} & = &
-\frac{1}{( \gamma_1 \gamma_2 \gamma_3)^{\nicefrac{3}{2}}}
\left[\gamma^2_2\left(2\gamma_2-\gamma_3-\gamma_1\right)
-\left(\gamma_3+\gamma_1\right)\left(\gamma_3-\gamma_1\right)^2 \right] , \\
{2 w_{\rm CS} \over \kappa^2 \gamma_3} {\mathrm{d} \gamma_3 \over \mathrm{d}t} & = &
-\frac{1}{( \gamma_1 \gamma_2 \gamma_3)^{\nicefrac{3}{2}}}
\left[\gamma^2_3\left(2\gamma_3-\gamma_1-\gamma_2\right)
-\left(\gamma_1+\gamma_2\right)\left(\gamma_1-\gamma_2\right)^2 \right] , \nonumber
\end{eqnarray}
and clearly they are much more complicated than the Darboux--Halphen system. It is
not known whether they are algebraically integrable when $\gamma_1 \neq \gamma_2 \neq \gamma_3$,
but it can be easily shown that, under any initial data $\gamma_i^{(0)}$, they flow exponentially
fast towards the fixed point which is the round metric on $S^3$ and it is conformally flat
\cite{Kisisel:2008jx}.
Since the Cotton tensor is odd under parity (because of the fully antisymmetric epsilon
symbol appearing in its definition), there is always an ambiguity in the overall sign of
the flow equations. Here, we have chosen the sign that takes any metric towards
the fixed point rather than away from it when $w_{\rm CS} > 0$.

The behavior of the Cotton flow is similar to the normalized Ricci flow, as they both preserve
the volume of space $V= 2\pi^2 L^3$, but the convergence rate is different. To compare the
two it is sufficient to linearize the corresponding equations around the fixed point
\begin{equation}
\gamma_1 = \gamma_2 = \gamma_3 = {L^2 \over 4}
\end{equation}
by considering small perturbations of the metric coefficients
\begin{equation}
\gamma_1 (t) = {L^2 \over 4} \left(1 + \delta x (t) \right) , \quad
\gamma_2 (t) = {L^2 \over 4} \left(1 + \delta y (t) \right) ,
\end{equation}
whereas $\gamma_3(t)$ changes accordingly,
\begin{equation}
\gamma_3 (t) = {L^2 \over 4} \left(1 - \delta x (t) - \delta y (t) \right) ,
\end{equation}
so that the volume of space is preserved. Then, the autonomous system of Cotton flow
equations \eqn{thecottfle} becomes to linear order
\begin{equation}
 {\mathrm{d} \over \mathrm{d}t} \begin{pmatrix}
\delta x \\
\delta y    \end{pmatrix}= -{12  \kappa^2\over w_{\rm CS} \ L^3}
\begin{pmatrix}
1 &   0 \\
0 &   1
    \end{pmatrix}
 \begin{pmatrix}
\delta x \\
\delta y    \end{pmatrix}
\end{equation}
and the two eigenvalues are equal and negative for $w_{\rm CS} > 0$,
\begin{equation}
\zeta_1 = \zeta_2 = -{12 \kappa^2 \over w_{\rm CS} L^3} ~,
\end{equation}
ensuring stability in all directions around the fixed point.

The perturbations diminish exponentially fast, as
\begin{equation}
\delta x (t) = A \mathrm{e}^{\nicefrac{-t }{ \tau_{\rm C}}} ~,\quad
\delta y(t) = B  \mathrm{e}^{\nicefrac{-t }{ \tau_{\rm C}}} ~,
\end{equation}
using arbitrary integration constants $A$, $B$ and the characteristic time scale of
dissipation
\begin{equation}
\tau_{\rm C} = {w_{\rm CS} L^3 \over 12 \kappa^2} ~.
\end{equation}
Thus, when $w_{\rm CS}$ is very small compared to
$\kappa_{\rm W}^2$ so that the Cotton tensor dominates the flow over the Ricci curvature
and the cosmological constant terms, or equivalently when $L$ is very small so that
the volume of space is tiny, $\tau_{\rm C}$ is very small and the metric
perturbations dissipate very fast at late times\footnote{We note here for completeness,
and it will be used in the next section, that the normalized Ricci flow
$\mathrm{d}\gamma_i /\mathrm{d}t = -R_{ii} + R\gamma_i/3$ can be similarly analyzed by
considering small perturbations around the fully isotropic fixed point.
The two eigenvalues also turn out to be equal and negative, but the corresponding
characteristic time scale for dissipation depends quadratically on $L$ as
$\tau_{\rm R} = L^2/4$ with respect to the appropriate time coordinate.
Comparison with the dissipation rate of the Cotton flow will become relevant in section 4.}.

There is an additional fixed point, which is unique, up to permutation of the axes,
that arises when $\gamma_1 = \gamma_2 = \infty$ and
$\gamma_3 = 0$  (correlated limit with $V$ held fixed). It corresponds to a squashed
$S^3$ that is completely flattened and has zero Cotton tensor, as it can be explicitly
checked. Although this configuration is degenerate in one principal direction, it has no
curvature singularities and it is legitimate to consider.

Axisymmetric solutions of the Cotton flow can be constructed in closed form, as for the
Ricci flow. Assuming
\begin{equation}
\gamma_1 = \gamma_2 \equiv x {L^2 \over 4} ~, \quad \gamma_3 = {L^2 \over 4x^2} ~,
\end{equation}
so that the volume of space is held fixed to $V=2\pi^2 L^3$, the Cotton flow equations
\eqn{thecottfle} reduce to a single equation
\begin{equation}
{\mathrm{d}x \over \mathrm{d}t} = {4 \kappa^2 \over w_{\rm CS} L^3} {1 - x^3 \over x^5}\ ,
\end{equation}
which is solved as follows,
\begin{equation}
\label{mourira}
{-t + t_{\star} \over \tau_{\rm C}} = x^3 -1 + {\rm log}\left|x^3 - 1\right|
\end{equation}
with integration constant $t_{\star}$.

This solution has two branches. For $x \geq 1$, $x(t)$ changes from $+\infty$ to $1$ as
$t$ varies from $-\infty$ to $+\infty$; the three-sphere deforms starting from the
singular configuration $\gamma_1 = \gamma_2 = \infty$, $\gamma_3 = 0$ and gradually
becomes rounder until it reaches the isotropic fixed point after infinitely long time.
For $x \leq 1$, $x(t)$
changes from $0$ to $1$ as $t$ varies from $t_0 = t_{\star} + \tau_{\rm C}$ to $+\infty$;
in this case, the three-sphere evolves from the
singular configuration $\gamma_1 = \gamma_2 = 0$, $\gamma_3 = \infty$ towards the
isotropic fixed point.

Finally, we point out that there is no known solution for the fully anisotropic
model geometry that is analogous to the general solution of the Darboux--Halphen
system. It remains an open question whether the system is algebraically
integrable and find its solution.

\section{Normalized Ricci--Cotton flow}\label{norma}
\setcounter{equation}{0}

We will now investigate the combined Ricci--Cotton flow when the effective speed of light
vanishes by letting $\lambda \rightarrow -\infty$. In this case, the flow equations take
the following general form,
\begin{equation}
\partial_t g_{ij}= - \frac{\kappa^2}{\kappa_W^2}\left(R_{ij} - {1 \over 3} R g_{ij}
\right)+\frac{\kappa^2 }{w_{\mathrm{CS}}} C_{ij}
\end{equation}
and they become independent of $\Lambda_W$. The driving curvature term is traceless and
the deformations preserve the volume of space. Thus, the resulting normalized Ricci--Cotton
flow is a superposition of the Cotton and the normalized Ricci flow with competing effects,
in general, that depend on the sign of $w_{\mathrm{CS}}$.

\subsection{The general system of Bianchi IX equations}

Using the Bianchi IX ansatz for the three-dimensional geometry, the normalized
Ricci--Cotton flow equations form an autonomous system of equations for the
coefficients $\gamma_1$, $\gamma_2$,
$\gamma_3$. Since the volume $V=2\pi^2 L^3$ is conserved, it is convenient to
use two independent variables $x(t)$ and $y(t)$,
\begin{equation}
\label{red}
\gamma_1=\frac{xL^2}{4}, \quad \gamma_2=\frac{yL^2}{4}, \quad \gamma_3=\frac{L^2}{4xy} ~,
\end{equation}
and also set
\begin{equation}
\tau = {4 \kappa^2 \over \kappa_W^2 L^2} t ~, \quad \mu = {w_{\rm CS} L \over
\kappa_W^2} ~.
\end{equation}
Then, the general system of Bianchi IX equations takes the form
\begin{eqnarray}
\frac{\mathrm{d}x}{\mathrm{d}\tau}&=&\frac{1}{\mu x^2y^3}
\bigg\{
\left(1+\frac{\mu}{3}xy +xy^2\right)\left(1-xy^2\right)^2\nonumber\\
&&+\left(\frac{\mu}{3}+x\right)\left(1+xy^2-2x^2y\right)x^3y^2
\bigg\}, \label{RCNx}
\\
\frac{\mathrm{d}y}{\mathrm{d}\tau}&=&\frac{1}{\mu x^3y^2}
\bigg\{
\left(1+\frac{\mu}{3}xy +x^2y\right)\left(1-x^2y\right)^2\nonumber\\
&&+\left(\frac{\mu}{3}+y\right)\left(1+x^2y-2xy^2\right)x^2y^3
\bigg\}.\label{RCNy}
\end{eqnarray}

It is not known whether this system of equations is integrable, in general.
However, we will be able to characterize its fixed points and study their stability
on general grounds in order to infer the qualitative behavior of the flow
lines. Note that the individual components of our system, the normalized Ricci flow
and the Cotton flow, can compete with each other. When $\mu > 0$ ($w_{\rm CS} > 0$),
they both work in the same direction, but when $\mu < 0$ ($w_{\rm CS} < 0$) they
work against each other and can affect the form and stability properties of the fixed
points.

\begin{figure}[!h]
\begin{center}
\includegraphics[height=9cm]{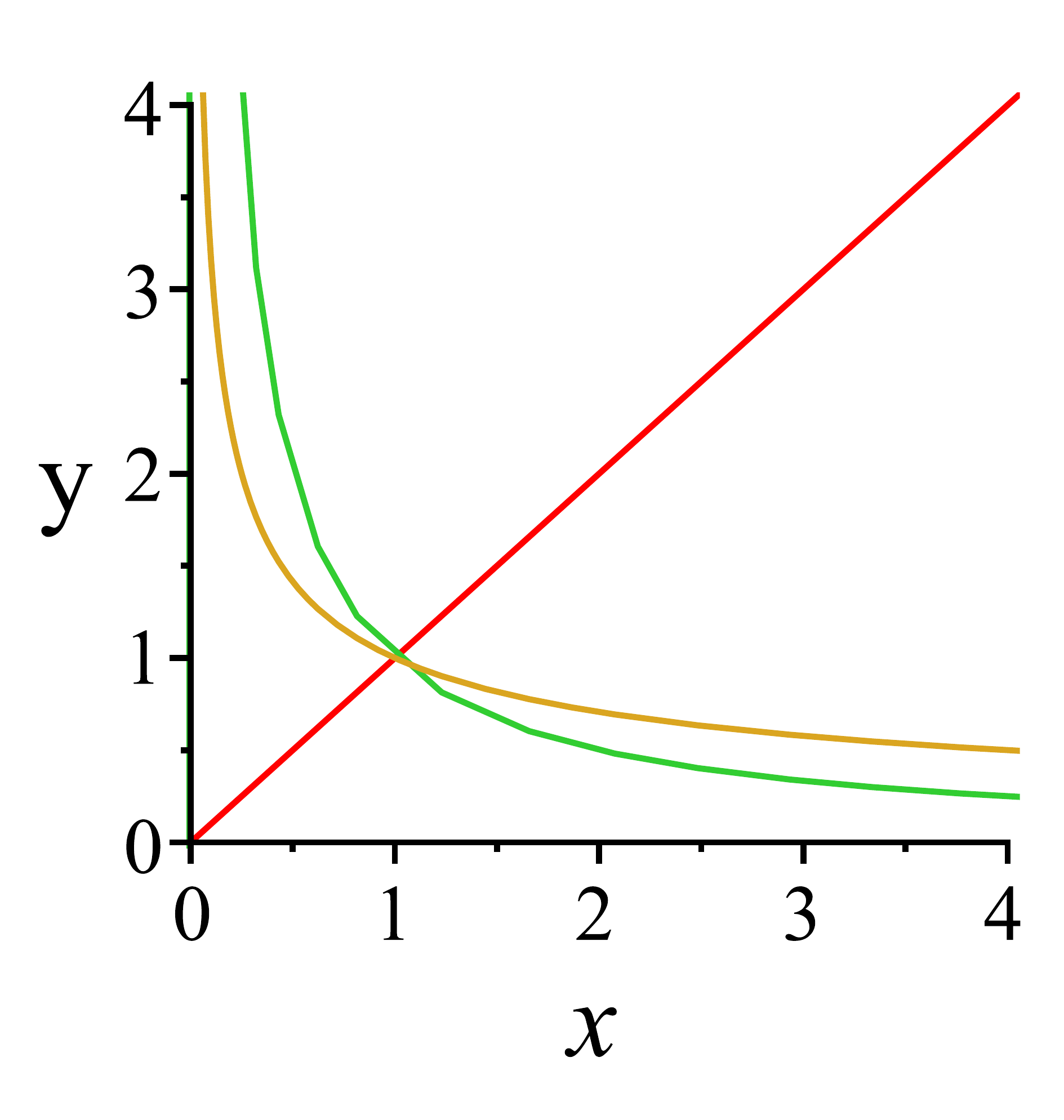}
\end{center}
\caption{The lines of enhanced $SU(2)\times U(1)$ symmetry in the $(x,y)$ plane.}
\label{xyplane}
\end{figure}

It will also be seen later that these equations can be solved exactly in the
axially symmetric case $x=y$.
Actually, there are three curves of axial symmetry in the problem, but, in practice,
it is sufficient to consider only one of these axially
symmetric cases, since the other two follow by permutation of the principal axes of
$S^3$. Thus, apart from  $x=y$, we also have $x^2y=1$ and
$xy^2=1$, depending on the pair of $\gamma_i$'s that become equal and reduce the
flow equations to a single one. They correspond to
metrics on $S^3$ with enhanced symmetry $SU(2) \times U(1)$ and they all intersect at the
fully isotropic point $x=y=1$. These curves are by themselves flow lines, which, however,
cannot be crossed by other flow lines; if any two
$\gamma_i$'s become equal at a given (finite) time they will remain equal for ever.
Therefore, these three curves provide the barriers for six regions in the
$(x,y)$ plane where the generic flow lines are confined depending on initial conditions.
The maximal time range of any given flow line also depends on the region in which the
flow is confined. Finally, the flow lines along the three curves of axial symmetry can
reach the fully isotropic point but cannot continue running beyond it. The three curves
of axial symmetry are depicted in figure \ref{xyplane}, which is
restricted to the first quadrant of the $(x, y)$ plane so that the metric has
physical signature, and they intersect at $(1,1)$.

These general
qualitative remarks are sufficient to illustrate the evolution of any given initial
data on the $(x,y)$ plane, provided that all fixed points are found and the
arrows of flow lines around them are correctly identified by stability analysis.
The same remarks apply equally well to the normalized Ricci and Cotton flows
which can arise separately as special cases.

\subsection{Classification of the fixed points} \label{normadyn}

We are going now to classify all fixed points of the system \eqn{RCNx} and
\eqn{RCNy} and find the critical values of $\mu$ that separate
their behavior into different phases of stability.

\paragraph{The isotropic fixed point.}

The metric on the round sphere is a fixed point of the normalized Ricci--Cotton flow
for all values of $\mu$ (positive and negative) and corresponds to the point
\begin{equation}
\label{isofp}
x_{\star}= y_{\star}=1
\end{equation}
on the $(x,y)$ plane. Its stability, however, depends on the values of the parameter $\mu$.

By considering small perturbations around this fixed point, as
\begin{equation}
x(t) = x_{\star} + \delta x(t) ~, \quad
y(t) = y_{\star} + \delta y(t) ~,
\end{equation}
we find that the linearized system of equations takes the form
\be
\label{linam}
{\mathrm{d} \over \mathrm{d} \tau} \begin{pmatrix}
\delta x \\
\delta y    \end{pmatrix} = -\left(1 + {3 \over \mu} \right)
 \begin{pmatrix}
1&0 \\
0&1
 \end{pmatrix}
 \begin{pmatrix}
\delta x \\
\delta y    \end{pmatrix} .
\ee
The two eigenvalues are equal
\begin{equation}
\label{lam}
\zeta_1 = \zeta_2 =
-\left(1 + \frac{3}{\mu} \right)
\end{equation}
and follow by linear superposition of the corresponding eigenvalues of the
normalized Ricci and Cotton flows discussed in the previous section.

The fixed point is absolutely stable when $\mu$ satisfies the bound
$3/\mu > -1$, i.e., $\mu > 0$ or $\mu <-3$. In these cases, the flow line converge
towards the fixed point from all directions. Otherwise,
for $-3 < \mu < 0$, the isotropic point is absolutely unstable and the flow lines
diverge away from it in all directions. Finally, when $\mu = -3$, the eigenvalues are zero
and all points in the vicinity of the fixed point are at equilibrium. Thus,

\begin{itemize}
\item $\mu > 0$ or $\mu < -3$:  \quad absolutely stable fixed point,
\item $-3 < \mu < 0$: \quad absolutely unstable fixed point.
\end{itemize}

These results are in
exact agreement with the competition between the Ricci and Cotton components of the flow and
can be understood by comparing the characteristic time scales $\tau_{\rm R}$ and
$\tau_{\rm C}$ that govern metric perturbations at late times\footnote{According to
definitions we have the following relations
\begin{equation}
\mu = {w_{\rm CS} L \over \kappa_W^2} = 3 {\tau_{\rm C} \over \tau_{\rm R}} \quad
{\rm with} \quad
\tau_{\rm C} = {w_{\rm CS} L^3 \over 12 \kappa^2} ~, \quad
\tau_{\rm R} = {\kappa_W^2 L^2 \over 4 \kappa^2} ~.
\nonumber
\end{equation}
The characteristic time scales $\tau_{\rm R}$ and $\tau_{\rm C}$ refer to the original
time coordinate $t$ but their ratio is the same in the time coordinate $\tau$.}.
For $\mu > 0$ both components
dissipate all metric perturbations exponentially fast, but for $\mu < 0$ the Cotton flow
contributes differently leading to exponential growth of the perturbations.
For the critical value $\mu = -3$, the dissipation of the normalized Ricci flow is canceled
by the exponential growth of the Cotton flow, making zero the characteristic matrix of metric
perturbations.

\paragraph{Anisotropic fixed points.}

For negative $\mu$ there are additional fixed points that correspond to particular
axisymmetric metrics. As such, they appear in three copies related by permutation of
the axes of $S^3$ and they are located on the lines of axial symmetry in the $(x,y)$ plane.
In particular, they arise
\begin{itemize}
\item on the diagonal $x=y$ with $x_{\star} = \sqrt{-3/ \mu}$ ~,
\item on the branch $x^2 y=1$ with $x_{\star} = \sqrt{-3 / \mu}$~,
\item on the branch $x y^2=1$ with $y_{\star} = \sqrt{-3 / \mu}$~.
\end{itemize}
Note that all these points coalesce with the fully isotropic point when $\mu$ assumes the
critical value $\mu= -3$.

It suffices to perform stability analysis around one of these fixed points, say the one
located on the diagonal line, since the results will be identical for all of them by the
symmetry of the problem. Using small fluctuations around the fixed point
\begin{equation}
x_{\star} = y_{\star} = \sqrt{-{3 \over \mu}} ~,
\end{equation}
so that $x(t) = x_{\star} + \delta x(t)$ and $y(t) = y_{\star} + \delta y(t)$, the
linearized system takes the form,
\be
{\mathrm{d} \over \mathrm{d} \tau}  \begin{pmatrix}
\delta x \\
\delta y    \end{pmatrix}={1 \over 2}
 \begin{pmatrix}
\zeta_1 + \zeta_2 &   \zeta_1 - \zeta_2 \\
\zeta_1 - \zeta_2 &   \zeta_1 + \zeta_2
 \end{pmatrix}
 \begin{pmatrix}
\delta x \\
\delta y    \end{pmatrix} ,
\ee
where
\ba
\zeta_1 & = & {2 \over 3} \sqrt{-{\mu  \over 3}} \left[
\left({-{\mu \over 3}} \right)^{\nicefrac{3}{2}} - 1 \right] , \label{lam1}
\\
\zeta_2 & = & {2 \over 3 } \sqrt{-{\mu  \over 3}} \left[
\left({-{3 \over \mu}} \right)^{\nicefrac{3}{2}} - 1 \right] \left[
4 \left({-{3 \over \mu }} \right)^{\nicefrac{3}{2}} - 1 \right] \label{lam2}.
\ea

The two eigenvalues are $\zeta_1$ and $\zeta_2$ and they are unequal offering various
possibilities in general. They both vanish for $\mu = -3$. Otherwise, we have the
following cases depending on the sign of the eigenvalues:
\begin{itemize}
\item $-3<\mu<0$: \quad saddle fixed point with $\zeta_1<0<\zeta_2$~,
\item $-6 \, \sqrt[3]{2}<\mu<-3$: \quad saddle fixed point with $\zeta_2<0< \zeta_1$~,
\item $\mu<-6 \, \sqrt[3]{2}$: \quad absolutely unstable fixed point with
$\zeta_i > 0$~.
\end{itemize}
Thus, the axisymmetric fixed points are never absolutely stable. They are saddle or
unstable depending on $\mu$.

\paragraph{Totally anisotropic fixed points.}

It is not obvious from the beginning whether there are any fixed points
with $\gamma_1 \neq \gamma_2 \neq \gamma_3$. Close inspection of the equations,
assisted by numerical scanning, reveals the presence of two totally anisotropic
fixed points that coexist with the axially symmetric anisotropic fixed
point\footnote{We thank Christos Sourdis for pointing out the presence
of these additional fixed points and thoroughly investigating their properties.
This analysis was missed in a previous version of our paper and we are indebted
to him for providing all the details. Similar results also appeared in \cite{nutku}.}
when $\mu<-6 \, \sqrt[3]{2}$. In fact, by the $\mathbb{Z}_3$ symmetry of the
problem, there are six such additional fixed points, but we only focus attention
on two of them appearing symmetrically left and right of the diagonal line $x=y$
in the lower two (out of the six disconnected) regions shown in figure 1; their
presence should not be confused with the mirror images of the axially symmetric
anisotropic fixed point discussed earlier.

The characteristic property of the totally anisotropic fixed points is that
their Ricci scalar curvature vanishes and their location on the $(x, y)$ plane
is given by
\begin{equation}
\label{zericc}
x + {1 \over \sqrt{x}} = -{\mu \over 4} =  y + {1 \over \sqrt{y}} \quad
{\rm with} \quad x \neq y ~.
\end{equation}
Of course, one can always find Bianchi IX metrics with zero scalar curvature
by imposing the appropriate algebraic condition on the metric coefficients, but
these are not fixed points of the
flow lines for general values of $\mu$. Remarkably, they are real solutions of
$\mathrm{d}x / \mathrm{d} \tau = 0 = \mathrm{d}y / \mathrm{d} \tau$ with $x \neq y$,
which coexist with the axially symmetric anisotropic fixed point when
$\mu<-6 \, \sqrt[3]{2}$. They are not present when $\mu > -6 \, \sqrt[3]{2}$, since
there are no real solutions to equation \eqn{zericc} in that case.
We also note that when $\mu = -6 \, \sqrt[3]{2}$ the totally anisotropic fixed
points coalesce with the axially symmetric anisotropic fixed point. Furthermore,
there are no other fixed points in the problem.

\begin{figure}[!h]
\begin{center}
\includegraphics[height=8cm]{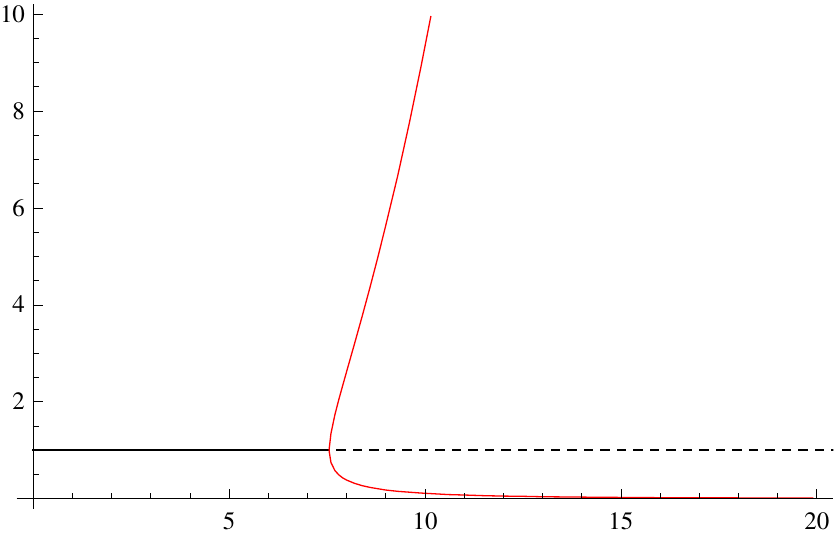}
\put(-106,-20){${\mathbf{\mbox{\boldmath${-\mu}$}}}$}
\put(-351,-4){\small{$0$}}
\put(-380,130){${\mathbf{\frac{x}{y}}}$}
\end{center}
\caption{Relative location of all anisotropic fixed points for $\mu < 0$. }
\label{totani}
\end{figure}

The location of these fixed points is depicted in figure \ref{totani}.
Here, we plot the ratio $x/y$ of the anisotropic fixed points as function of $\mu$
(actually $-\mu$). The horizontal line represents the axially symmetric anisotropic
fixed point that exists below and above the critical value of $\mu$.

At this point, it is instructive to consider the Ricci scalar curvature of the
axially symmetric anisotropic fixed point $x_{\star} = y_{\star} = \sqrt{-3/ \mu}$,
which turns out to be
\begin{equation}
R = -{2 \mu \over 9L^2} \left(\mu + 12 \sqrt{-{3 \over \mu}} \right) ~.
\end{equation}
When $\mu = 0$ this point is at infinity and the curvature is zero.
As $\mu$ varies from $0$ to $-3$ the axially symmetric anisotropic fixed point
is approaching the isotropic fixed point and the curvature increases monotonically.
The curvature becomes maximal when these two points coincide at the critical value
$\mu = -3$, and, then it decreases monotonically as $\mu$ varies from $-3$ to
$-6 \, \sqrt[3]{2}$. It becomes zero at the other critical value
$\mu = -6 \, \sqrt[3]{2}$, and, then, it turns negative for
$\mu < -6 \, \sqrt[3]{2}$. The value $\mu = -6 \, \sqrt[3]{2}$ is also critical for the
creation of the totally anisotropic fixed points, which pop out symmetrically from
the diagonal and have zero curvature for all lower values of the parameter $\mu$.

These additional fixed points appear to be saddle points, as can be verified by
numerical investigation for different values of $\mu$. It is not easy to obtain
closed formulas for the eigenvalues of the characteristic matrix describing small
perturbations around them. However, their stability properties are important for
constructing instanton solutions of Ho\v{r}ava--Lifshitz gravity, as will be seen
in detail later in section 6.  Another characteristic property of these points that
will also be used later is the universal value of the gravitational Chern--Simons
functional. Explicit calculation shows that $W_{\rm CS}$ for the fully anisotropic
fixed points is
\begin{equation}
W_{\rm CS} = {80 \pi^2 \over w_{\rm CS}} ~,
\end{equation}
which is independent of $\mu$! It actually coincides with the value of $W_{\rm CS}$
for the axially symmetric anisotropic fixed point when $\mu = -6 \, \sqrt[3]{2}$ and
it is ten times larger than the value of $W_{\rm CS}$ evaluated at the totally
isotropic fixed point.

An important remark is in order at this point. The equations that determine the fixed
points of the normalized Ricci--Cotton flow provide only the traceless part of the
classical equations of motion of topologically massive gravity, leaving the trace
undetermined. Then, depending on the value of their Ricci scalar curvature, these points
also satisfy the trace equation $R = 6 \Lambda_{\rm W}$ for appropriately chosen
effective value of $\Lambda_{\rm W}$. Fixed points with positive, negative or zero
Ricci scalar curvature are vacua of topologically massive gravity with positive, negative
or zero cosmological constant, respectively. As a result, the fixed points of the
Ricci--Cotton flow with general couplings, which are the vacua of topologically
massive gravity, are expected to be less than the fixed points of the normalized flow,
and, in fact, they can be obtained from them in certain ways\footnote{The mathematical
problem one has to solve to determine the set of fixed points of massive gravity for given
$\Lambda_{\rm W}$ is to fix $R$ instead of the volume of the normalized fixed points,
as functions of $\mu$ and $L$,  and deduce from it the allowed range of $\mu$ for the
selected set of fewer points.}. This, however, does not make
our analysis in section 5 redundant since their location, volume and stability properties
also depend crucially on the flow equations we are considering in each case.

Summarizing the results for the normalized Ricci--Cotton flow, we note that the fully
isotropic metric on $S^3$ is the unique fixed point for $\mu > 0$ that attracts all
trajectories starting from any point in the first quadrant of the $(x,y)$ plane.
For $\mu < 0$, there are various
possibilities that result to attractive or repelling directions around the fixed points.
Note that the isotropic point becomes absolutely unstable when $-3 < \mu < 0$, in which
case the anisotropic fixed point is a saddle that attracts partially the flow lines.
For $-6 \, \sqrt[3]{2} \leq \mu < 0$ there are four in total fixed points, including
their $\mathbb{Z}_3$ mirrors, whereas for $\mu < -6 \, \sqrt[3]{2}$ the total number
of fixed points is ten.

\subsection{Phase portraits of the flow}

The qualitative behavior of the flow lines is illustrated by the phase portraits
shown below for all possible values of $\mu$.

\begin{figure}[!h]
\begin{center}
\includegraphics[height=10cm]{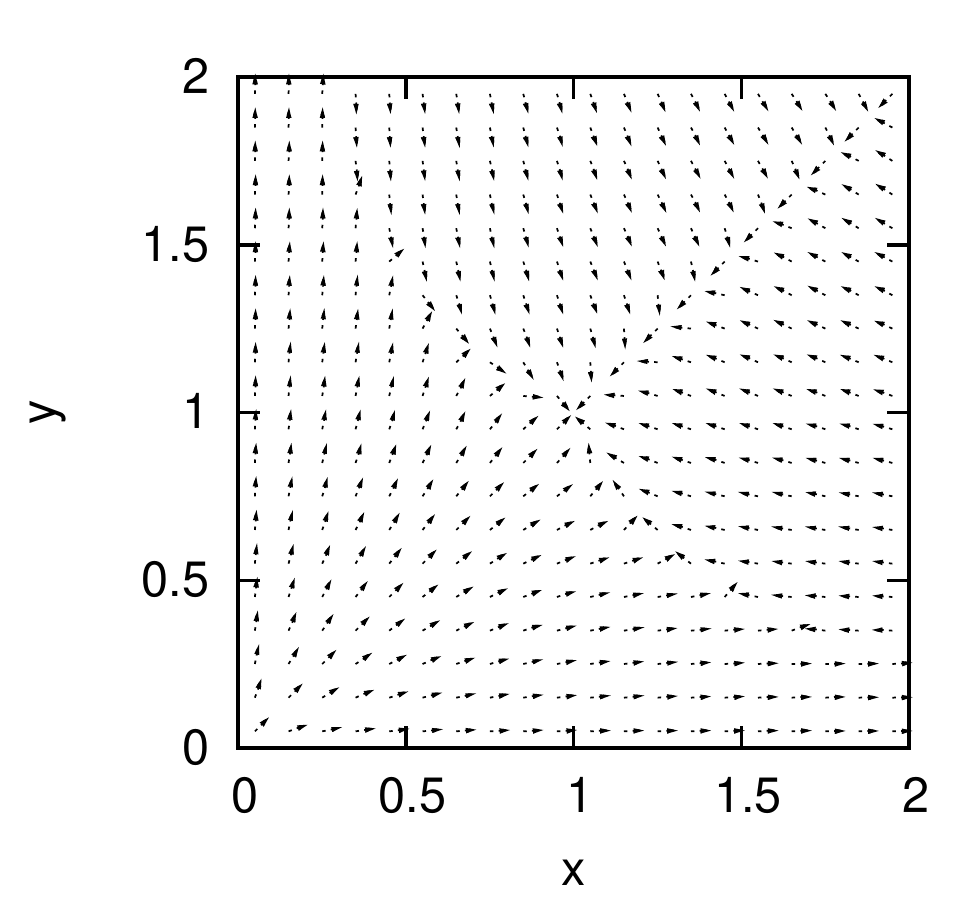}
\end{center}
\caption{The flow lines for $\mu>0$; here, $\mu=2$.}\label{xy0}
\end{figure}

For positive $\mu$, which is qualitatively the same as for the
normalized Ricci flow and the Cotton flow, the phase portrait is given in figure
\ref{xy0}.

\begin{figure}[!h]
\begin{center}
\includegraphics[height=10cm]{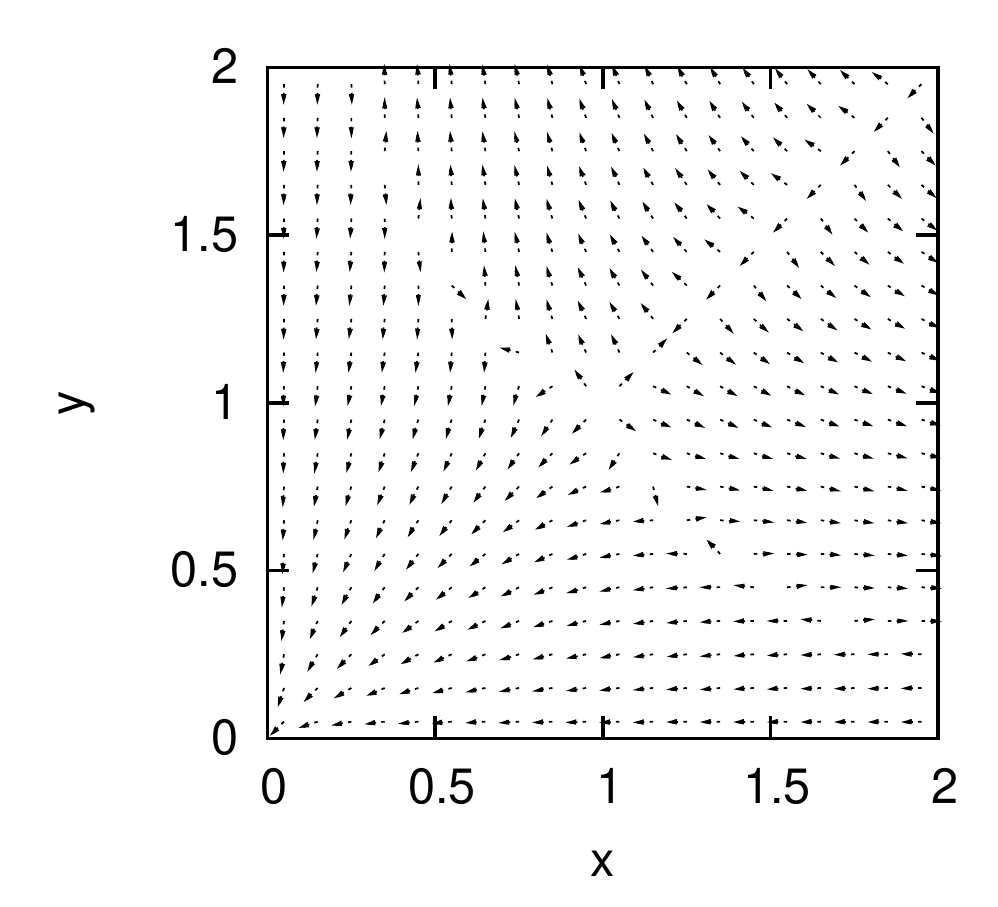}
\end{center}
\caption{The flow lines for $-3<\mu<0$; here, $\mu=-2$.}\label{xy1}
\end{figure}

\begin{figure}[!h]
\begin{center}
\includegraphics[height=10cm]{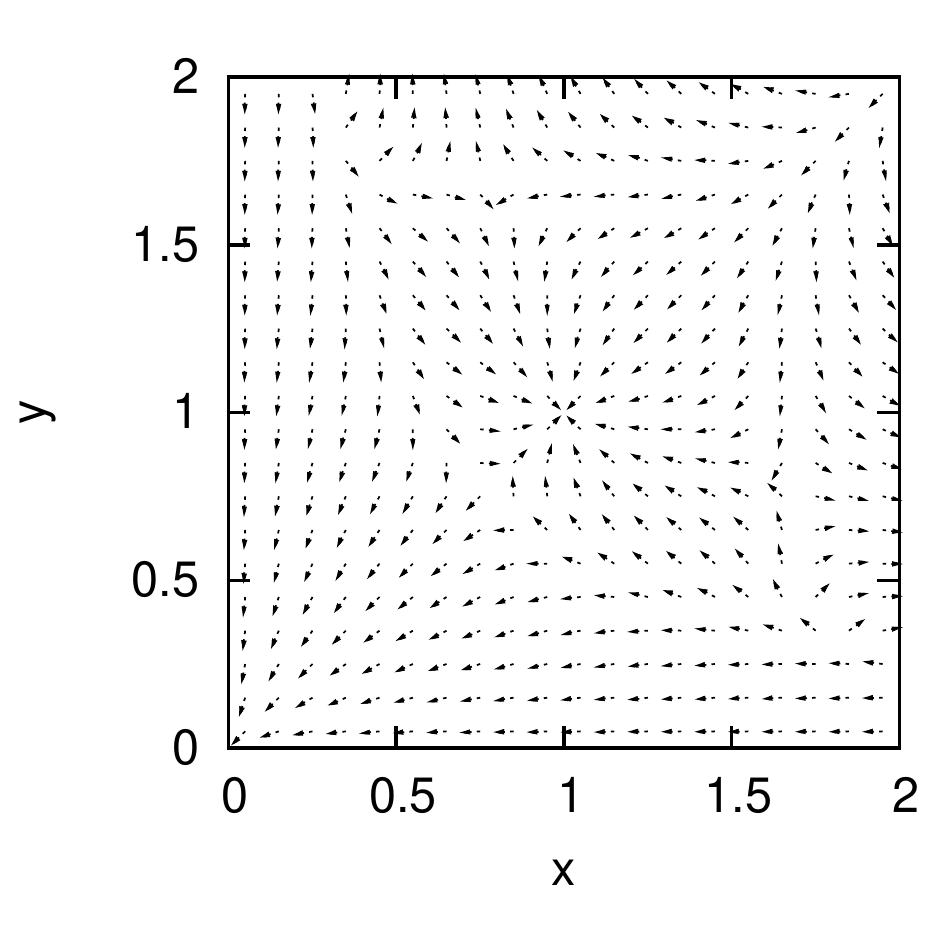}
\end{center}
\caption{The flow lines for  $-6 \, \sqrt[3]{2}<\mu<-3$; here, $\mu=-5$.}\label{xy2}
\end{figure}

\begin{figure}[!h]
\begin{center}
\includegraphics[height=10cm]{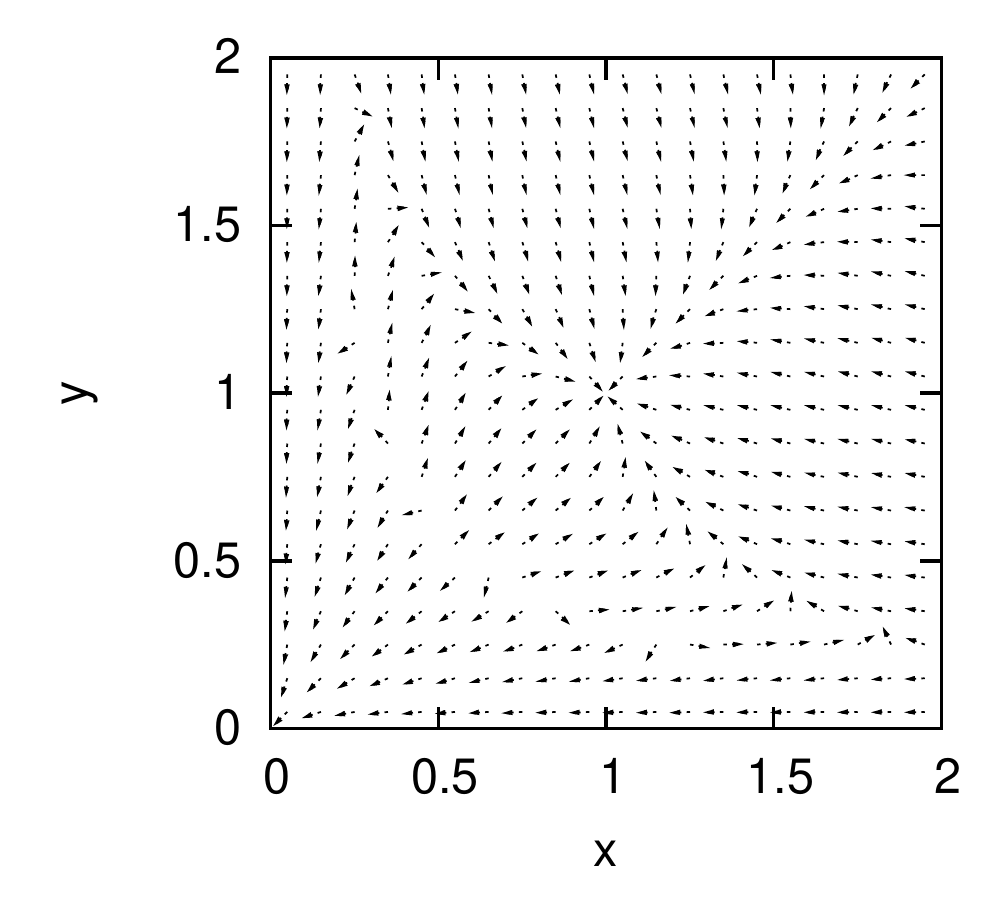}
\end{center}
\caption{The flow lines for $\mu<-6 \, \sqrt[3]{2}$; here,  $\mu=-10$.}\label{xy3}
\end{figure}

For negative $\mu$ the stability properties of the fixed points is different in the
intervals $-3 <\mu < 0$, $-6 \, \sqrt[3]{2}<\mu<-3$  and $\mu<-6 \, \sqrt[3]{2}$ and all
these possibilities are represented in figures \ref{xy1}, \ref{xy2} and \ref{xy3},
respectively. For later reference, it is important to realize
the existence of trajectories interpolating between different fixed points. In
figure \ref{xy1} there is only one such flow line since one fixed point is unstable
and the other a saddle. The same applies to figure \ref{xy2} where the interpolating
flow line connects a stable fixed point with a saddle. The picture changes drastically
in figure \ref{xy3} since there are infinitely many flow lines interpolating
between an unstable and a stable fixed point. Also, in this case, there are additional
fixed points away from the diagonal, which are saddle points.
Figure \ref{xy3} contains two such totally anisotropic fixed points located at
$x \simeq 0.19$, $y \simeq 1.75$ (and $x \simeq 1.75$, $y \simeq 0.19$) for
$\mu = -10 < -6 \, \sqrt[3]{2} \simeq -7.56$. There are flow lines
connecting these fixed points with the other two lying on the diagonal, but they
are not easily seen on the phase portrait due to numerical deficiency.

\subsection{Axisymmetric solutions}

Here, we present the exact axially symmetric solution of the normalized
Ricci--Cotton flow, setting $x=y$. As such, it generalizes the axially
symmetric solutions of the normalized Ricci and the Cotton flows discussed
in the previous section.
The system (\ref{RCNx}) and (\ref{RCNy}) reduces to a single equation
\begin{equation}
\label{axieqRC}
\frac{\mathrm{d}x}{\mathrm{d}\tau}= \frac{1}{\mu}\left(\frac{1}{x^3}-1\right)
\left(\frac{1}{x^2}+\frac{\mu}{3}\right),
\end{equation}
and the flow takes place on the diagonal line of the $(x,y)$ plane. The flow
connects either the origin or infinity with a fixed point or it can extend
between two different fixed points.
These regions do not communicate with each other and the time interval that
supports the solutions depends on $\mu$ and the choice of trajectory.

The behavior around $x=0$ and $x\to\infty$ is universal and can be extracted directly
from equation (\ref{axieqRC}),
\begin{itemize}
\item  $x\approx \left({6 \over \mu}(\tau-\tau_0)\right)^{\nicefrac{1}{6}}$,  \quad as
$x \to 0$,
\item  $x\approx -{1 \over 3}\tau$,  \quad as $x \to \infty$.
\end{itemize}
Around the fixed points the time dependence is exponential and determined by the eigenvalues
$\zeta$ as $x-x_{\star} \approx {\rm exp} (\zeta\tau)$. The relevant eigenvalue for the
isotropic point is given by (\ref{lam}) and for the anisotropic by (\ref{lam1}).

\paragraph{Case I: $\mu > 0$.}

\begin{figure}[!h]
\begin{center}
\includegraphics[height=6.5cm]{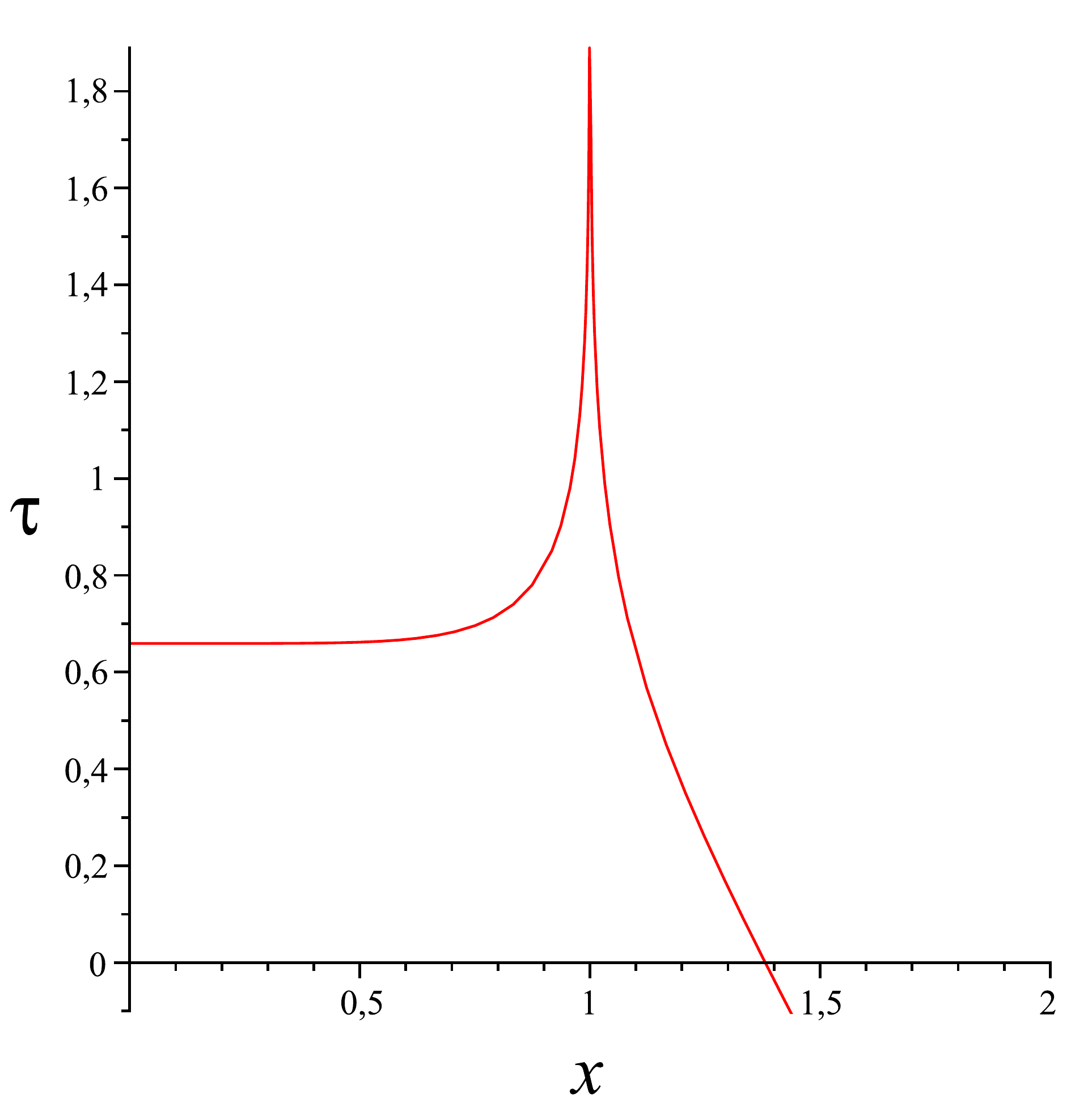}
\end{center}
\caption{The dependence $\tau(x)$ for positive $\mu$; here, $\mu=1$.}\label{mupos}
\end{figure}

The solution behaves similarly for all positive values of $\mu$, but it
looks different on the two sides of the isotropic fixed point $x_{\star}=1$. We find that
\begin{description}
\item[\underline{$x>1$}]: \quad  $-\infty <\tau <+\infty$ as $x$ decreases
from $+\infty$ to $1$~,
\item[\underline{$x<1$}]: \quad $\tau_0<\tau<+\infty$ as $x$ increases from  $x(\tau_0)=0$
to $1$~.
\end{description}
The exact solution reads
\begin{eqnarray}
\tau-\tau_{\star}&= &-3x + \frac{\mu\left(\mu-6\right)}{2\left(\mu^2-3\mu+9\right)}
\log \left(x^2+x+1\right)+\frac{\sqrt{3}\mu^2}{\mu^2-3\mu+9}\arctan\frac{2x+1}{\sqrt{3}}
\nonumber\\
&&-\frac{\mu}{\mu+3}\log\vert x-1\vert +\frac{27 \mu}{2\left(\mu^3+27\right)}\log
\left(\mu x^2+3\right)\nonumber\\
&&+\frac{81}{\mu^3+27} \sqrt{{3 \over \mu}} \arctan \left(x\sqrt{\frac{\mu}{3}}
\right)
\label{eq:mpos}
\end{eqnarray}
and it is represented by figure \ref{mupos} with appropriately chosen
integration constant $\tau_{\star}$.

\paragraph{Case II: $\mu<0$.}

In this case the solution depends on the particular value of $\mu$.
First, we present the result for the critical value $\mu=-3$, which is simpler,
\begin{eqnarray}
\tau-\tau_{\star} & = & - 3x + \frac{1}{2(x-1)}+{1 \over \sqrt{3}} \arctan
\frac{2x+ 1}{\sqrt{3}} -\frac{7}{4}\log |x-1| \nonumber\\
& & + \frac{3}{4}\log (x+1) + {1 \over 2} \log\left(x^2 + x + 1\right) ~.
\label{eq:mucrit}
\end{eqnarray}

\begin{figure}[!h]
\begin{center}
\includegraphics[height=6.5cm]{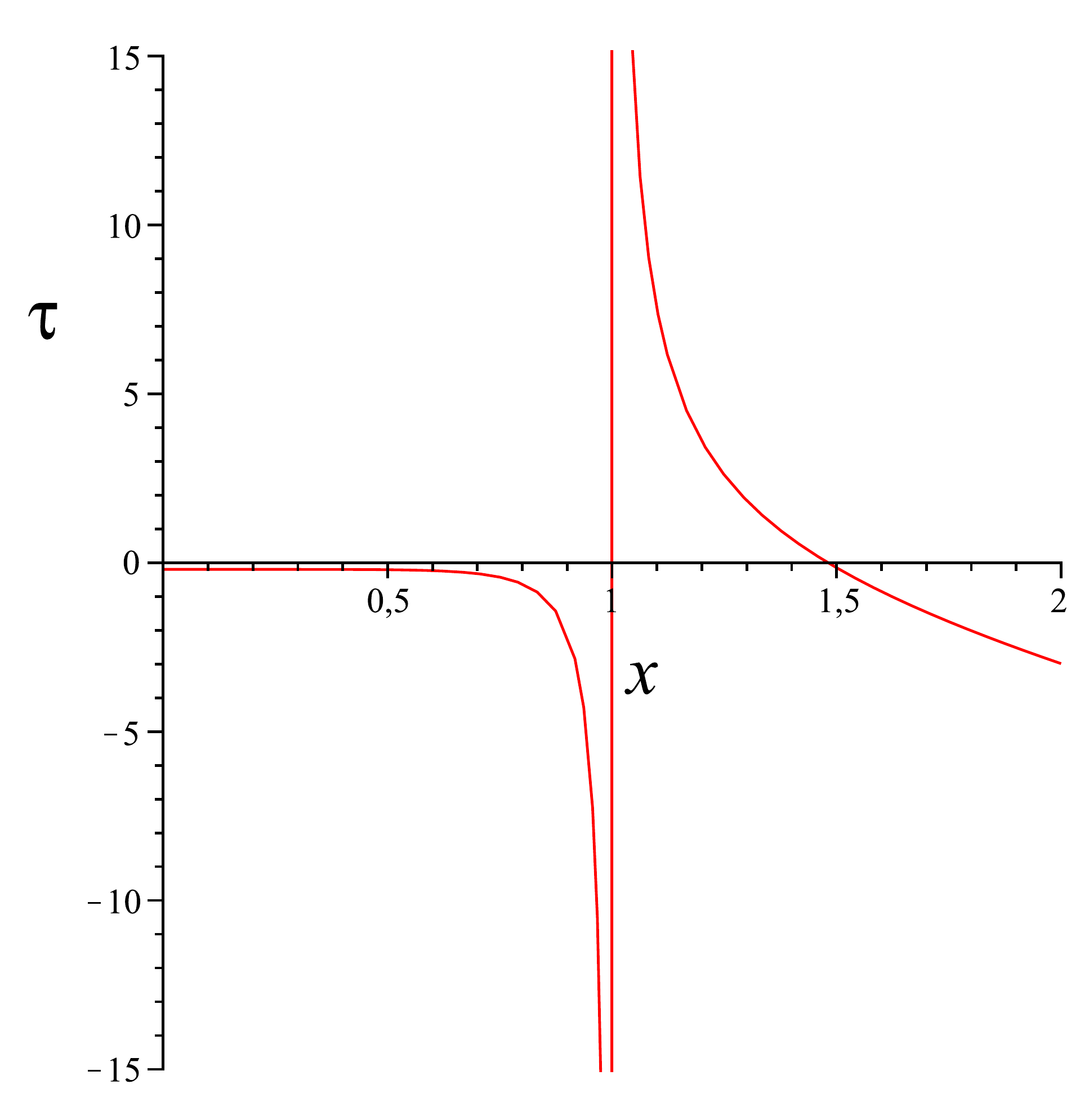}
\end{center}
\caption{The dependence $\tau(x)$ for the critical value critical $\mu=-3$.}\label{mucrit}
\end{figure}
All fixed points coalesce to the isotropic and the eigenvalues vanish, so that there are
no arrows infinitesimally close to this point. However, this behavior is lifted at second
order in perturbation theory and there are arrows pointing from large to small values of
$\tau$. This particular case is depicted in figure \ref{mucrit}, which also shows
the range of $\tau$ in the two branches.

For $\mu\neq -3$ the expression becomes much more involved and reads
\bea \label{formulamuneg}
\tau-\tau_{\star}&= &-3x + \frac{\mu\left(\mu-6\right)}{2(\mu^2-3\mu+9)}
\log \left(x^2+x+1\right)+\frac{\sqrt{3}\mu^2}{\mu^2-3\mu+9}\arctan\frac{2x+1}{\sqrt{3}}
\nonumber\\
&&-\frac{\mu}{\mu+3}\log\vert x-1\vert +\frac{27 \mu}{2(\mu^3+27)}\log
\left|\mu x^2+3\right|\nonumber\\
& &- {81 \over 2 ({\mu}^3+27)}
\sqrt{-{3 \over \mu}} \left[\log \left| 1- x\sqrt{-{\mu \over 3}} \right|
-\log \left|1+  x\sqrt{-{\mu \over 3}} \right| \right].
\eea
Then, depending on whether $-3<\mu<0$ or $\mu<-3$, the function $\tau(x)$ looks
different. In all these cases the solution has three branches but
the range of time is not the same. We have, in particular, the following behavior
depending on $\mu$:
\begin{itemize}
\item For $-3<\mu<0$ the two fixed points are ordered as $x_{\star}^{\mathrm{iso}}
=1<x_{\star}^{\mathrm{aniso}}=\sqrt{-3/\mu}$ with the isotropic being repulsive and
the anisotropic attractive. The exact solution is represented by figure \ref{mugm3}.

\begin{figure}[!h]
\begin{center}
\includegraphics[height=6.5cm]{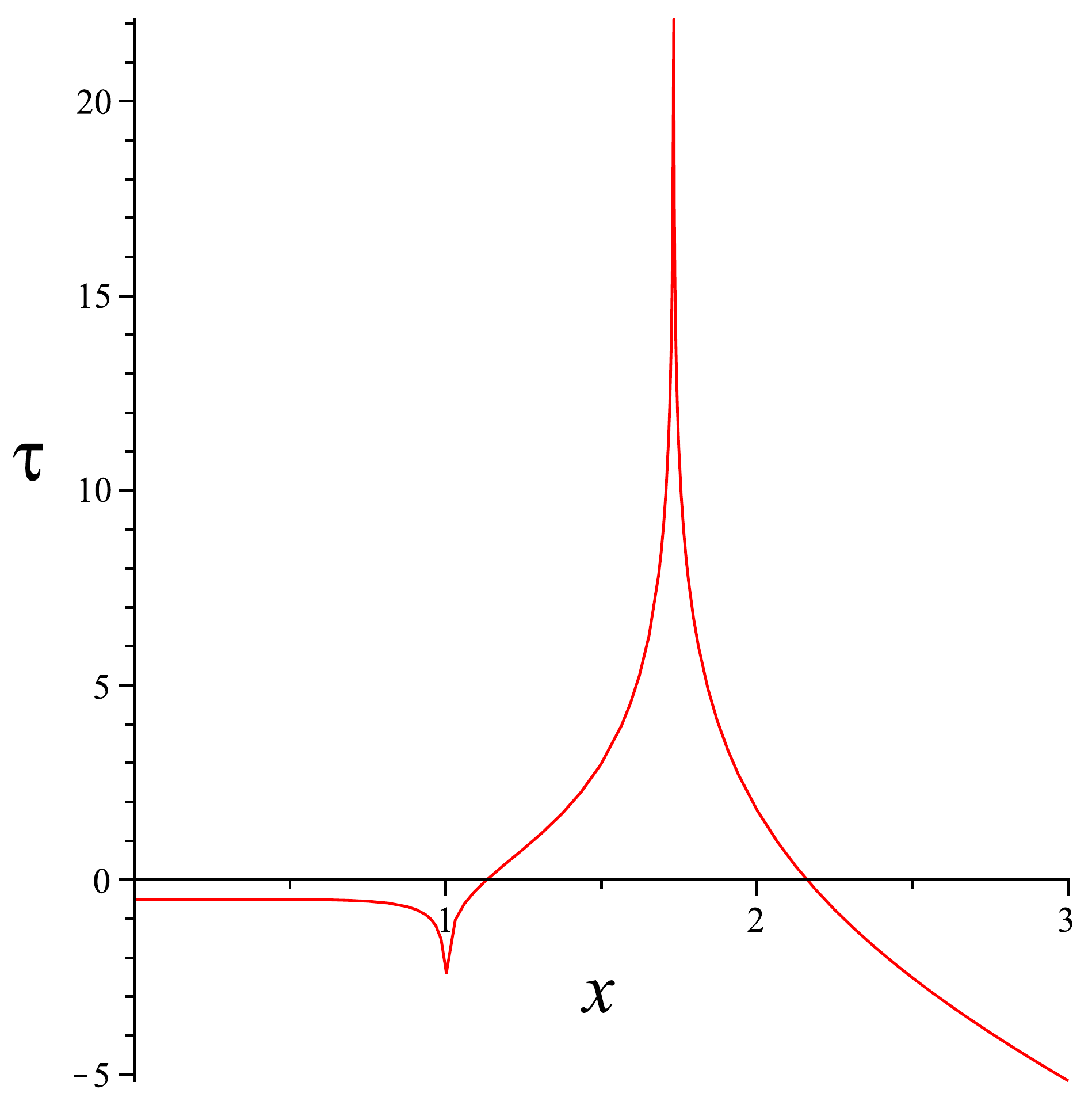}
\end{center}
\caption{The dependence $\tau(x)$ for $-3<\mu<0$; here, $\mu=-1$.}\label{mugm3}
\end{figure}

\item For $\mu<-3$ the two fixed points are ordered differently,
as $x_{\star}^{\mathrm{aniso}}=\sqrt{-3/\mu}<x_{\star}^{\mathrm{iso}}=1$. The
isotropic point is now attractive whereas the anisotropic is repulsive. Then, the exact
solution along the diagonal is represented by figure \ref{mulm3} and there is no
essential difference for $\mu$ below or above the value $-6 \, \sqrt[3]{2}$~.

\begin{figure}[!h]
\begin{center}
\includegraphics[height=6.5cm]{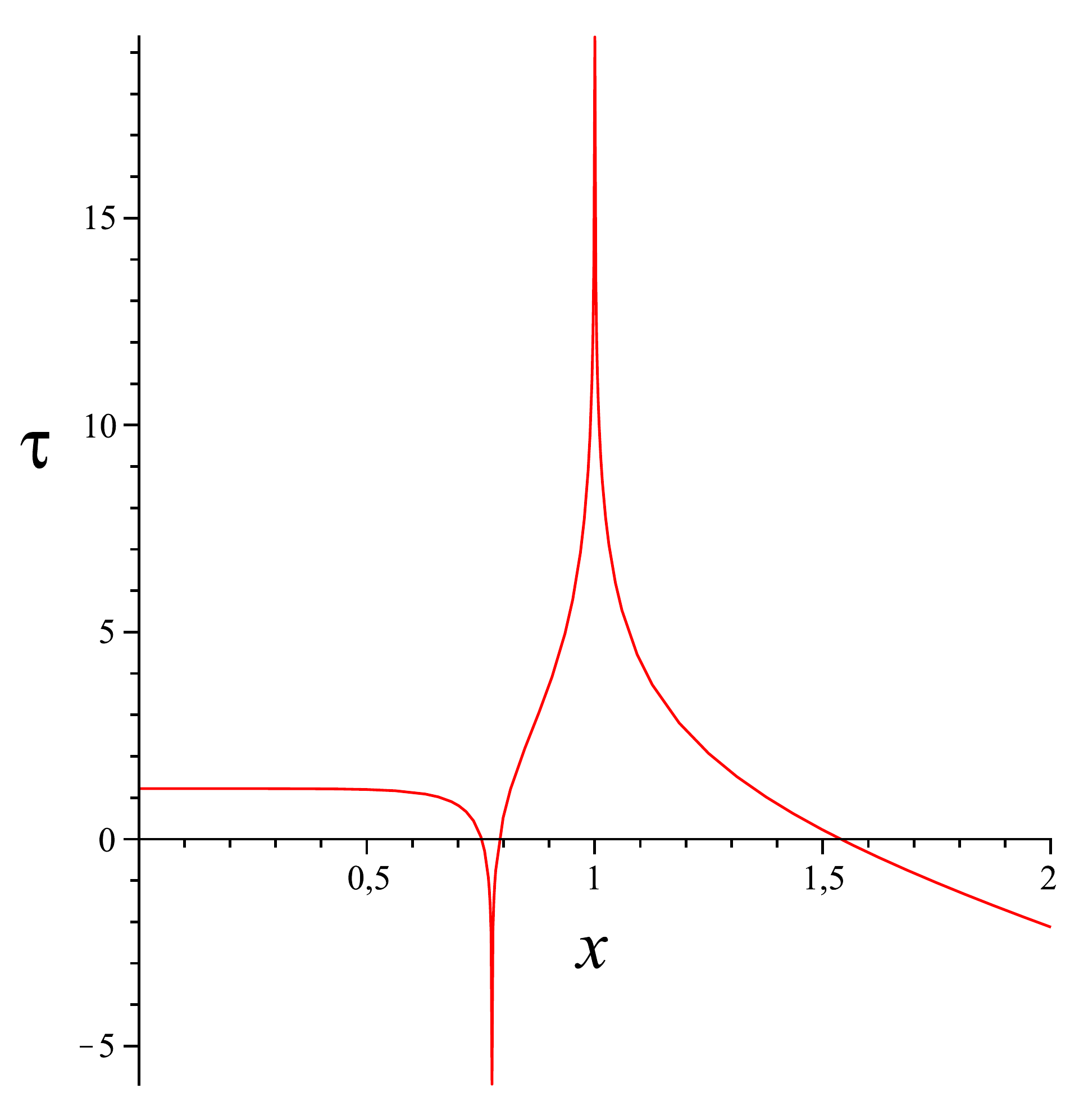}
\end{center}
\caption{The dependence $\tau(x)$ for $\mu<-3$; here, $\mu=-5$.}\label{mulm3}
\end{figure}

Note that the three branches shown in figures 9 and 10 range differently:
the first has support on a semi-infinite time interval, whereas the second and third to
the right are eternal solutions that exist for all time $-\infty < \tau < + \infty$.
(Despite appearances, caused by numerical deficiency, the spike at $x=1$ extends to
infinity in figure 9).
\end{itemize}

\section{Ricci--Cotton flow with general couplings} \label{nonormalisedsol}
\setcounter{equation}{0}

Let us now examine the Ricci--Cotton flow equations for Bianchi IX model geometries
with arbitrary couplings by letting $\lambda$ and $\Lambda_W$ take arbitrary values.
The system of equations that needs to be studied is provided by (\ref{m4dHLfullflow})
\begin{equation}
\label{miromesnil}
{\mathrm{d} \gamma_i \over \mathrm{d} \tau} = -R_{ii} + {2 \lambda -1 \over 2(3\lambda -1)}
R \gamma_i + {\Lambda_W \over 3 \lambda -1} \gamma_i + {1 \over \mu} C_{ii}
\end{equation}
with
\begin{equation}
\tau = {\kappa^2 \over \kappa_W^2} t ~, ~~~~~~ \mu = {w_{\rm CS} \over \kappa_W^2} ~.
\end{equation}
The definition of $\tau$ and $\mu$ resembles that for the normalized flow in section 4,
but it does not include the rescaling by the characteristic length of space.

Since $\lambda < 1/3$, we will confine our discussion to the case of
non-negative cosmological constant, $\Lambda_W \geq 0$, so that the effective speed of
light is real,
and investigate the structure of the fixed points and their stability properties.
The choices $\Lambda_W > 0$ and $\Lambda_W = 0$ will be discussed separately, although the
latter can be obtained as limiting case of the former.
Mathematically it is also interesting to consider the more general situation, without
imposing any restrictions on $\lambda$ and $\Lambda_W$, but these cases will not be included
here\footnote{The nature of the corresponding fixed points
changes drastically compared to the cases that are discussed in this paper. However, when
$\Lambda_W < 0$ it is more appropriate to consider Bianchi type VIII model geometries
rather than Bianchi IX.}.

\subsection{The axisymmetric Bianchi IX model}\label{negcos}

The flow of the metric coefficients $\gamma_i (t)$ does not preserve the volume of $S^3$ in
this case, and, therefore, the three coupled equations are rather difficult to
investigate in all generality with $\gamma_1 \neq \gamma_2 \neq \gamma_3$.
Restricting attention to axially symmetric configurations
simplifies matters without shadowing too much the rich structure of the
system. Our analysis will be based on the qualitative theory of dynamical systems, as in
previous sections, but because of the mathematical complexity of the equations it is not
possible to obtain explicit solutions in closed form, apart from the isotropic solution.
Also, the reader should be aware of the limitations: the fixed points can become unstable
in other directions, when axial symmetry is relaxed, and the conclusions drawn here may
be altered and not be as general. Of course,
this is part of a more general criticism for mini-superspace models compared to metric
deformations with an infinite number of moduli taking place in the entire superspace.

With these explanations in mind, we are going to study the system of two equations
\begin{eqnarray}
\label{totiral}
{1 \over \gamma_1} {\mathrm{d} \gamma_1 \over \mathrm{d} \tau} & = & - {\sqrt{\gamma_1}
\over \mu ~ \gamma_2^3}
(\gamma_1 - \gamma_2) - {8 \lambda -3 \over 4(3 \lambda - 1)} {\gamma_1 \over \gamma_2^2}
+ {2 \lambda -1 \over 3 \lambda - 1} {1 \over \gamma_2} + {\Lambda_W \over 3 \lambda - 1} ~, \\
{1 \over \gamma_2} {\mathrm{d}  \gamma_2 \over \mathrm{d} \tau} & = & {\sqrt{\gamma_1}
\over 2 \mu ~ \gamma_2^3}
(\gamma_1 - \gamma_2) + {4 \lambda -1 \over 4(3 \lambda - 1)} {\gamma_1 \over \gamma_2^2}
- {\lambda \over 3 \lambda - 1} {1 \over \gamma_2} + {\Lambda_W \over 3 \lambda - 1} ~,
\end{eqnarray}
which is obtained from (\ref{miromesnil}) by setting $\gamma_2 = \gamma_3$.
The existence and properties of the fixed points depends crucially on the values of $\mu$,
as in other examples considered so far. In particular, for $\mu < 0$, an anisotropic
fixed point will coexist with the isotropic one.

There is an exact solution of these equations which is available for all $\Lambda_W \geq 0$
and describes the evolution of the isotropic metric on $S^3$. This possibility does not
arise for the normalized Ricci--Cotton flow, since the isotropic metric is a fixed point.
In particular, setting all $\gamma_i \equiv \gamma$, we have the following result:
\begin{equation}
\label{exactama}
\gamma (t) = A ~ {\rm exp} \left({\Lambda_W \over 3\lambda -1} \tau \right) +
{1 \over 4\Lambda_W} \quad {\rm for}\   \Lambda_W > 0
\end{equation}
and
\begin{equation}
\label{exactama2}
\gamma (t) = -{1 \over 4(3\lambda -1)} (\tau- \tau_0) \quad {\rm for} \ \Lambda_W = 0
\end{equation}
with $A$ and $\tau_0$ being arbitrary integration constant. Since $\lambda < 1/3$,
the metric flows to the isotropic fixed point (to be
discussed next in detail for more general trajectories) as $\tau \rightarrow + \infty$.

The difference between $\Lambda_W >0$ and $\Lambda_W = 0$ is reflected in the
life-time of the solutions. When $\Lambda_W > 0$ the solution has two branches: on the
first branch $A>0$ and the solution is
eternal existing for all $-\infty < \tau < + \infty$; it describes a round sphere with
infinite radius in the infinite past flowing towards a round sphere with radius
set by $\Lambda_W$ in the infinite future. The second branch corresponds to $A<0$ in
which case the solution exists for $\tau_{\star} \leq t < + \infty$, with appropriately
chosen $\tau_{\star}$, so that the sphere starts from zero radius and reaches the fixed
point as $t \rightarrow + \infty$.  On the other hand, when $\Lambda_W = 0$, there is
only one branch as the solution exists for $\tau_0 \leq \tau < +\infty$, interpolating
between a fully collapsed configuration at $\tau=\tau_0$ to a sphere of infinite radius
in the infinite future.

\subsection{Classification of the fixed points}

First, we consider the case of non-vanishing cosmological constant and reserve the
last subsection to study $\Lambda_W = 0$ separately.

\paragraph{The isotropic fixed point.}

For $\Lambda_W > 0$, there is a natural length scale in the problem that gives rise to the
isotropic fixed point of the flow, irrespective of the sign of $\mu$,
\be
\gamma_1 = \gamma_2 = \gamma_3 = {1 \over 4\Lambda_W} ~.
\ee
This follows easily from the system of equations without assuming any restrictions on
$\gamma_1$ and $\gamma_2$; it can also be shown that it is a fixed point of the more
general system of equations with $\gamma_1 \neq \gamma_2 \neq \gamma_3$.

Linearizing around it as
\be
\gamma_1 (t) = {1 \over 4\Lambda_W} ~ \left(1 + \delta x(t) \right) , \quad
\gamma_2 (t) = {1 \over 4\Lambda_W} ~ \left(1 + \delta y(t) \right) ,
\ee
we find that the small perturbations satisfy the characteristic matrix
equation
\be
{\mathrm{d} \over \mathrm{d} \tau}  \begin{pmatrix}
\delta x \\
\\
\delta y    \end{pmatrix}
=\Lambda_W
\begin{pmatrix}
\displaystyle{-\frac{8}{\mu} \sqrt{\Lambda_W} - \frac{8 \lambda -3}{3 \lambda -1}}
&   \displaystyle{ \frac{8}{\mu} \sqrt{\Lambda_W} + 2 \frac{4 \lambda -1}{ 3 \lambda -1}} \\
&\\
\displaystyle{\frac{4}{\mu} \sqrt{\Lambda_W} + \frac{4 \lambda -1}{3 \lambda -1}  }&
\displaystyle{-\frac{4}{\mu} \sqrt{\Lambda_W} - 2 \frac{2 \lambda -1}{ 3 \lambda -1}}
\end{pmatrix}
\begin{pmatrix}
\delta x \\
  \\
\delta y    \end{pmatrix}.
\ee
The eigenvalues are
\be
\zeta_1 = {\Lambda_W \over 3 \lambda - 1} ,\quad
\zeta_2 = -4\Lambda_W \left(1 + {3 \over \mu} \sqrt{\Lambda_W} \right)
\ee
and so $\zeta_1 < 0$ whereas $\zeta_2$ can take all values, positive or negative,
depending on $\mu$.

Keeping $\Lambda_W$ fixed and varying $\mu$ we obtain the following characterization
of the isotropic fixed point:
\begin{itemize}
\item $\mu > 0$  \ or \   $\mu < -3 \sqrt{\Lambda_W}$: \quad  absolutely  stable,
\item $-3 \sqrt{\Lambda_W} < \mu < 0$: \quad saddle  point.
\end{itemize}
Note the emergence of a critical value,  $\mu = -3 \sqrt{\Lambda_W}$,
where the two eigenvalues $\zeta_1$ and $\zeta_2$ vanish, separating
stability from instability along the corresponding eigen-directions\footnote{The
critical value of $\mu$ is similar to that found for the normalized Ricci--Cotton
flow; direct comparison can be made by replacing $\sqrt{\Lambda_W}$ with $1/L$ and
rescaling $\mu$ with the characteristic length of space.
In both cases, the critical value of $\mu$ occurs when the competing
effects of the Ricci and Cotton deformations are balanced exactly.}.
Also note for completeness that if we were allowing $\lambda > 1/3$, the
isotropic point would never be absolutely stable (it would be absolutely unstable or
a saddle point in the respective intervals of the $\mu$-line.).

Actually, one can go further and investigate whether the exact isotropic running
solution $\gamma(t)$, given by \eqn{exactama}, is stable against small fluctuations,
\begin{equation}
\gamma_i (t) = \gamma(t) + \delta \gamma_i (t) ~,
\end{equation}
acting as {\em attractor} of nearby trajectories. Thus, given a small tube around
the trajectory $\gamma (t)$, one is interested to know if any other trajectory with
initial conditions inside this tube will remain there after some time and what is
the size of tube that guarantees this attractor property. The differential equations
for $\delta \gamma_i (t)$ are most conveniently stated using $\gamma (t)$ rather
that $t$ as flow time. Then, within the axially symmetric ansatz \eqn{totiral}, the
linearized equations take the following form,
\begin{eqnarray}
{\mathrm{d} \over \mathrm{d} \gamma (t)} \delta \gamma_1 (t) & = & -{2 (\delta \gamma_1
- \delta \gamma_2) \over \gamma (4 \Lambda_W \gamma -1)}
\left[{2(3\lambda -1) \over \mu \sqrt{\gamma}} + 4\lambda -1\right]
+ {4\Lambda_W \delta \gamma_1 \over 4\Lambda_W \gamma - 1} ~, \\
{\mathrm{d} \over \mathrm{d} \gamma (t)} \delta \gamma_2 (t) & = & {\delta \gamma_1
- \delta \gamma_2 \over \gamma (4 \Lambda_W \gamma -1)}
\left[{2(3\lambda -1) \over \mu \sqrt{\gamma}} + 4\lambda -1\right]
+ {4\Lambda_W \delta \gamma_2 \over 4\Lambda_W \gamma - 1} ~,
\end{eqnarray}
generalizing the characteristic matrix equations of small perturbations around the
isotropic fixed point. These equations apply for all $\Lambda_W$ including the
special case $\Lambda_W =0$ that will be discussed separately.

Solutions of this system can be expressed as functions of $t$ through $\gamma (t)$.
Since $\delta \gamma_i$ are required to be small, for validity of the linearized
analysis, the attractor property of the isotropic trajectory appears to be very
limited. This behavior can be seen schematically in the phase portraits of the
flow that will appear in the next two subsections.

\paragraph{Anisotropic fixed point.}

When $\mu <0$, there is an additional fixed point associated with the axially
symmetric metric with coefficients
\be\label{Lamaxifp}
\gamma_1 = {36 \mu^2 \over \left(\mu^2 + 27 \Lambda_W \right)^2} ,\quad
\gamma_2 = \gamma_3 = {9 \over \mu^2 + 27 \Lambda_W} .
\ee
There are no other restrictions on the values of $\mu$ for the existence of this
second fixed point.

Notice that this new fixed point and the isotropic one will coalesce if $\mu =
-3 \sqrt{\Lambda_W}$. For $-3 \sqrt{\Lambda_W} < \mu < 0$, the anisotropic
fixed point has $\gamma_1 < \gamma_2$, whereas for $\mu < -3 \sqrt{\Lambda_W}$
it has $\gamma_1 > \gamma_2$. Thus, $-3 \sqrt{\Lambda_W}$ appears as a critical
value of $\mu$.

By considering small perturbations around the anisotropic fixed point, as
\be
\gamma_1 (t) = {36 \mu^2 \over \left(\mu^2 + 27 \Lambda_W \right)^2}
\left(1 + \delta x(t)\right) , \quad
\gamma_2 (t) = {9 \over \mu^2 + 27 \Lambda_W} \left(1 +  \delta y(t) \right),
\ee
we find the characteristic matrix of the linearized system with respect to $\tau$
\be
{1 \over 27 (3\lambda -1)}
\begin{pmatrix}
(9\lambda - 2)\mu^2 - 27\Lambda_W (3\lambda - 1) &   -(18\lambda - 5) \mu^2
+ 27 \Lambda_W (6\lambda -1) \\
-{1 \over 2} (9\lambda - 5) \mu^2 + {27 \over 2} \Lambda_W (3\lambda -1) &
(9\lambda - 4)\mu^2 - 27\Lambda_W (3\lambda - 2)
\end{pmatrix}.
\ee
The corresponding eigenvalues are
\be
\zeta_{\pm} = {1 \over 18 (3\lambda -1)} \left[2(3\lambda -1) \mu^2 - 27 \Lambda_W
(2\lambda -1) \pm \sqrt{\Delta} \right] ,
\ee
where
\be
\Delta = 6(3\lambda -1) (2 \lambda -1) \mu^4 - 72 \Lambda_W \mu^2 (3\lambda -1)^2
+ 243 \Lambda_W^2 \left(12\lambda^2 -6\lambda +1\right) ~ .
\label{descrim}
\ee

The eigenvalues $\zeta_{\pm}$ are real\footnote{It follows by noting that the two roots
of $\Delta$ occur at
\be
\mu_{\pm}^2 = {3 \Lambda_W \over 2 (3\lambda -1) (2\lambda -1)} \left[4(3\lambda -1)^2
\pm \sqrt{2(3\lambda -1)} \right] \nonumber
\ee
and they are complex for $\lambda < 1/3$. Thus, $\Delta$ has the same sign as the
coefficient of its $\mu^4$-term, which is positive.}. Since their product is given by
\be\label{eigepro}
\zeta_+ \zeta_- = {\left(\mu^2 - 9 \Lambda_W\right)\left (\mu^2 + 27 \Lambda_W\right) \over
162 (3\lambda -1)} ,
\ee
we note the appearance of a critical value $\mu$, which is the same as for the isotropic
fixed point, $\mu = -3 \sqrt{\Lambda_W}$. Then, for $\mu < -3 \sqrt{\Lambda_W}$ the
anisotropic point is saddle. On the other hand, in order to
examine the stability of this fixed point for $-3 \sqrt{\Lambda_W} < \mu < 0$, we consider
the sum of the two eigenvalues,
\be
\zeta_+ + \zeta_- = {1 \over 9 (3\lambda -1)} \left[2 (3\lambda -1) \mu^2 -
27 \Lambda (2\lambda -1) \right] ~,
\label{eigencha}
\ee
which is now negative. Therefore, for $-3 \sqrt{\Lambda_W} < \mu < 0$, the anisotropic
fixed point is absolutely stable.
Summarizing all results obtained above, we have the following:
\begin{itemize}
\item An isotropic fixed point exists for all $\mu$ and it is absolute stable when $\mu > 0$
or $\mu < -3 \sqrt{\Lambda_W}$. For $-3 \sqrt{\Lambda_W} < \mu < 0$ it is a saddle point.
\item An anisotropic fixed point exists for all $\mu<0$. It is absolutely stable for
$-3 \sqrt{\Lambda_W} < \mu < 0$ and saddle for $\mu < -3 \sqrt{\Lambda_W}$, which
is reverse to the behavior of the isotropic fixed point.
\end{itemize}

As can be seen there are similarities as well as some differences with the classification
of fixed points of the normalized Ricci--Cotton flow.

\subsection{Phase portraits of the flow}

A qualitative picture of the flow lines is provided by three consecutive phase
portraits for different values of the parameter $\mu$.
In all drawings we choose
$\Lambda_W = 0.25$ and $\lambda = 0.1$, and so the three regimes are $\mu > 0$ or
$-1.5 < \mu <0$ or $\mu < -1.5$. The isotropic fixed point appears at $\gamma_1
= \gamma_2 = 1/4\Lambda_W = 1$.

\begin{figure}[!h]
\begin{center}
\includegraphics[height=10cm]{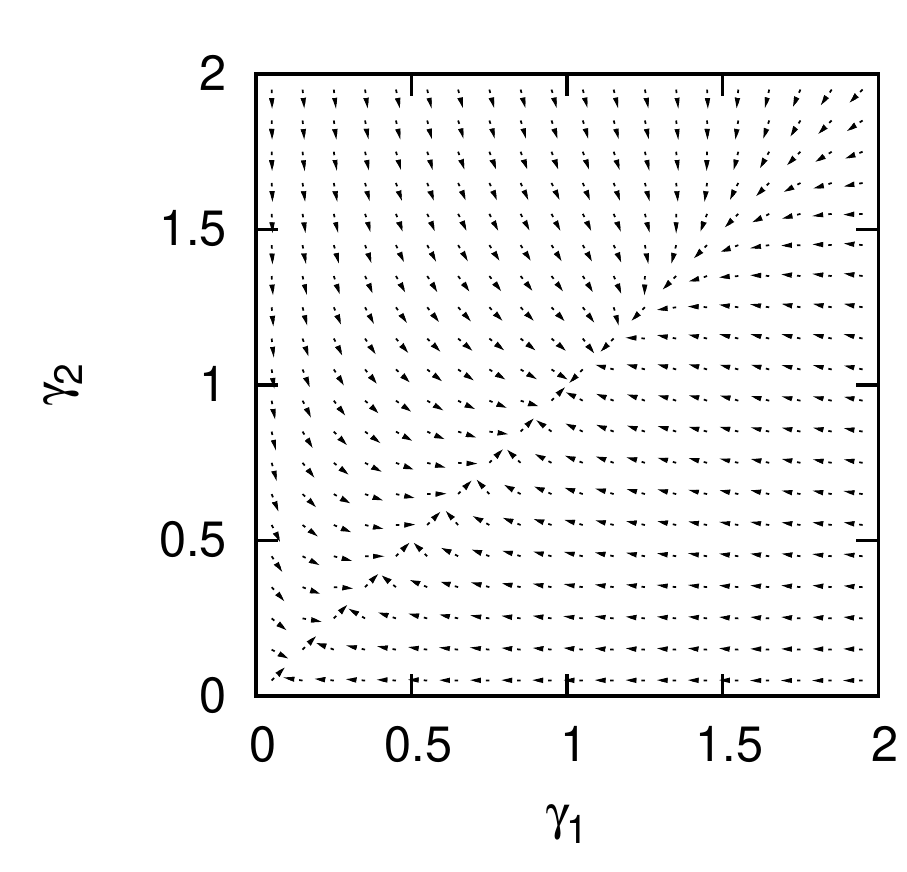}
\end{center}
\caption{The flow lines for $\mu>0$; here, $\mu=1$.}\label{fig10}
\end{figure}

First, we consider the case $\mu >0$ that exhibits only one (isotropic) fixed point,
as shown in figure \ref{fig10}.

Next, we consider the case $\mu <0$ that exhibits an additional (anisotropic) fixed
point and make the following choices for the plots shown in figures  \ref{fig11} and
\ref{fig12}:

\begin{figure}[!h]
\begin{center}
\includegraphics[height=10cm]{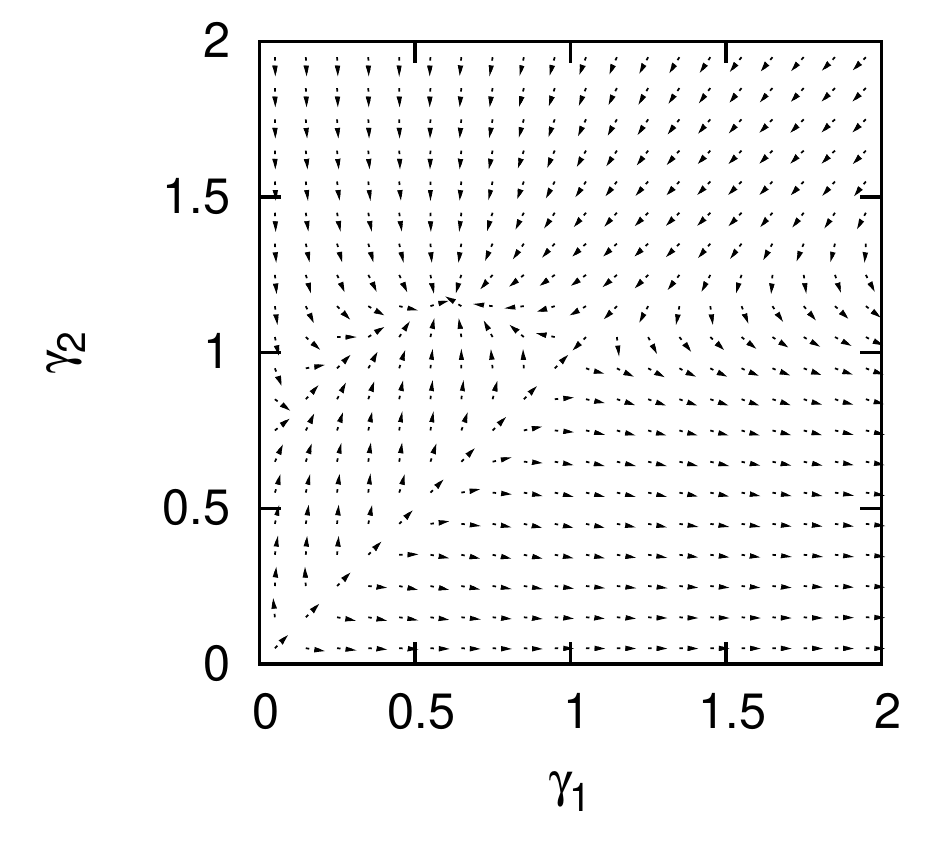}
\end{center}
\caption{The flow lines for $-1.5 = -3 \sqrt{\Lambda_W} <\mu<0$; here, $\mu=-1$.}\label{fig11}
\end{figure}

\begin{figure}[!h]
\begin{center}
\includegraphics[height=10cm]{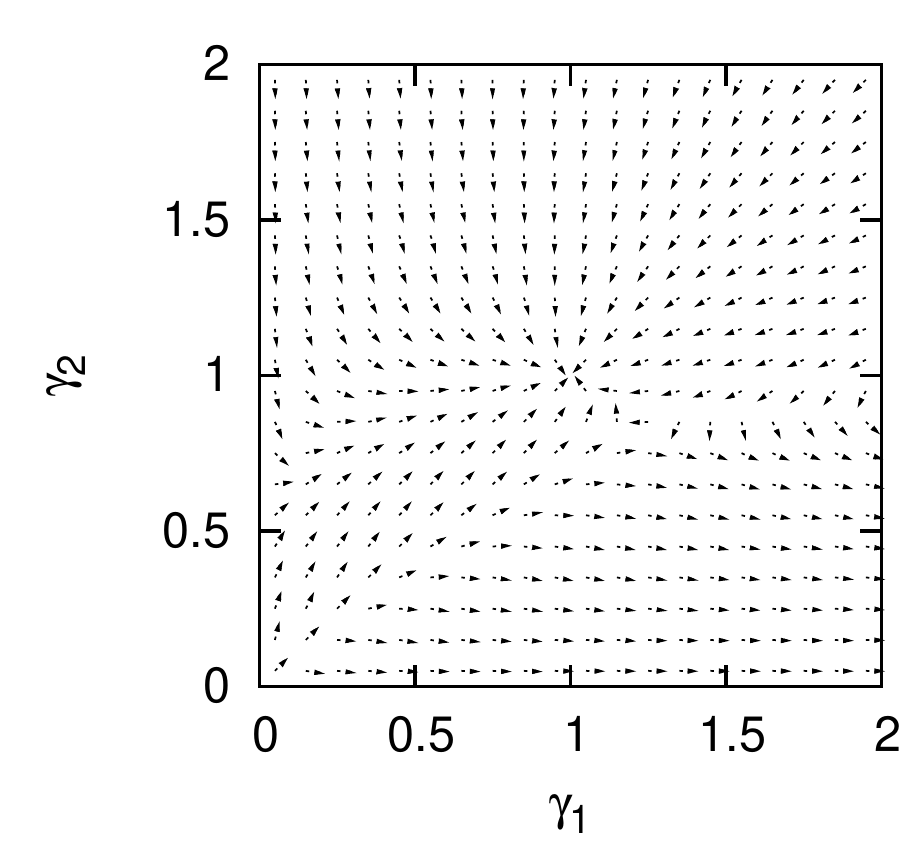}
\end{center}
\caption{The flow lines for $\mu<-3 \sqrt{\Lambda_W} =-1.5$; here, $\mu=-2$.}\label{fig12}
\end{figure}

\begin{itemize}
\item $\mu = -1$: the anisotropic fixed point occurs at $\gamma_1 \simeq 0.60$, $\gamma_2 =
\gamma_3 \simeq 1.16$.
\item $\mu = -2$: the anisotropic fixed point occurs at $\gamma_1 \simeq 1.24$, $\gamma_2 =
\gamma_3 \simeq 0.84$.
\end{itemize}

\boldmath
\subsection{The special case $\Lambda_W = 0$}
\unboldmath

Setting $\Lambda_W = 0$ corresponds to taking the effective speed of light equal to zero,
while keeping $\lambda < 1/3$ arbitrary. It should be contrasted with the normalized
Ricci--Cotton flow, which is independent of $\Lambda_W$ and has also zero effective speed
of light.

The isotropic fixed point is now pushed to infinity and corresponds to a round $S^3$
with infinite radius. This is also apparent from the exact isotropic solution
\eqn{exactama2} that converges to it after infinitely long time. However, it is not
strictly speaking a fixed point of the flow lines, since $\mathrm{d}\gamma_1/\mathrm{d}t$ and
$\mathrm{d}\gamma_2/\mathrm{d}t$ do not vanish there when $\Lambda_W = 0$.

When $\mu < 0$ there is an anisotropic fixed point of the axially symmetric
flow for
\begin{equation}
\gamma_1 = {36 \over \mu^2} = 4\gamma_2 = 4\gamma_3 ~.
\end{equation}
This is always a saddle point because the corresponding eigenvalues of the characteristic
matrix of small perturbations are
real for $\lambda <1/3$ and their product is negative. The results follow setting
$\Lambda_W = 0$ in the expressions we had before (see e.g. (\ref{eigepro})).

\begin{figure}[!h]
\begin{center}
\includegraphics[height=9cm]{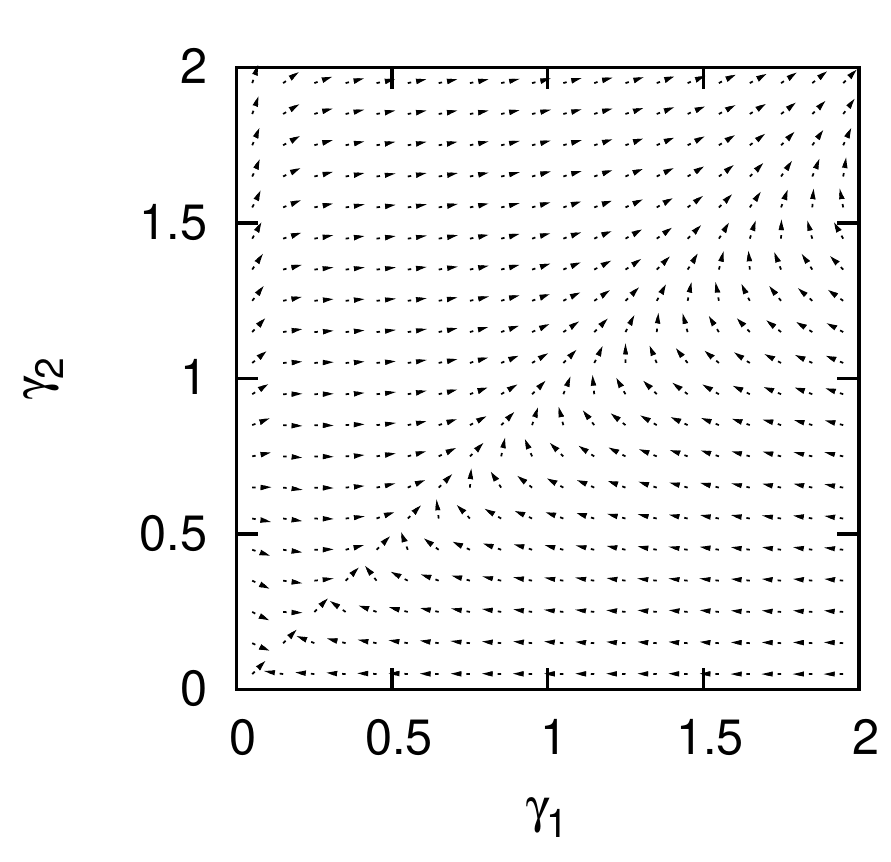}
\end{center}
\caption{The flow lines for $\mu > 0$; here, $\mu=1$.}\label{fig13}
\end{figure}

\begin{figure}[!h]
\begin{center}
\includegraphics[height=10cm]{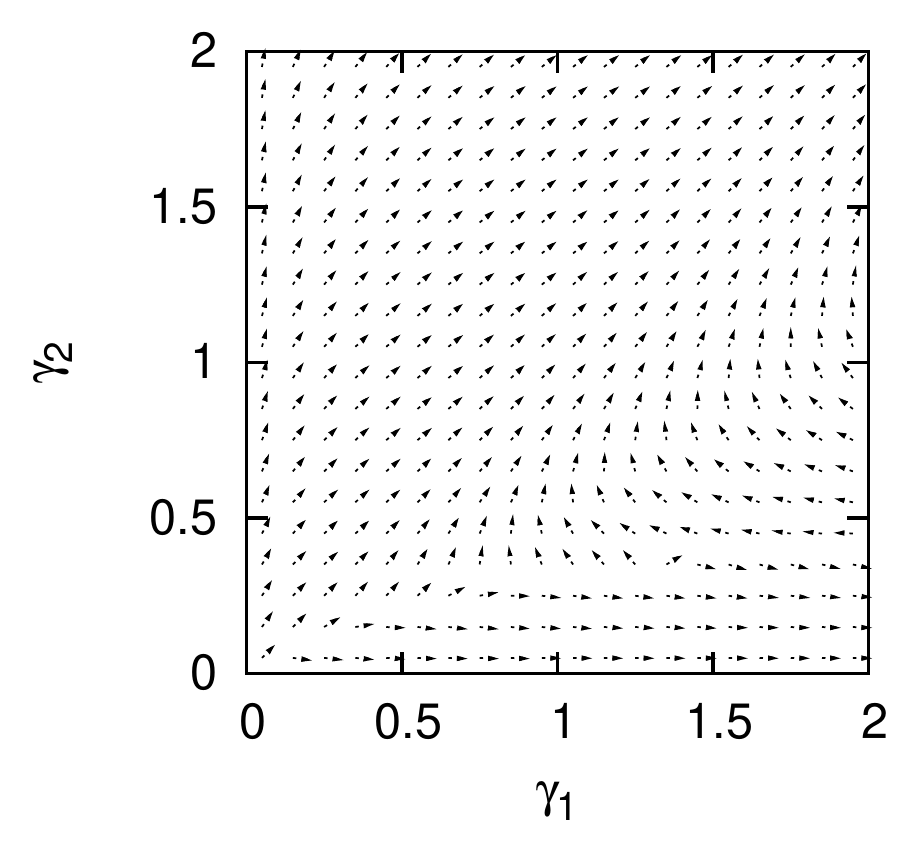}
\end{center}
\caption{The flow lines for $\mu < 0$; here, $\mu=-5$.}\label{fig14}
\end{figure}

We include two phase portraits of the flow lines that are characteristic for
$\mu >0$ and $\mu <0$, respectively, choosing $\lambda = 0.1$.
The case $\mu >0$ is shown first in figure \ref{fig13}.
Next, we consider
the case $\mu <0$ that exhibits an additional (anisotropic) fixed
point and make the choice $\mu = -5$. The anisotropic fixed point occurs at $\gamma_1
\simeq 1.44$, $\gamma_2 = \gamma_3 \simeq 0.36$ as shown in figure  \ref{fig14}.

\begin{figure}
\begin{center}
\includegraphics[scale=1]{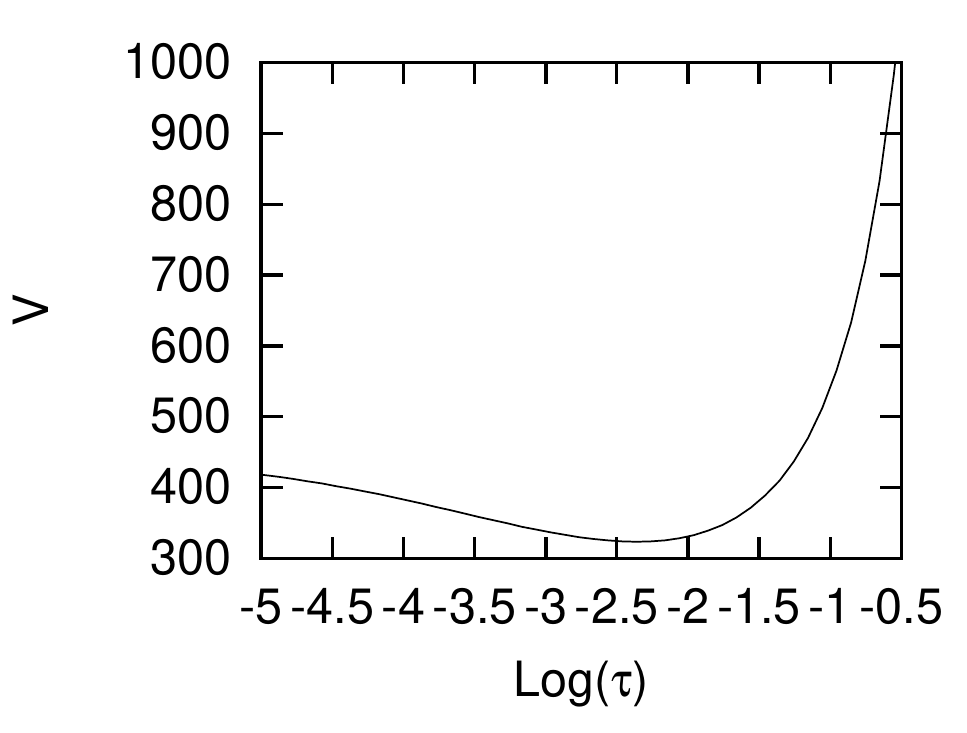}
\caption{Bouncing solution with initial data $\gamma_1^{(0)}=28$, $\gamma_2^{(0)} =
\gamma_3^{(0)} =4$ for $\lambda =0.26$, $\Lambda_W = 0$ and $\mu=-5$.\label{bounces}}
\end{center}
\end{figure}

Finally, concluding this section, we end up with an interesting observation that
arose in our study of the flow lines. Along these lines, the volume of space changes, but
the dependence on $t$ is not monotonic in general. It is therefore
interesting to inquire in this context for the existence of bouncing solutions for which the
volume reaches a minimum and then increases in time. Although this behavior is not
generic, it seems to arise along particular flow lines that can be found by
numerical scanning. An example of this kind is provided in figure \ref{bounces}
for an appropriate choice of initial data and couplings.
Cases with full anisotropy, but with initial conditions close to axial symmetry, also
seem to lead to bounces, including minima with very small volume.
Similar conclusions hold when $\Lambda_W > 0$, with or without axial symmetry, but we have
not been able to obtain any quantitative characterization of the phenomenon so far.

The bouncing solutions
can be regarded (in some sense) as the Euclidean space analogue of bouncing
models in standard cosmology, which provide a viable alternative to inflation.
Matter bounces have already appeared in studies of Ho\v{r}ava--Lifshitz cosmology
(see, for instance, \cite{calca1, calca2, calca3}), but they are also
non-generic. It remains to be seen whether they have any special meaning and
implications for the models we study here, although their occurrence
does not require any matter couplings, as they are purely geometric, and, hence,
different from those arising in cosmology.

\section{Space-time interpretation of the flow lines}
\setcounter{equation}{0}

The solutions of geometric flows are shown as flow lines in the
various phase portraits that have been drawn. Explicit solutions were
obtained in special cases, whereas more general solutions can only be described
pictorially. The problems that will be addressed in this section are the selection
of flow lines that can qualify as regular gravitational-instanton solutions in four
dimensions and the completeness of the corresponding space-time metrics. It
will also be useful in this context to compare $SU(2)$ instanton solutions
of Ho\v{r}ava--Lifshitz theory with those of ordinary Einstein gravity.

According to the analysis of section \ref{sec23}, only the  flow lines that
interpolate between two fixed points (when more than
one fixed point is present in our models) qualify as instantons. They are indeed
finite-action solutions, and this property is sufficient
to determine the global structure of space--time and its asymptotic behavior,
and render the corresponding metrics complete. Our main result in a nut-shell
is that all gravitational instantons with $SU(2)$ isometry
are globally $\mathbb{R} \times S^3$ describing a smooth deformation of $S^3$ as
$t$ runs from $-\infty$ to $+\infty$, without ever encountering a singularity.
The details are given below together with the {\em complete classification} of
gravitational instantons with $SU(2)$ isometry in Ho\v{r}ava--Lifshitz gravity
with anisotropy parameter $z=3$.
We will also compute their Euclidean action and determine their moduli spaces.
In most cases we have explicit solutions. There are also a few other solutions
that are shown to exist, but we have not (yet) been able to obtain expressions
for their metric in closed form.

The variant of the theory with anisotropy parameter $z=2$ will not be especially
discussed, since it is clear that it exhibits no instanton solutions (of the type
we are considering here) with $SU(2)$ isometry. Recall in this case that the relevant
equations are provided by the Ricci flow on $S^3$, which takes the form \eqn{halfi}
in proper time with arbitrary parameters $\lambda < 1/3$ and $\Lambda_W > 0$. Its
fixed points are determined by the equations
\begin{equation}
R_{ii} -\frac{2\lambda-1}{2(3\lambda-1)}R \gamma_{i} +\frac{\Lambda_W}{1-3\lambda}
\gamma_{i} = 0 \ , \quad i = 1, 2, 3
\end{equation}
and clearly there is a unique solution given by the constant-curvature metric with
$R = 6 \Lambda_W$. The absence of other fixed points, which is also implied on more
general grounds by Poincar\'e's conjecture for $S^3$, shows that there can be no
finite-action instanton solutions in this case.

Thus, in the following, we focus on instanton solutions of the $z=3$ theory and
explain their properties, as outlined above.

\subsection{Global structure and completeness of the metrics}

First of all we examine the occurrence of singularities that can appear at finite
proper time $t$ (or $\tau$, since the two are simply related by rescaling) and
render the Euclidean space--time manifold incomplete. Singularities
arise when some or all of the metric coefficients of the Bianchi IX
model geometry vanish and they are classified, in general, as {\em nuts} and
{\em bolts}. Such singularities are intimately related to the fixed points of
Killing vector fields by geometrical reasoning and they are independent of the
gravitational equations. Following \cite{nuts, Eguchi:1980jx} we recall that the
structure of the fixed point set of a Killing vector field $\xi_{\mu}$ acting on any
four-dimensional Riemannian manifold with metric $g_{\mu \nu}$ is determined
by the rank of the $4 \times 4$ matrix $\nabla_{\mu} \xi_{\nu}$. This is an
anti-symmetric matrix (since its symmetric part vanishes identically by definition
of a Killing vector) which can have rank $4$ or $2$; rank $0$ is excluded for,
otherwise, the vector field vanishes everywhere.
In the former case there are no directions left invariant at the tangent space
of the fixed point, which, thus, appears to be isolated
and it is called nut. In the latter case only a two-dimensional subspace of
the tangent space at the fixed point remains invariant under the action of the
Killing vector field, whereas the
two-dimensional orthogonal complement rotates into itself. Then, the fixed point
set is provided by this invariant two-dimensional subspace and it is naturally
called bolt (it is typically a two-sphere, as in Bianchi IX geometries).

Nuts and bolts lead to incomplete manifolds, in general, but in certain cases the
apparent singularities can be removed and provide regular and complete metrics with
no curvature singularities. It all depends on the form of the metric as these
singularities are approached. A removable nut singularity contributes one unit to
the Euler number $\chi$ of the four--manifold and a removable $S^2$ bolt singularity
contributes two units \cite{nuts}, following the theorems on fixed points.
This counting applies to compact four--manifolds without boundary,
but it also generalizes to non-compact spaces when the Killing vector field is
either everywhere tangential to the boundary (as in space--times with homogeneous
spatial sections that we are considering here) or is everywhere transverse.
Thus, if $\chi \neq 0$, any Killing vector field will have at least one fixed
point. No fixed points imply that $\chi = 0$. Instanton solutions of Einstein
and Ho\v{r}ava--Lifshitz gravity are quite different in this respect, as will be
seen shortly, having important implications for their global topological structure.

Let us briefly review when such singularities can be removed from a
Riemannian four-manifold without referring to any specific theory or any solutions
at the moment. Using locally the Bianchi IX form of the metric (\ref{GRgravinst})
\begin{equation}
\label{mirmigki}
\mathrm{d}s^2 = \mathrm{d}t^2 + a^2 (t) \left(\sigma^1\right)^2 + b^2 (t) \left(\sigma^2\right)^2
+ c^2 (t) \left(\sigma^3\right)^2 ,
\end{equation}
we suppose that a singularity (nut or bolt) occurs at some finite proper
distance, say $t=0$. It is well known that the metric has a {\em removable nut
singularity} provided that near $t=0$ all metric coefficients vanish as
\begin{equation}
a^2(t) = b^2(t) = c^2(t) = {1 \over 4} t^2 \quad {\rm as} \quad  t \rightarrow 0 ~.
\end{equation}
In this case we have a coordinate singularity of the polar coordinate system in
$\mathbb{R}^4$ centered at $t=0$, which is simply removed by changing to a local
Cartesian coordinate system near the point $t=0$ and adding it to the manifold.
Also, it is well known that the metric has a {\em removable bolt singularity}
provided that near $t=0$ two of the metric coefficients (say $a^2$ and $b^2$)
become equal and the third vanishes as
\begin{equation}
a^2(t) = b^2(t) = {\rm finite} ~, \quad c^2(t) = {1 \over 4} n^2 t^2 \quad
 {\rm as} \quad t \rightarrow 0 \quad \mathrm{with} \quad n  \in \mathbb{Z} ~.
\end{equation}
Then, $a^2 = b^2 = R_0^2$ implies $a^2 (\sigma^1)^2 + b^2 (\sigma^2)^2 = R_0^2
(\mathrm{d}\vartheta^2 + \sin^2 \vartheta \mathrm{d} \varphi^2)$, which is the canonical $S^2$ metric,
while the $\mathrm{d}t^2 + c^2 (\sigma^3)^2$ part of the four-dimensional metric becomes
$\mathrm{d}t^2 + (n^2 t^2 / 4) \mathrm{d}\psi^2$ near $t=0$, keeping $\vartheta$ and $\varphi$ constant.
In this case, the topology of the manifold is locally $\mathbb{R}^2 \times S^2$
and the $\mathbb{R}^2$ factor shrinks to a point on $S^2$ as $t \rightarrow 0$.
By adjusting the range of $\psi$ so that $n\psi /2$ runs from $0$ to $2\pi$,
the apparent singularity at $t=0$ becomes a coordinate singularity of the polar
system in $\mathbb{R}^2$ and can be removed as before. In all other cases the
singularities cannot be removed and the manifold is incomplete.

The above reasoning is purely geometrical without reference to any field
equations. Thus, different gravitational theories for Euclidean space--times of the
form \eqn{mirmigki} may or may not lead to removable singularities at the fixed points
of a Killing vector field. This depends on the way that the metric coefficients
approach zero in the vicinity of a singularity and it is sensitive to the dynamics.
The space--time singularities of Euclidean Einstein and Ho\v{r}ava--Lifshitz gravity
(if they are present) will follow different power-law behavior, which, in turn, will
affect the completeness of the corresponding metrics. Thus, the absence of non-removable
singularities provides a natural selection for the physically admissible solutions
in those theories.

In Euclidean Ho\v{r}ava--Lifshitz gravity,  a singularity can only arise if an
eligible flow line reaches the boundaries -- including the origin -- of the physical
parameter space, namely the two wedges of the first quadrant in the $(x, y)$ (or
$(\gamma_1, \gamma_2)$) plane used in the phase portraits. Then, the flow becomes
extinct as it cannot be continued beyond that point. Such singularities, if
they are present, will arise at finite proper time (say $t=t_0$, but we can always
set $t_0 = 0$ without loss of generality). This is obviously so because such
singular points can also act as initial data for the time-reversed flow at a given
finite instance of (proper) time\footnote{Proving finiteness of extinction time for
the solutions of geometric flow on certain three-manifolds is an intricate mathematical
problem that will not be addressed in all generality, since we are only considering
homogeneous geometries on $S^3$. It is a key point in Perelman's proof of the Poincar\'e
conjecture based on Ricci flow  \cite{perel} and it is not yet clear how it may generalize
to the Ricci--Cotton flow.}. Such possibilities should be ruled out by the theory, unless
the singularities are removable nuts or bolts, for, otherwise, the space--time metric
will be singular. Using the Bianchi IX form of the metric \eqn{mirmigki} with
$a^2 = \gamma_1$, $b^2 = \gamma_2$ and $c^2 = \gamma_3$ (like in \eqn{m4dHL}), which is the appropriate
choice in this case, we may set (as in nuts)
\begin{equation}
\gamma_1 = \beta_1 t^{p_1} ~, \quad \gamma_2 = \beta_2 t^{p_2} ~,  \quad
\gamma_3 = \beta_3 t^{p_3}
\end{equation}
or alternatively (as in bolts)
\begin{equation}
\gamma_1 = \beta_1 ~,  \quad  \gamma_2 = \beta_2 ~,  \quad
\gamma_3 = \beta_3 t^{p_3}
\end{equation}
and determine the allowed values of the coefficients $\beta_i$ and the exponents
$p_i$ as $t \rightarrow 0$.
In all cases we find that the first-order system of Bianchi IX equations for
Ho\v{r}ava--Lifshitz gravity with general couplings does not lead to removable
nuts or bolts. Only non-removable singularities can arise along the
flow lines and they seem troublesome\footnote{The power-law behavior of solutions as
the flow lines approach the origin can also be read off from the exact solutions we have
presented in various cases.}. Recall, however, that instanton solutions are
rather restrictive, since they are described only by those trajectories that
interpolate between two fixed points. Therefore, their metrics would be singular if
any one of the fixed points were singular.
But this is a contradiction of terms and cannot happen, since a fixed point,
unlike a singularity, is
only reached at infinite proper time, $t \rightarrow \pm \infty$ (with sign that
depends on the direction of the flow), and not at
at finite time\footnote{This is also implied by the stability analysis around the
fixed points, which shows that the time dependence of small fluctuations varies
exponentially as $t$ goes to infinity.}. Thus, the instanton solutions protect
themselves from the singularities that may otherwise arise by moving along generic
flow lines. Whenever instanton solutions exist, their spaces will be always complete
without any singularities.

In Euclidean Einstein gravity  nuts and bolts are important
elements in the theory of gravitational instantons, since all known solutions exist
thanks to their presence. In this case, there is a certain class of gravitational instantons
\eqn{mirmigki} that follows from the Ricci flow equations \eqn{halphen} in proper time
$t$ with $\gamma_1 = a$, $\gamma_2 = b$ and $\gamma_3 = c$, as explained in section 3.2.
They include the trivial flat--space metric associated with the isotropic solution
\eqn{triviasola}, having $a^2 = b^2 = c^2 = t^2/4$ everywhere (with $t_0 = 0$),
as well as the Taub--NUT metric
as the next non-trivial example with an additional axial symmetry
$a(t) = b(t)$ and a removable nut singularity at the origin\footnote{The
gravitational field equations determine the Taub--NUT metric in the form shown in
section 3.2,
\begin{equation}
{a(t) \over m} + {\rm arcsinh} {a(t) \over m} = {-t + t_0 \over m} =
{\rm log} {2m + c(t) \over 2m - c(t)} - 2m \left({1 \over 2m + c(t)}
- {1 \over 2m - c(t)} \right) ~. \nonumber
\end{equation}
Setting $t_0 = 0$ for convenience, the power-law behavior of the coefficients
close to the origin $t=0$ turns out to be $a^2(t) = b^2(t) = c^2(t) = t^2 /4$
and describes a removable nut singularity.}. The Atiyah--Hitchin metric provides
an even more complicated solution, which is fully anisotropic and complete
\cite{Atiyah1, Atiyah2}. It exhibits a removable bolt singularity at the origin,
where $b=-c$ and it comes asymptotically close to Taub--NUT metric with $a=b$ as
$t \rightarrow \infty$; we skip the details as they are not important for the
present work. It turns out that these are the only complete gravitational
instantons with $SU(2)$ isometry that satisfy the Ricci flow equations
\eqn{halphen}; there is another complete metric with $SU(2)$ isometry, the
Eguchi--Hanson instanton, which has $a=b$ everywhere and a removable bolt
singularity at the origin, but its coefficients satisfy a different system of
first-order equations. Finally, we note for completeness, that exactly the
same reasoning applies to gravitational instantons of Einstein gravity
with cosmological constant \cite{nuts, Eguchi:1980jx} that have removable nut
and bolt singularities (e.g., $\mathbb{C}P^2$ as a gravitational instanton).

Thus, on the one hand, in Euclidean Einstein gravity the instanton spaces have non-vanishing
Euler number and, in many cases, they also have non-vanishing signature (given
roughly, but without any further explanation here, by the number of nuts minus
the number of anti-nuts \cite{nuts} that may be present). On the other hand,
the instanton solutions of Ho\v{r}ava--Lifshitz gravity are globally
$\mathbb{R} \times S^3$ having zero Euler number and signature.
They simply describe the evolution of a three-sphere from $t = -\infty$ to
$t = + \infty$ which deforms geometrically by the flow without ever becoming
singular along the way; as such they resemble closer the behavior of ordinary
instantons in particle theories rather than the instantons of Einstein
gravity. This is not surprising in retrospect, since consistency of the
Ho\v{r}ava--Lifshitz gravity is not questionable in the projectable case
for space--times with global cross-product foliation structure.
In either case, the corresponding metrics are regular everywhere and complete
and their Euclidean gravitational action is finite -- though the reasoning is
different for each theory. The finiteness of the action, which will be
discussed more extensively shortly, makes these solutions mostly relevant in
the quantum theory using, for instance, the path integral approach.

Let us also discuss the asymptotic structure of the solutions and compare them
to those of ordinary gravity, since there are also important differences between
the two theories. In Einstein gravity, the physical boundary conditions are
largely determined by the positive-action conjecture that requires that the action
of any asymptotically Euclidean four-metric be positive, vanishing if and only if
the space is flat \cite{perry,Eguchi:1980jx, Gibbons:1979xn}.
Then, using the Bianchi IX form of the metric \eqn{mirmigki},
the following possibilities arise at infinite proper distance $t$: either there
is a Euclidean infinity
\begin{equation}
a^2(t) = b^2(t) = c^2(t) = {1 \over 4} t^2 \quad {\rm as} \quad t \rightarrow \pm \infty
\end{equation}
when $0 \leq \psi \leq 4\pi$ (it is a conical infinity when $0 \leq \psi \leq 2\pi$)
or a Taubian infinity
\begin{equation}
a^2(t) = b^2(t) = t^2 ~, \quad c^2 = {\rm finite} \quad {\rm as} \quad t \rightarrow \pm
\infty
\end{equation}
that encompasses the Taub--NUT metric. Combining all distinct boundary conditions
that are available at $t = 0$ and $t = \pm \infty$, one ends up with a few viable
solutions that provide the list of all complete gravitational instanton metrics
with $SU(2)$ isometry. Similar considerations may apply to solutions with cosmological
constant.

In Ho\v{r}ava--Lifshitz theory the situation is different. The Euclidean action is
always positive-definite (at least for $\lambda < 1/3$ that we are considering here)
and vanishes when the three-dimensional metrics are vacua of topologically massive
gravity without any time dependence. Thus, there are no a priori conditions on the
asymptotic structure of instantons other than the mere existence of multiple vacua
in three dimensions that serve as fixed points of the flow.
Then, the asymptotic structure of space--time as $t \rightarrow \pm \infty$ is simply
determined by the specific form of the metric coefficients $\gamma_1$, $\gamma_2$
and $\gamma_3$ at the initial and final fixed points, respectively. Their time
dependence is exponential and it is completely determined by the eigenvalues of the
characteristic matrix of small fluctuations around these fixed points. The departure
from usual asymptotics (with zero or positive cosmological constant) is inherited to
the solutions from the detailed balance condition and seems to be rather universal.
It inflicts other classes of solutions, such as the construction
of black-hole solutions whose right asymptotic structure requires departure from
detailed balance using more general couplings \cite{black1, black2, black3} (otherwise
there is no match with observations at large distances).
This is also closely related to the
problem of obtaining ordinary gravity by arguing (naively) that all higher-order
curvature terms are suppressed in the infrared regime of the theory\footnote{For
the same reason we cannot obtain the instantons of Einstein gravity from
those of Ho\v{r}ava--Lifshitz theory. The first arise by dropping all higher-curvature
terms and setting $\lambda = 1$, whereas the latter exist in the full
theory only for $\lambda < 1/3$. Bianchi type IX models may offer a glimpse at
this problem since the gravitational potential is derived from a superpotential
in both cases (see, for instance, \cite{Gibbons:1979xn} for the derivation of
the superpotential that governs $SU(2)$ instantons in ordinary gravity; this
reference also provides a neat qualitative picture for the completeness of their
metrics using Hamiltonian methods). It should be easier to explore the
renormalization of the coefficients of the superpotential and the
parameter $\lambda$ for this class of mini-superspace models, as the theory is
taken from the ultra-violet to the infrared domain.}. However, it
is not necessarily a big problem in the ultra-violet regime relevant to early time
cosmology, where our discussion
is applicable keeping $\lambda < 1/3$. Abandoning detailed balance will ruin our
general construction of instanton solutions.

\boldmath
\subsection{The action and moduli of $SU(2)$ instanton metrics}
\unboldmath

Let us now give some examples of instantons, based on the results described in previous
sections, and compute their action $S_{\rm instanton} = |\Delta W|/2$ in each case
separately (see \eqn{finitac}). In general, the superpotential consists of two terms
$W = W_{\rm CS} + W_{\rm EH}$,
which are given by the following expressions for Bianchi IX model geometries,
\begin{equation}
W_{\rm CS} = {16 \pi^2 \over w_{\rm CS}} \left[1 + {1 \over 2 \gamma_1 \gamma_2 \gamma_3}
(\gamma_1 + \gamma_2 - \gamma_3) (\gamma_1 - \gamma_2 + \gamma_3) (\gamma_1 - \gamma_2
- \gamma_3) \right]
\end{equation}
and
\begin{equation}
W_{\rm EH} = - {16 \pi^2 \over \kappa_W^2} \left[{1 \over \sqrt{\gamma_1 \gamma_2 \gamma_3}}
(\gamma_1^2 + \gamma_2^2 + \gamma_3^2 - 2 \gamma_1 \gamma_2 - 2 \gamma_2 \gamma_3 -
2 \gamma_3 \gamma_1) + 4 \Lambda_W \sqrt{\gamma_1 \gamma_2 \gamma_3} \right]
\end{equation}
and will be used next to evaluate the instanton action.

The cases below refer to instantons constructed from interpolating trajectories of the
Ricci--Cotton flow for different couplings, changing as the complexity of the equations
increases.

\paragraph{Cotton flow.}

For the pure Cotton flow treated in section 3.3 there are two
fixed points: the isotropic point with $\gamma_1 = \gamma_2 = \gamma_3 = L^2/4$ and the
anisotropic fixed point with $\gamma_1 = \gamma_2 = xL^2/4$, and $\gamma_3 = L^2/4x^2$
that arises for $x = \infty$; two more anisotropic fixed points are obtained from it by
permuting the three principal axes of $S^3$. The
corresponding instanton, which is given in closed form by \eqn{mourira}, describes the
evolution of a fully squashed (flattened) sphere towards the round sphere as $t$ varies
from $-\infty$ to $+\infty$; the anti-instanton follows by reversing the time direction.

Note that a natural entropy function associated with the volume-preserving deformation of
$S^3$ (other than $W$) can be defined in this case\footnote{For Bianchi IX
model geometries one can define in general an additional function (other than
$W$) that changes monotonically under the Cotton flow. We consider
\begin{equation}
F(t) = {1 \over \gamma_1^2} + {1 \over \gamma_2^2} + {1 \over \gamma_3^2} ~,
\nonumber
\end{equation}
which is bounded from below by $3/(\gamma_1 \gamma_2 \gamma_3)^{\nicefrac{2}{3}} =
3 (16 \pi^2 / V)^{\nicefrac{4}{3}}$ for a three-sphere with volume $V$. The lower
bound is attained in the fully isotropic case $\gamma_1 = \gamma_2 = \gamma_3$.
$F(t)$ becomes infinite when the sphere is completely squashed in one or more
directions; as such, it is a measure of the ``shape entropy'' of $S^3$.
Using the Cotton flow \eqn{thecottfle}, we obtain
\begin{equation}
{\mathrm{d}F \over \mathrm{d}t} = -{\kappa^2 \over w_{\rm CS} (\gamma_1 \gamma_2 \gamma_3)^{3/2}}
\left[{(\gamma_2 + \gamma_3) (\gamma_2 - \gamma_3)^2 \over \gamma_1^2}
+ {(\gamma_3 + \gamma_1) (\gamma_3 - \gamma_1)^2 \over \gamma_2^2} +
{(\gamma_1 + \gamma_2) (\gamma_1 - \gamma_2)^2 \over \gamma_3^2} \right]
\nonumber
\end{equation}
and, therefore, $F(t)$ changes monotonically. For $w_{\rm CS} > 0$, these
properties of $F(t)$ suffice to prove the convergence of the flow lines to the fully
isotropic fixed point regardless of initial conditions \cite{Kisisel:2008jx}.}. It
is important for the mathematics of the Cotton flow, but, unlike $W$ that determines
the action of the instanton, this entropy has no deeper meaning in space--time (as far
as we can tell now). Also, there is no (yet) known analogue of it for the combined
Ricci--Cotton flow.

The instanton solution has enhanced isometry $SU(2) \times U(1)$, since the deformation
line possesses axial symmetry. Also, since the interpolating trajectory is unique, the
instanton has no moduli other than the radius of the sphere at the fixed point.
In this case, $W = W_{\rm CS}$ and one finds that the superpotential takes the following
values at the two fixed points,
\begin{equation}
W^{\rm iso} = {8 \pi^2 \over w_{\rm CS}} ~, ~~~~~~
W^{\rm aniso} = {16 \pi^2 \over w_{\rm CS}} ~.
\end{equation}
Therefore, the action is
\begin{equation}
S_{\rm instanton} = {4 \pi^2 \over |w_{\rm CS}|}
\end{equation}
and it is independent of the modulus $L$. Obviously, there are no other instantons
derived from the Cotton flow equations.

\paragraph{Normalized Ricci--Cotton flow.}

The normalized Ricci--Cotton flow has more than one fixed points when $\mu < 0$,
in which case there are instantons with $SU(2)$
isometry interpolating between them as $t$ varies for $-\infty$ to $+\infty$.

Let us first consider the instantons connecting the two axially symmetric
fixed points. Recall that the isotropic fixed point appears at $\gamma_1 = \gamma_2
= \gamma_3 = L^2/4$ and the anisotropic point at $\gamma_1 = \gamma_2 = L^2/4a$ and
$\gamma_3 = a^2 L^2/4$ (up to permutations of the axes of $S^3$), setting for
notational convenience
\begin{equation}
a = \sqrt{-{\mu \over 3}} ~.
\end{equation}
Then, explicit calculation shows that $W = W_{\rm CS} + W_{\rm EH}$ (with $\mu =
w_{\rm CS} L / \kappa_W^2$) takes the following form at the two fixed points,
\begin{equation}
W^{\rm iso} = {8 \pi^2 \over w_{\rm CS}} \left(1 - 9a^2 + 3a^2 \Lambda_W L^2\right)
\end{equation}
and
\begin{equation}
W^{\rm aniso} = {8 \pi^2 \over w_{\rm CS}} \left(2 + 4a^6 -14a^3 + 3a^2 \Lambda_W L^2\right) ~.
\end{equation}
Although $\Lambda_W$ does not appear in the normalized Ricci--Cotton flow equations,
it enters into $W$ by contributing the same at all points (recall that the volume $V$ is
preserved in this case). Consequently, the instanton action takes the value
\begin{equation}
S_{\rm instanton} = {4 \pi^2 \over |w_{\rm CS}|} (a-1)^2 \left(4a^4 + 8a^3 + 12a^2 + 2a +1\right)
\end{equation}
and it is independent of $L$, as expected. Notice that it vanishes when $a=1$ ($\mu = -3$),
as required, since the two fixed point coalesce and there is no instanton in this case.
The action is non-zero and positive for all other values $\mu < 0$.

The axisymmetric solutions of the normalized Ricci--Cotton flow have been constructed
explicitly in section 4.4, but one should only use those branches that interpolate
between the two fixed points. The stability analysis performed in section 4.2
shows that for $-3 < \mu < 0$ the isotropic fixed point is absolutely unstable and the
anisotropic is a saddle point. Therefore, there can be only one flow line interpolating
between the two fixed points (corresponding to the axisymmetric solution we obtained)
and the instanton has no moduli other than $L$. Exactly the same conclusion holds for
$-6 \, \sqrt[3]{2}<\mu<-3$, since the isotropic fixed point is now absolutely stable and
the anisotropic is a saddle point. The absence of moduli in these cases can also be seen
schematically
in figures 4 and 5, respectively. Thus, for all $-6 \, \sqrt[3]{2}<\mu<0$ the instantons
have enhanced $SU(2) \times U(1)$ isometry. The situation changes drastically when
$\mu < -6 \, \sqrt[3]{2}$, since the isotropic fixed point is absolutely stable and
the anisotropic is absolutely unstable. In this case, we have several flow lines
interpolating between the two fixed points, as can also be seen schematically in figure 6,
and the instantons have an additional (real) modulus that labels these trajectories.
The physical interpretation of this modulus is nothing else but the geometric shape of
$S^3$ (there is only one shape modulus, since the volume of space is held fixed by
specifying $L$). Of course, the instanton action is independent of all moduli.

The axisymmetric solution we have obtained in this case has enhanced $SU(2) \times U(1)$
isometry, whereas the other ones should only have an $SU(2)$ isometry group. They all
correspond to regular and complete metrics on $\mathbb{R} \times S^3$, but we have not
been able to find them in closed form. They should be the analogue
of the Atiyah--Hitchin metric for Ho\v{r}ava--Lifshitz gravity when $\lambda =
- \infty$. Their explicit construction is an interesting open mathematical problem.

Finally, we turn to instantons that owe their existence to the presence of
totally anisotropic fixed points in the problem when $\mu < -6 \, \sqrt[3]{2}$.
They have no moduli (other than their volume) since they connect a saddle point with
a stable or an unstable fixed point. These instantons also have $SU(2)$ isometry
but no higher symmetry.

The value of the superpotential for the totally anisotropic fixed points (see also
section 4.2) turns out to be
\begin{equation}
W^{\rm total ~ aniso} = {8 \pi^2 \over w_{\rm CS}} (10 + 3a^2 \Lambda_W L^2) ~.
\end{equation}
Therefore, the instanton that interpolates between these points and the totally
isotropic fixed point has action
\begin{equation}
S_{\rm instanton} = {72 \pi^2 \over |w_{\rm CS}|} (a^2 + 1) ~.
\end{equation}
It never becomes zero because these points cease to exist before they have the
chance to meet with the isotropic point.
Similarly, the instanton that interpolates between the totally anisotropic and the
axially symmetric anisotropic fixed points has action
\begin{equation}
S_{\rm instanton} = {16 \pi^2 \over |w_{\rm CS}|} (a^3-4) (2a^3 + 1) ~.
\end{equation}
The latter vanishes when $a^3 = 4$ ($\mu = -6 \, \sqrt[3]{2}$), as the end-points
coalesce in this case, and it is positive definite otherwise.

\paragraph{General Ricci--Cotton flow.}

The Ricci--Cotton flow with general couplings
(provided that $\lambda < 1/3$ and $\Lambda_W$ is non-negative) was found to exhibit
two fixed points when $\mu < 0$, in which case there are instanton solutions in
Ho\v{r}ava--Lifshitz gravity. Recall that the isotropic point appears at $\gamma_1 =
\gamma_2 = \gamma_3 = 1/4\Lambda_W$ and the anisotropic point at $\gamma_1 = 36 \mu^2
/\left(\mu^2 + 27 \Lambda_W\right)^2$ and $\gamma_2 = \gamma_3 = 9/\left(\mu^2 + 27 \Lambda_W\right)$,
assuming the presence of an axial symmetry $\gamma_2 = \gamma_3$ for all time.

Taking $\mu < 0$, we define, for notational convenience, the non-negative number
\begin{equation}
a = - {\mu \over 3 \sqrt{\Lambda_W}}
\end{equation}
and evaluate the superpotential $W = W_{\rm CS} + W_{\rm EH}$ at the two fixed points.
Using $\mu = w_{\rm CS}/\kappa_W^2$, as defined in section 5, we obtain the following
results
\begin{equation}
W^{\rm iso} = {8 \pi^2 \over w_{\rm CS}} (1 - 6a) ~, \quad
W^{\rm aniso} = {16 \pi^2 \left(5a^4 - 54a^2 + 9\right) \over w_{\rm CS} \left(a^2 + 3\right)^2} ~.
\end{equation}
Therefore, the instanton action turns out to be
\begin{equation}
S_{\rm instanton} = {12 \pi^2 \over |w_{\rm CS}|} {(a-1)^2 \over \left(a^2 + 3\right)^2}
\left(2a^3 + 7a^2 + 24a + 3\right) ~.
\end{equation}
Note that the action is manifestly positive-definite, as required, and vanishes
when $a=1$ ($\mu = -3 \sqrt{\Lambda_W}$), in which case the two fixed points
coalesce and there is no instanton.

Even in the presence of axial symmetry, which was used to simplify the analysis
of the general Ricci--Cotton flow equations, we have not been able to obtain the
interpolating
solutions in closed form. Nevertheless, it is clear that a unique solution exists in
this case, for all $\mu <0$, which interpolates between the two fixed points.
For $0 < a < 1$ ($-3 \sqrt{\Lambda_W} < \mu < 0$) the isotropic fixed point is
saddle and the anisotropic is absolutely stable, and, therefore, there is a single
flow line that connects the two. For $a>1$ ($\mu < -3 \sqrt{\Lambda_W}$)
the isotropic fixed point is now absolutely stable and the anisotropic is a saddle
point and, therefore, the interpolating flow line is again unique. This can also be seen
by inspecting figures 12 and 13.

The solutions at hand have no moduli at all\footnote{Even the
size of $S^3$ at the isotropic fixed point is not free, as $L$ was free to vary
in the normalized flow, but it is determined by the parameters of the differential
equations. Thus, it is not surprising that the instanton action depends on $\Lambda_W$
(through $a$) in this case.}. They correspond
to instantons with $SU(2) \times U(1)$ isometry. It will be very interesting to
construct them explicitly. Also, other more general solutions with strict $SU(2)$ isometry
are expected to exist in the general case, with $\gamma_1 \neq \gamma_2 \neq \gamma_3$,
but their investigation will not be pursued in the present work. We only note here that
all anisotropic fixed points of the Ricci--Cotton flow equations with general couplings
seem to be axially symmetric even when $\gamma_1 \neq \gamma_2 \neq \gamma_3$ at generic
points. Thus, we expect to have instanton solutions without axial symmetry that
interpolate between these fixed points, serving as the Ho\v{r}ava--Lifshitz analogue
of the Atiyah--Hitchin metric. They should depend only on one free parameter.

\paragraph{The special  case  $\Lambda_W = 0$.}

Finally, note that as $\Lambda_W$ is
taken to zero, while keeping $\mu$ fixed in the general system of Ricci--Cotton flow equations,
$W^{\rm iso}$ blows up to infinity, whereas $W^{\rm aniso}$ remains finite, tending to
the value $80 \pi^2/w_{\rm CS}$. Consequently, $S_{\rm instanton}$ becomes infinite
and one may consider it a problem, since instantons must have finite action.
However, in this case, there is no contradiction, since the isotropic configuration
ceases to be (strictly speaking) a fixed point when it is pushed away to infinity
by setting $\Lambda_W = 0$ and, therefore, the flow line that interpolates between
the two points (see figure 15) does not qualify as instanton solution of the theory.
In conclusion, there are no instanton solutions when $\Lambda_W = 0$.

This completes our analysis of $SU(2)$ gravitational instantons of
Ho\v{r}ava--Lifshitz theory with anisotropy scaling parameter $z=3$.
We have obtained complete classification of all
explicit and implicit solutions that exist for all different couplings of the theory
satisfying the detailed balance condition, provided that $\lambda < 1/3$ and
$\Lambda_W > 0$. By the same token, the variant of the theory with scaling parameter
$z=2$ does not exhibit any such instanton solutions. The results are on par with
the classification of instantons with $SU(2)$ isometry in Einstein gravity. The only
missing technical part is the explicit construction of some of these instanton metrics.

We end this section with some general remarks concerning the existence and description
of instanton metrics in Ho\v{r}ava--Lifshitz gravity without relying on isometry groups,
such as $SU(2)$.
According to definition, they should be trajectories of the Ricci--Cotton flow equations
interpolating between any two solutions of three-dimensional topologically massive gravity
that provide the fixed points. The landscape of vacua of topologically massive gravity
is not known completely\footnote{Note, however, the recent work \cite{gurses}
that develops techniques to solve the field equations of topologically massive
gravity (and other massive-gravity models) for three-dimensional geometries admitting
a Killing vector field. Older results in this direction are neatly summarized in
\cite{classif} although most of them focus on vacua with negative cosmological
constant.} and, therefore, it is difficult to make explicit general constructions. Also,
it is rather difficult to investigate the general behavior of the flow equations by
standard mathematical techniques, since they are third-order in space derivatives and
even the short-time existence of solutions is difficult to establish in all generality.
The formation and characterization of singularities is another related general open problem
for these flow equations. Addressing these issues successfully will lead to further advances.

\section{Generalization to higher dimensions}
\setcounter{equation}{0}

In this section we make a few remarks concerning higher-dimensional generalizations
of Ho\v{r}ava--Lifshitz theory and the correspondence of its instanton solutions
to the theory of higher-order geometric flows.

\boldmath
\subsection{Ho\v{r}ava--Lifshitz gravity in $4+1$ dimensions}
\unboldmath

The general aspects have been reviewed in section 2 for all space--time dimensions.
The theory is power-counting renormalizable when $z=D$ using the appropriate
superpotential $W$. Let us concentrate on $D=4$ for definiteness, so that $W$ is
the action of four-dimensional gravity with higher-order corrections of the
general form \cite{Horava:2009uw}
\begin{equation}
\label{kokorikos}
W[g] = \int \mathrm{d}^4x \sqrt{g} \left(\alpha C_{ijk\ell} C^{ijk\ell} + \beta R^2 +
\gamma (R-2\Lambda_W) \right) .
\end{equation}
Here, $C_{ijk\ell}$ is the Weyl tensor and $R$ is the Ricci scalar curvature
of a four-dimensional Riemannian metric $g$ that describes the geometry of
spatial slices in a five-dimensional space--time with topology
$\mathcal{M}_5 = \mathbb{R} \times \mathcal{M}_4$. Here, there is no need to include the
term $R_{ij}R^{ij}$ because it can be removed by a Gauss--Bonnet topological
term, adjusting the coefficients $\alpha$ and $\beta$.

Thus, Ho\v{r}ava--Lifshitz gravity in $4+1$ dimensions with anisotropic scaling
$z=4$ is defined by the action
\begin{equation}
S=\frac{2}{\kappa^2}\int \mathrm{d}t\,\mathrm{d}^4 x\,\sqrt{g}N K_{ij}G^{ijk\ell}
K_{k\ell} - \frac{\kappa^2}{8}\int \mathrm{d}t\,\mathrm{d}^4 x\,\sqrt{g}N
\left({1 \over \sqrt{g}} {\delta W \over \delta g_{ij}}\right) \mathcal{G}_{ijk\ell}
\left({1 \over \sqrt{g}} {\delta W \over \delta g_{k\ell}}\right)
\end{equation}
using the extrinsic curvature $K_{ij}$ of $\mathcal{M}_4$ and the metric
$G^{ijk\ell}$ of superspace with parameter $\lambda$. Also, following the
general discussion of section 2, we will also take $\lambda < 1/4$ so that the
Euclidean counterpart of this action is manifestly bounded from below.

The theory with detailed balance is completely specified by the choice of $W[g]$.
It is given by the general expression \eqn{kokorikos} in $D=4$; other appropriate
choices of $W$ should be made in higher dimensions to render the theory power-counting
renormalizable. We also note for completeness that if higher-order curvature functionals are
chosen in $D$ spatial dimensions so that $z>D$, the resulting gravitational theory
will be power-counting superenormalizable \cite{Horava:2009uw}. Such generalizations
will not be considered at all in the present work.

Next, we illustrate the structure of the resulting equations by considering the
simplest higher-dimensional case with $z=D=4$.

\subsection{Bach flow and its variants}

Solutions of the Euclidean five-dimensional Ho\v{r}ava--Lifshitz
gravity can be obtained from the geometric-flow equation
\begin{equation}
{1 \over N(t)} \partial _t g_{ij}=\pm \frac{\kappa^2}{2\sqrt{g}}\mathcal{G}_{ijk\ell}
\frac{\delta W[g]}{\delta g_{k\ell}} + \nabla_i\xi_j+\nabla_j\xi_i
\end{equation}
that describes deformations of the four-dimensional Riemannian metric $g_{ij}$.
The lapse function $N(t)$ can be set equal to $1$ by time redefinition.

The details can be worked out using the following identity, which is well known among
people working in conformal Weyl gravity,
\be
B^{ij} = - {1 \over \sqrt{ g}} {\delta W_{\rm Weyl} \over
\delta g_{ij}} ~,
\ee
where
\be
W_{\rm Weyl} = \int \mathrm{d}^4 x \sqrt{ g} ~ C_{ijk\ell}
C^{ijk\ell}
\ee
is the quadratic Weyl tensor action functional and
\be
B^{ij} = \nabla_{k} \nabla_{\ell} C^{ikj\ell}
+ {1 \over 2} R_{k \ell} C^{ikj\ell}
\ee
is the so called {\em Bach tensor} \cite{bach}. It is a fourth-order symmetric
and traceless tensor that clearly vanishes when the four-dimensional metric is
conformally flat. The Bach tensor provides the analogue of the Einstein
tensor in the field equations of conformal Weyl gravity, and, as such, it is
also covariantly conserved.

Thus, for this particular choice of superpotential $W$, the corresponding
geometric flow takes the form
\begin{equation}
\partial _t g_{ij}=\mp \frac{\kappa^2}{2} B_{ij} + \nabla_i\xi_j+\nabla_j\xi_i ~,
\end{equation}
and it can be naturally called Bach flow. Its fixed points (modulo reparametrizations)
are the vacuum solutions of conformal Weyl gravity and include the isotropic (constant
curvature) metric on $S^4$. It is mathematically more interesting to pick the sign
that drives the evolution towards the fixed points rather that away from them.
Although this is a higher-order flow, it is
better behaved mathematically than the third-order Cotton flow. Thus, one should
investigate it in detail and attempt to construct solutions. It is a new geometric
flow that has not appeared in the mathematics literature before, to the best of our
knowledge. By restricting it to K\"ahler manifolds, it might be also interesting to
compare it (and the variants which are discussed below) with other well known geometric
flows of fourth-order, such as the Calabi flow \cite{calabi}.

If there is an additional contribution to $W$ given by the quadratic Ricci scalar
curvature action,
\be
W_{R^2} = \int \mathrm{d}^4 x \sqrt{g} ~ R^2 ~,
\ee
it will account for the gradient term
\be
H^{ij} = - {1 \over \sqrt{g}} {\delta W_{R^2} \over
\delta g_{ij}}
\ee
with
\be
H^{ij} = 2 g^{ij} \nabla_{k} \nabla^{k} R - 2 \nabla^{i}
\nabla^{j} R - 2 R R^{ij} + {1 \over 2} g^{ij} R^2 ~.
\ee
This tensor is symmetric but not traceless. Then, the complete flow equation will
be a variant of the Bach flow receiving contributions from $B^{ij}$ and $H^{ij}$,
which are both fourth-order. Of course, in the general case, there will also be
subleading curvature terms associated with the Einstein tensor $G^{ij}$ by adding
the four-dimensional Einstein--Hilbert action (possibly with a cosmological constant)
to the superpotential $W$.

Instanton solutions will correspond to flow lines interpolating between different
vacua of four-dimensional conformal Weyl gravity (and its deformations thereof),
but again it seems rather difficult to derive explicit general results. Using
four-dimensional model geometries may provide some simple and tractable
mini-superspace models that are worth studying in the future.

Similar considerations apply to all higher-dimensional generalizations of
Ho\v{r}ava--Lifshitz gravity. In $D+1$ dimensions, the non-relativistic gravitational
theory becomes power-counting renormalizable when the anisotropic scaling parameter
is $z=D$. Then, for the appropriate choice of $W$, we obtain geometric flows of
order $D$ that describe instanton-like configurations of the Euclidean
($D+1$)-dimensional theory when $\lambda < 1/D$. This framework hosts very
naturally a whole hierarchy of geometric flows and provides a reason to study them.

\section{Conclusions and discussion}
\setcounter{equation}{0}

We examined the Euclidean version of Ho\v{r}ava--Lifshitz gravity satisfying the
detailed balance condition and described its instanton solutions as flow lines
interpolating between different fixed points of
a new class of geometric evolution equations, which are first-order in time.
Although the specific couplings implied by detailed balance are rather restrictive
(and sometimes appear to be problematic), the general connection between instanton
solutions and geometric flows is rather interesting in many respects. Focusing
to $3+1$ dimensions, where the potential term is derived from a superpotential
$W$ given by the action functional of three-dimensional topologically massive gravity
and the anisotropy scaling parameter of the theory is $z=3$, the driving curvature
terms are provided by a certain combination of the Cotton and Ricci tensors as well as
the cosmological constant term. The geometric-flow equations, called Ricci--Cotton flow,
were shown to exhibit an entropy functional that is given by $W$ and can be used to
put a lower bound on the Euclidean Ho\v{r}ava--Lifshitz gravitational action.

Our construction requires $\lambda < 1/3$ and $\Lambda_W > 0$, but otherwise the
parameters of the theory can be arbitrary within the class of detailed balance couplings.
Fixed points of the flow are provided by classical solutions of the three-dimensional
topologically massive gravity and they correspond to static solutions of the
($3+1$)-dimensional theory. As such, they include constant-curvature isotropic metrics in three
dimensions as well as anisotropic configurations obtained by balancing the deformation
effect of the Cotton and Ricci tensors. Since there is no general classification of these
metrics, the landscape of fixed points remains largely unexplored
to the best of our knowledge. Running solutions represent genuine time-dependent
configurations, but they are even more difficult to investigate in exact terms.
Thus, the Ricci--Cotton flow appears to be a rather complex system of equations
that deserves proper mathematical study on general grounds. Addressing these
problems in all generality remains out of reach at the moment, but some simple
mini-superspace truncations of the equations help to obtain concrete results in
simple cases that are interesting both physically and mathematically.

We found that the homogeneous model geometries on three-manifolds provide consistent
truncation of the Ricci--Cotton flow equations. In particular, focusing on the
Bianchi IX model geometries on $S^3$, so that the corresponding gravitational
instantons exhibit $SU(2)$ group of isometries, we were able to classify the
fixed points of the flow (isotropic as well as anisotropic) and study their
stability properties for a variety of different couplings. Some special solutions
with axial symmetry (associated with $SU(2) \times U(1)$ isometry group) were
constructed explicitly and their space--time interpretation was discussed in analogy
with the gravitational instanton solutions of ordinary gravity. In particular, we
have arrived at complete classification of the instanton solutions with $SU(2)$
isometry. It remains to be
seen whether more general running solutions can be constructed explicitly beyond
their qualitative description based on the phase portraits of the flow. Also, it
will be interesting to find other consistent reductions of the flow equations
beyond the class of homogeneous geometries, but we have not yet been able to obtain
any concrete results in this direction.

Another possibility that has not been discussed at all in this paper is to consider
super-renormalizable versions of Ho\v{r}ava--Lifshitz gravity in $3+1$ dimensions
with anisotropic scaling $z=4$. These are generated by a superpotential $W$
-- other than the action functional of topologically massive gravity -- which
contains higher-order Ricci curvature terms such as $R_{ij} R^{ij}$ and $R^2$
on top of the cosmological Einstein--Hilbert action in three dimensions
\cite{Horava:2009uw}. In this case, the instanton solutions will be described
by geometric flows in three dimensions with fourth-order derivatives in their
driving curvature terms. The resulting equations appear to have some nice
mathematical properties (compared to the third-order Ricci--Cotton flow) and they
also seem to admit consistent reduction to an autonomous system of ordinary differential
equations for homogeneous model geometries. A particular choice of such $W$ is
provided by the action of the so called ``new massive gravity'' in three dimensions
that contains both terms $R_{ij} R^{ij}$ and $R^2$ with relative coefficient
$-3/8$ \cite{townsend}. We intend to investigate elsewhere the corresponding
fourth-order flows \cite{sourd}, together with the associated instanton solutions,
and examine the privileged role (if any) of new massive gravity in this context.

Higher-dimensional generalizations were also briefly discussed. It was pointed out
that instanton solutions exist in all dimensions and their defining equations
provide new classes of geometric-flow equations, such as the Bach flow in four
dimensions. In general, the driving curvature terms of such flows contain
spatial derivatives of order $z$ (equal to the anisotropy scale parameter that
renders the higher-dimensional Ho\v{r}ava--Lifshitz gravity power-counting
renormalizable) and they describe metric deformations on spatial slices of
dimension $D=z$. The hierarchy of such flows has not been considered in the
literature before and they certainly pose several interesting questions that are
worth studying in the details. They should also be of interest to the mathematics
community working on geometric analysis. In all cases,
the non-relativistic theory of gravity
provides a general framework to embed geometric evolution equations. The situation
should be compared to general relativity and string theory, where such embedding
is only possible in some very special cases, such as the Ricci flow on homogeneous
three-geometries that can be interpreted as self-dual gravitational instantons in four
dimensions or using some appropriately chosen higher-dimensional plane-wave
gravitational backgrounds.

The off-shell formulation of string theory based on the world-sheet renormalization
group equations provides a natural framework for the appearance of the Ricci flow
(and other closely related geometric-flow equations) in gravitational physics. In this
context, closed string tachyon condensation is described by transitions from one
fixed point to another more stable fixed point. Thus, the lines of the Ricci
flow resemble instanton transitions among different vacua of the string
landscape. In Ho\v{r}ava--Lifshitz gravity, on the other hand, the Ricci flow
(and its variants) describe the instantons of the theory when the anisotropic
scaling is $z=2$. Therefore, it seems interesting to investigate further this
aspect while searching for possible embedding of the non-relativistic theory of
gravity into a more fundamental theory. Likewise, non-relativistic theories
of gravity with higher anisotropic scaling, in particular $z=D$, and their
instanton solutions may admit a similar description and interpretation in terms
of a more fundamental theory. It is not yet known, however, whether the geometric
evolution equations we are considering here can also arise as renormalization-group
equations in a class of quantum field theories.

Finally, another interesting problem is the use of instantons for the quantization
of Ho\v{r}ava--Lifshitz gravity. One possible line of work in this direction is the
path integral approach over Euclidean space--times with applications to quantum
cosmology in the spirit of Hartle--Hawking proposal. The quantization of
mini-superspace models appears to be tractable, at least for homogeneous
(but generally non-isotropic) geometries, and requires special attention. They
can also provide some non-perturbative information about the quantum theory and
a testing bed for comparison with the quantization of ordinary gravity.

It remains to be seen whether the non-relativistic theory of gravity is a
viable alternative to Einstein gravity at very short distances. However, the
simplified version of the theory with detailed balance can also play another
role in physics, serving as landscape explorer of the vacuum structure of
relativistic field
theories determined by $W$ (with topologically massive gravity being just an
example). It provides an effective particle model to describe transitions
among different vacua through instantons. It also
offers a dynamical principle for vacuum selection that is worth exploring
further in all generality using the powerful tools of geometric flows and
associated entropy functions. Advocating this point of view introduces a new
twist to the subject and departs from the idea (and the problems that seem to
accompany it) that Ho\v{r}ava-Lifshitz gravity is the ultra-violet completion
of a fundamental theory. It could have also been used from the very beginning
as an alternative motivation for the present work.

We hope to return to these topics elsewhere in the near future.

%\vskip1cm

\section*{Acknowledgements}

The authors would like to thank  G. Huisken, A. Petkou, C. Sourdis and M. Taylor for
stimulating discussions. Many of the ideas developed in the present work were triggered
during the 2009 GGI workshop \textsl{New Perspectives in String Theory}.  I. Bakas,
F. Bourliot and M. Petropoulos would like to thank the LMU for kind
hospitality. This research was supported by the Cluster of Excellence
\textsl{Origin and the Structure of the Universe} in Munich, Germany, the
French Agence Nationale pour la Recherche, contract  05-BLAN-NT09-573739
\textsl{String Cosmology}, the ERC Advanced Grant  226371
\textsl{Mass Hierarchy and Particle Physics at the TeV Scale},
the ITN programme PITN-GA-2009-237920  \textsl{Unification in the LHC Era}, the IFCPAR 
programme 4104-2 and the GRC APIC PICS-Gr\`ece  3747.

%\newpage


\vskip1cm


\begin{thebibliography}{99}

\bibitem{lifshitz:1941}
E.M. Lifshitz, ``On the theory of second order phase transitions I \& II'',
Zh. Eksp. Teor. Fiz. \underline{11} (1941) 255 \& 269.

\bibitem{Chadha:1982qq}
S.~Chadha and H.B.~Nielsen,
``Lorentz invariance as a low-energy phenomenon'',
Nucl. Phys. \underline{B217} (1983) 125.

\bibitem{Iliopoulos:1980zd}
J.~Iliopoulos, D.V.~Nanopoulos and T.N.~Tomaras,
``Infrared stability or anti grand unification'',
Phys.\ Lett.\   \underline{94B} (1980) 141.

\bibitem{Antoniadis:1983ek}
I.~Antoniadis, J.~Iliopoulos and T.~Tomaras,
``On the infrared stability of gauge theories'',
Nucl.\ Phys.\  \underline{B227} (1983) 447.

\bibitem{Petrini:1997kk}
M.~Petrini,
``Infrared stability of N = 4 super Yang--Mills theory'',
Phys.\ Lett.\   \underline{B404} (1997) 66 [arXiv:hep-th/9704004].

\bibitem{Horava:2008ih}
P.~Ho\v{r}ava,
``Membranes at quantum criticality'',
JHEP \underline{0903} (2009) 020 [arXiv:0812.4287 [hep-th]].

\bibitem{Horava:2009uw}
P.~Ho\v{r}ava,
``Quantum gravity at a Lifshitz point'',
Phys.\ Rev.\   \underline{D79} (2009) 084008 [arXiv:0901.3775 [hep-th]].

\bibitem{Orlando:2009en}
D.~Orlando and S.~Reffert,
``On the renormalizability of Ho\v{r}ava--Lifshitz-type gravities'',
Class.\ Quant.\ Grav.\  \underline{26} (2009) 155021
[arXiv:0905.0301 [hep-th]].

\bibitem{wu}
F.-W. Shu and Y.-S. Wu, ``Stochastic quantization of the Ho\v{r}ava
gravity'' [arXiv:0906.1645 [hep-th]].

\bibitem{Charmousis:2009tc}
C.~Charmousis, G.~Niz, A.~Padilla and P.M.~Saffin,
``Strong coupling in Ho\v{r}ava gravity'',
JHEP \underline{0908} (2009) 070
[arXiv:0905.2579 [hep-th]].

\bibitem{pang}
M. Li and Y. Pang, ``A trouble with Ho\v{r}ava--Lifshitz gravity'',
JHEP \underline{0908} (2009) 015
[arXiv:0905.2751 [hep-th]].


\bibitem{blas}
D. Blas, O. Pujolas and S. Sibiryakov, ``On the extra mode and inconsistency
of Ho\v{r}ava gravity'', JHEP \underline{0910} (2009) 029
[arXiv:0906.3046 [hep-th]];
``A healthy extension of Ho\v{r}ava gravity'' [arXiv:0909.3525 [hep-th]];
``Comment on strong coupling in extended Ho\v{r}ava--Lifshitz gravity''
[arXiv:0912.0550 [hep-th]].

\bibitem{koyama}
K. Koyama and F. Arroja, ``Pathological behavior of the scalar graviton in
Ho\v{r}ava--Lifshitz gravity'' [arXiv:0910.1998 [hep-th]].

\bibitem{hamilt}
R. Hamilton, ``Three-manifolds with positive Ricci curvature'', J. Diff. Geom.
\underline{17} (1982) 255.

\bibitem{perel}
G. Perelman, ``The entropy formula for the Ricci flow and its geometric applications''
[math.DG/0211159]; ``Ricci flow with surgery on three-manifolds''
[math.DG/0303109]; ``Finite extinction time for the solutions to the Ricci flow
on certain three-manifolds'' [math.DG/0307245].

\bibitem{yau}
H.-D. Cao, B. Chow, S.-C. Chu and S.-T. Yau eds., {\em Collected Papers on Ricci Flow},
Series in Geometry and Topology, \underline{37}, International Press, Somerville, 2003.

\bibitem{tian}
J.W. Morgan and G. Tian, {\em Ricci Flow and the Poincar\'e Conjecture}, Clay
Mathematics Monographs, Amer. Math. Soc., Cambridge, 2007 [math.DG/0607607].

\bibitem{Friedan:1980jm}
D.H.~Friedan,
``Nonlinear models in two + epsilon dimensions'',
Phys. Rev. Lett. \  \underline{45} (1980) 1057;
``Nonlinear models in two + epsilon dimensions'',
Ann. Phys.\  \underline{163} (1985) 318.

\bibitem{Schmidhuber:1994bv}
C.~Schmidhuber and A.A.~Tseytlin,
``On string cosmology and the RG flow in 2-d field theory'',
Nucl.\ Phys.\   \underline{B426} (1994) 187
[arXiv:hep-th/9404180].

\bibitem{Bakas:2006bz}
I.~Bakas, D.~Orlando and P.M.~Petropoulos,
``Ricci flows and expansion in axion-dilaton cosmology'',
JHEP \underline{0701} (2007) 040
[arXiv:hep-th/0610281].

\bibitem{Sfetsos:2006}
  M.~Cvetic, G.W.~Gibbons, H.~Lu and C.N.~Pope,
  ``Cohomogeneity one manifolds of $\mathrm{Spin}(7)$ and $G(2)$ holonomy'',
  Phys.\ Rev.\   \underline{D65} (2002) 106004
  [arXiv:hep-th/0108245]; I. Bakas and K. Sfetsos, unpublished work (2004); G.W. Gibbons, private
communication (2005).
%; K. Sfetsos, unpublished work (2006), see http://www.cc.uoa.gr/~papost/SFETSOS.pdf

\bibitem{Bourliot:2009fr}
F.~Bourliot, J.~Estes, P.M.~Petropoulos and Ph.~Spindel,
``Gravitational instantons, self-duality and geometric flows''
[arXiv:0906.4558 [hep-th]].

\bibitem{harhaw}
J.B. Hartle and S.W. Hawking, ``Wave function of the universe'',
Phys, Rev. \underline{D28} (1983) 2960.

\bibitem{Misner:1969}
C.W.~Misner, ``Mixmaster universe'',
Phys.\ Rev.\ Lett.\  \underline{22} (1969) 1071;
``Quantum cosmology 1'',
Phys.\ Rev.\  \underline{186} (1969) 1319.

\bibitem{bkl}
V.A. Belinskii, I.M. Khalatnikov and E.M. Lifshitz, ``Oscillatory approach to a
singular point in the relativistic cosmology'', Adv. Phys. \underline{19} (1970) 525;
``A general solution of the Einstein
equations with a time singularity'', Adv. Phys. \underline{31} (1982) 639.

\bibitem{Barrow:1981sx}
J.D.~Barrow,
``Chaotic behaviour in general relativity'',
Phys.\ Rept.\  \underline{85} (1982) 1.

\bibitem{Bakas:2009ku}
I.~Bakas, F.~Bourliot, D.~L\"ust and M.~Petropoulos,
``Mixmaster universe in Horava--Lifshitz gravity'', Class. Quant. Grav. \underline{27} (2010) 045013
[arXiv:0911.2665 [hep-th]].

\bibitem{Myung:2009if}
Y.S. Myung, Y.W. Kim, W.S. Son and Y.J. Park,
``A non-chaotic mixmaster universe in the $z=2$ Ho\v{r}ava--Lifshitz gravity''
[arXiv:0911.2525 [gr-qc]].

\bibitem{Carroll}
L. Carroll, \emph{Alice's Adventures in Wonderland}, MacMillan, London, 1865.

\bibitem{misner6}
C.W. Misner, K.S. Thorne and J.A. Wheeler, {\em Gravitation}, Freeman, San Francisco,
1973.

\bibitem{henneaux}
M. Henneaux, A. Kleinschmidt and G. Lucena Gomez, ``A dynamical inconsistency of
Ho\v{r}ava gravity'' [arXiv:0912.0399 [hep-th]].

\bibitem{cs}
S. Deser, R. Jackiw and S. Templeton, ``Topologically massive gauge theories'',
Ann. Phys. \underline{140} (1982) 372; Erratum-ibid. \underline{185}
(1988) 406; ``Three-dimensional massive gauge theories'', Phys. Rev. Lett.
\underline{48} (1982) 975.

\bibitem{Eguchi:1980jx}
T.~Eguchi, P.B.~Gilkey and A.J.~Hanson,
``Gravitation, gauge theories and differential geometry'',
Phys.\ Rept.\  \underline{66} (1980) 213.

\bibitem{Gibbons:1979xn}
G.W.~Gibbons and C.N.~Pope,
``The positive action conjecture and asymptotically Euclidean metrics In
quantum gravity'',
Commun.\ Math.\ Phys.\  \underline{66} (1979) 267.

\bibitem{Hamilton}
R.S. Hamilton, ``The Ricci flow on surfaces'' in {\em Mathematics and General
Relativity}, Contemp. Math. \underline{71},
p. 237--262, Amer. Math. Soc., Providence, 1988.

\bibitem{Kisisel:2008jx}
A.U.O.~Kisisel, O.~Sarioglu and B.~Tekin,
``Cotton flow'',
Class.\ Quant.\ Grav.\  \underline{25} (2008) 165019
[arXiv:0803.1603 [hep-th]].

\bibitem{Isenberg:1992}
J. Isenberg and M. Jackson, ``Ricci flow of locally homogeneous geometries on closed
manifolds'', J. Diff. Geom. \underline{35} (1992) 723.


\bibitem{Darboux}
G. Darboux, ``M\'emoire sur la th\'eorie des coordonn\'ees curvilignes et des
syst\`emes orthogonaux'', Ann. Ec. Normale Sup\'er. \underline{7} (1878) 101.

\bibitem{halph}
G.-H. Halphen, ``Sur un syst\`eme d'\'equations diff\'erentielles'',
C.R. Acad. Sc. Paris \underline{92} (1881) 1001;
``Sur certains syst\`emes d'\'equations diff\'erentielles'',
C.R. Acad. Sc. Paris \underline{92} (1881) 1004.

\bibitem{Takhtajan:1992qb}
L.A.~Takhtajan,
``A simple example of modular forms as tau functions for integrable
equations'',
Theor.\ Math.\ Phys.\  \underline{93} (1992) 1308.

\bibitem{Atiyah1}
M.F.~Atiyah and N.J.~Hitchin,
``Low energy scattering of non-abelian monopoles'',
Phys.\ Lett.\  \underline{A107} (1985) 21;
{\em The Geometry and Dynamics of Magnetic Monopoles}, Porter Lectures,
Princeton University Press, Princeton, 1988.

\bibitem{Atiyah2}
G.W. Gibbons and N.S. Manton, ``Classical and quantum dynamics of BPS
monopoles'', Nucl. Phys. \underline{B274} (1986) 183.

\bibitem{nutku}
Y. Nutku and P. Baekler, ``Homogeneous, anisotropic three--manifolds of
topologically massive gravity'', Ann. Phys. \underline{195} (1989) 16.

\bibitem{calca1}
G. Calcagni, ``Cosmology of the Lifshitz universe'', JHEP \underline{0909}
(2009) 112 [arXiv:0904.0829 [hep-th]].

\bibitem{calca2}
E. Kiritsis and G. Kofinas, ``Ho\v{r}ava--Lifshitz cosmology'', Nucl. Phys.
\underline{B821} (2009) 467 [arXiv:0904.1334 [hep-th]].

\bibitem{calca3}
R.~Brandenberger, ``Matter Bounce in Ho\v{r}ava--Lifshitz Cosmology'',
Phys.\ Rev.\   \underline{D80} (2009) 043516
[arXiv:0904.2835 [hep-th]].

\bibitem{nuts}
G.W. Gibbons and S.W. Hawking, ``Classification of gravitational instanton symmetries'',
Commun. Math. Phys. \underline{66} (1979) 291.

\bibitem{perry}
G.W. Gibbons, S.W. Hawking and M.J. Perry, ``Path integrals and the indefiniteness of
the gravitational action'', Nucl. Phys. \underline{B138} (1978) 141.

\bibitem{black1}
A. Kehagias and K. Sfetsos, ``The black hole and FRW geometries in non-relativistic
gravity'', Phys. Lett. \underline{B678} (2009) 123 [arXiv:0905.0477 [hep-th]].

\bibitem{black2}
H. Lu, J. Mei and C.N. Pope, ``Solutions to Ho\v{r}ava gravity'', Phys. Rev. Lett.
\underline{103} (2009) 091301 [arXiv:0904.1595 [hep-th]].

\bibitem{black3}
E. Kiritsis and G. Kofinas, ``On Ho\v{r}ava--Lifshitz black holes''
[arXiv:0910.5487 [hep-th]].

\bibitem{gurses}
M. G\"urses, ``Killing vector fields in three dimensions: a method to solve
massive gravity field equations'' [arXiv:1001.1039 [gr-qc]].

\bibitem{classif}
D.K. Chow, C.N. Pope and E. Sezgin, ``Classification of solutions in topologically
massive gravity'' [arXiv:0906.3559 [hep-th]].

\bibitem{bach}
R. Bach, ``Zur Weylschen Relativit\"atstheorie und der Weylschen Erweiterung des
Kr\"ummungsbegriffs'', Math. Zeitschr. \underline{9} (1921) 110.

\bibitem{calabi}
E. Calabi, ``Extremal K\"ahler metric'' in {\em Seminar on Differential Geometry},
ed. S.-T. Yau, Annals of Mathematical Studies, \underline{102}, Princeton University Press,
Princeton, 1982.

\bibitem{townsend}
E.A. Bergshoeff, O. Holm and P.K. Townsend, ``Massive gravity in three dimensions'',
Phys. Rev. Lett. \underline{102} (2009) 201301 [arXiv:0901.1766 [hep-th]].

\bibitem{sourd}
I. Bakas and C. Sourdis, work in progress.

\end{thebibliography}
\end{document}